\documentclass[supplementary,11pt]{article}
\usepackage{amsmath,bbm,tikz,xr}

\usepackage[square,sort,comma,numbers]{natbib}
\usepackage{times} 
\usepackage{latexsym,epsfig,amssymb,amsmath,amsfonts,graphicx,amsthm,ifthen,pifont,comment,enumerate,setspace,multirow,color}
\usepackage{mhequ}
\usepackage[ruled,vlined]{algorithm2e}
\SetAlFnt{\small}
\SetAlCapFnt{\large}
\SetAlCapNameFnt{\large}
\usepackage{algorithmic}
\algsetup{linenosize=\tiny}

\usepackage{tikz}
\usepackage{natbib}
\usepackage{epsfig}
\usepackage{bm}
\usepackage{makeidx}
\usepackage{natbib}
\usepackage{color}
\usepackage{framed}
\usetikzlibrary{bayesnet}
\usepackage{float}
\usepackage{multicol,booktabs,colortbl,tabularx}
\usepackage{JASA_manu}
\usepackage{bm}

\usepackage[numbers]{natbib}
\usepackage[final]{hyperref} 
\hypersetup{
	colorlinks=true,       
	linkcolor=blue,        
	citecolor=blue, 	     						  
	filecolor=magenta,     
	urlcolor=blue         
}
\def\begmat{\left(\begin{array}}\def\endmat{\end{array}\right)}

\def\bi{\begin{itemize}\setlength{\itemsep}{0pt}} \def\ei{\end{itemize}}

\def\bl{\begin{list}{\labelitemi}{\leftmargin=1em}\setlength{\itemsep}{-2.5pt}}  \def\el{\end{list}}
\def\bn{\begin{enumerate}} \def\en{\end{enumerate}}
\def\bt{\begin{table}[h]} \def\et{\end{table}}
\def\bc{\begin{center}} \def\ec{\end{center}}

\theoremstyle{plain}

\theoremstyle{plain}

\theoremstyle{remark}

\theoremstyle{plain}

\newcommand \be{\begin{equs}}
\newcommand \ee{\end{equs}}

\begin{document}
\thispagestyle{empty}



\title{
\textbf{Bayesian Nonlinear Models for Repeated Measurement Data: An Overview, Implementation, and Applications}
}
\author{Se Yoon Lee\\
\begin{small}
stat\_math@tamu.edu
\end{small}
\\
\begin{small}
Department of Statistics, Texas A\&M University, College Station, Texas, USA
\end{small}
}
\date{}

\maketitle
\begin{abstract}
\noindent 
\baselineskip=17pt
Nonlinear mixed effects models have become a standard platform for analysis when data is in the form of continuous and repeated measurements of subjects from a population of interest, while temporal profiles of subjects commonly follow a nonlinear tendency. While frequentist analysis of nonlinear mixed effects models has a long history, Bayesian analysis of the models has received comparatively little attention until the late 1980s due primarily to the time-consuming nature of Bayesian computation. Since the early 1990s Bayesian approaches for the models began to emerge to leverage rapid developments in computing power, and recently, have received significant attention due to (1) superiority to quantify the uncertainty of parameter estimation; (2) utility to incorporate prior knowledge into the models; and (3) flexibility to match exactly the increasing complexity of scientific research arising from diverse industrial and academic fields. This review article presents an overview of modeling strategies to implement Bayesian approaches for the nonlinear mixed effects models, ranging from designing a scientific question out of real-life problems to practical computations.
\end{abstract}
\vspace*{.3in}

\newpage

\section{Introduction}\label{sec:Introduction}
One of the common challenges in biological, agricultural, environmental, epidemiological, financial, and medical applications is to make inferences on characteristics underlying profiles of continuous, repeated measures data from multiple individuals within a population of interest \citep{sterba2014fitting,mcardle1986latent,cook1983design,mehta2000putting}. By `repeated measures data' we mean the data type generated by observing a number of individuals repeatedly under differing experimental conditions where the individuals are assumed to constitute a random sample from a population of interest. A common type of repeated measures data is longitudinal data such that the observations are ordered by time \citep{zeger1992overview,diggle2002analysis}.

Linear mixed effects models for repeated measures data have become popular due to their straightforward interpretations, flexibility allowing correlation structure among the observations, and utility accommodating unbalanced and multi-level data structure (i.e., clustered designs that vary among individuals) \citep{demidenko2013mixed,snijders2011multilevel}. The modeling framework is also intuitively appealing: the central idea that individuals' responses are governed by a linear model with slope or intercept parameters that vary among individuals seems to be appropriate in many scientific problems (for e.g., see  \citep{goldstein2011multilevel,raudenbush2002hierarchical}). It also allows practitioners to test and evaluate multivariate causal relationships by conducting regression analysis at the population level. By preserving the multi-level structure in a single model, estimation or prediction for the analyses can take advantage of information borrowing \citep{efron2010future}. 

For many applications, researchers often want to theorize that time courses of individual response commonly follow a certain nonlinear function dictated by a finite number of parameters \citep{sheiner1972modelling}. These nonlinear functions are based on reasonable scientific hypotheses, typically represented as a differential equation system. By tuning the parameters, the shape of the function in terms of curvature, steepness, scale, height, etc, may change, which is used as the rationale behind describing heterogeneity between subjects. Nonlinear mixed effects models, also referred to as hierarchical nonlinear models, have gained broad acceptance as a suitable framework for these purposes \citep{lindstrom1990nonlinear,davidian2003nonlinear,davidian2017nonlinear}. Analyses based on this model are now routinely reported in various industrial problems, which is, in part, enabled by the breakthrough development of software \citep{beal1980nonmem,stan,fidler2019nonlinear,wang2016tutorial,stegmann2018nonlinear}. The excellent books and review papers were published by  \citep{davidian2003nonlinear,davidian2017nonlinear,vonesh1996lin}. Although their works were published more than 20 years ago, they still provide statisticians, programmers, and researchers with many pedagogical insights about the modeling framework, implementations, and practical applications of using the nonlinear mixed effects models.

While frequentist analysis of nonlinear mixed effects models has a long history, Bayesian analysis for the models was a relatively dormant field until the late 1980s. This is due primarily to the time-consuming nature of the calculations required for Bayesian computation to implement a Bayesian model \citep{lee2007structural}. Since the early 1990s Bayesian approaches began to re-emerge, motivated both by exploitation of rapid developments in computing power and by the growing desire to quantify the uncertainty associated with parameter estimation and prediction \citep{dellaportas1993bayesian,bush1996semiparametric,zeger1991generalized}. Since then, Bayesian nonlinear mixed effects models, also called Bayesian hierarchical nonlinear models, have been extensively used in diverse industrial and academic researches, endowed with new computational tools providing a far more flexible framework for statistical inference matching exactly the increasing complexity of scientific research \citep{brooks2003bayesian,bennett1996comparison,wakefield1996bayesian,gelman1996physiological,lee2020estimation,lee2021bayesian}.

The objective of this article is to present an updated look at the Bayesian nonlinear mixed effects models. Although the works of \citep{davidian2003nonlinear,davidian2017nonlinear} discuss some of the Bayesian approaches for the nonlinear mixed effects models, the main perspective adopted in the works is much more oriented to the frequentist framework, and prior distributions and Bayesian computing strategy explained in the works are quite outdated. In the literature, it is striking that very few research works provide an updated overview of the Bayesian methodologies on the nonlinear mixed effects models. Motivated this, in this article, we endeavor to presents an overview of modeling strategies to implement Bayesian approaches for the nonlinear mixed effects models, ranging from designing a scientific question out of real-life problems to practical computations. The novelty of this paper is as follow:

\begin{itemize}
\item[I.]  Guidance for Bayesian workflow to solve a real-life problem is provided for domain experts to facilitate efficient collaboration with quantitative researchers;
\item[II.]  Recently developed prior distributions and Bayesian computation techniques for a basic model and its extensions are illustrated for statisticians to develop more complex models built on the basic model;
\item[III.] Illustrated methodologies can be directly exploited in diverse applications, ranging from small data to big data problems, for quantitative researchers, modeling scientists, and professional programmers working in diverse industries.
\end{itemize}

This article is organized as follows. In Section \ref{sec:Trend and workflow}, we explore trends and workflow on the use of Bayesian nonlinear mixed effects. In Section \ref{sec:Bayesian Nonlinear mixed effects model}, we motivate readers to understand why it is necessary to use the Bayesian nonlinear mixed effects model by illustrating four real-life problems, which will be conceptualized as a statistical problem. To solve the statistical problem, we suggest a basic version of the Bayesian nonlinear mixed effects models in Section \ref{sec:The model}, and its likelihood is analyzed in Section \ref{sec:Likelihood} wherein frequentist computations are briefly discussed. Section \ref{sec:Bayesian inference and implementation} describes modern Bayesian computation strategies to implement the basic model. Popularly used prior distributions are presented in Section \ref{sec:Prior options}. Section \ref{sec:Model selection} discusses model selection, and Section \ref{sec:Extensions and recent developments} reviews recent advances and extensions that build on the basic model. Finally, Section \ref{sec:Discussion} concludes the article.

\section{Trends and workflow of Bayesian nonlinear mixed effects models}\label{sec:Trend and workflow}
\subsection{Rise in the use of Bayesian approaches for the nonlinear mixed effects models}\label{subsec:A rise of Bayesian approaches}
\begin{figure}[h!]
\centering
\includegraphics[width=\textwidth]{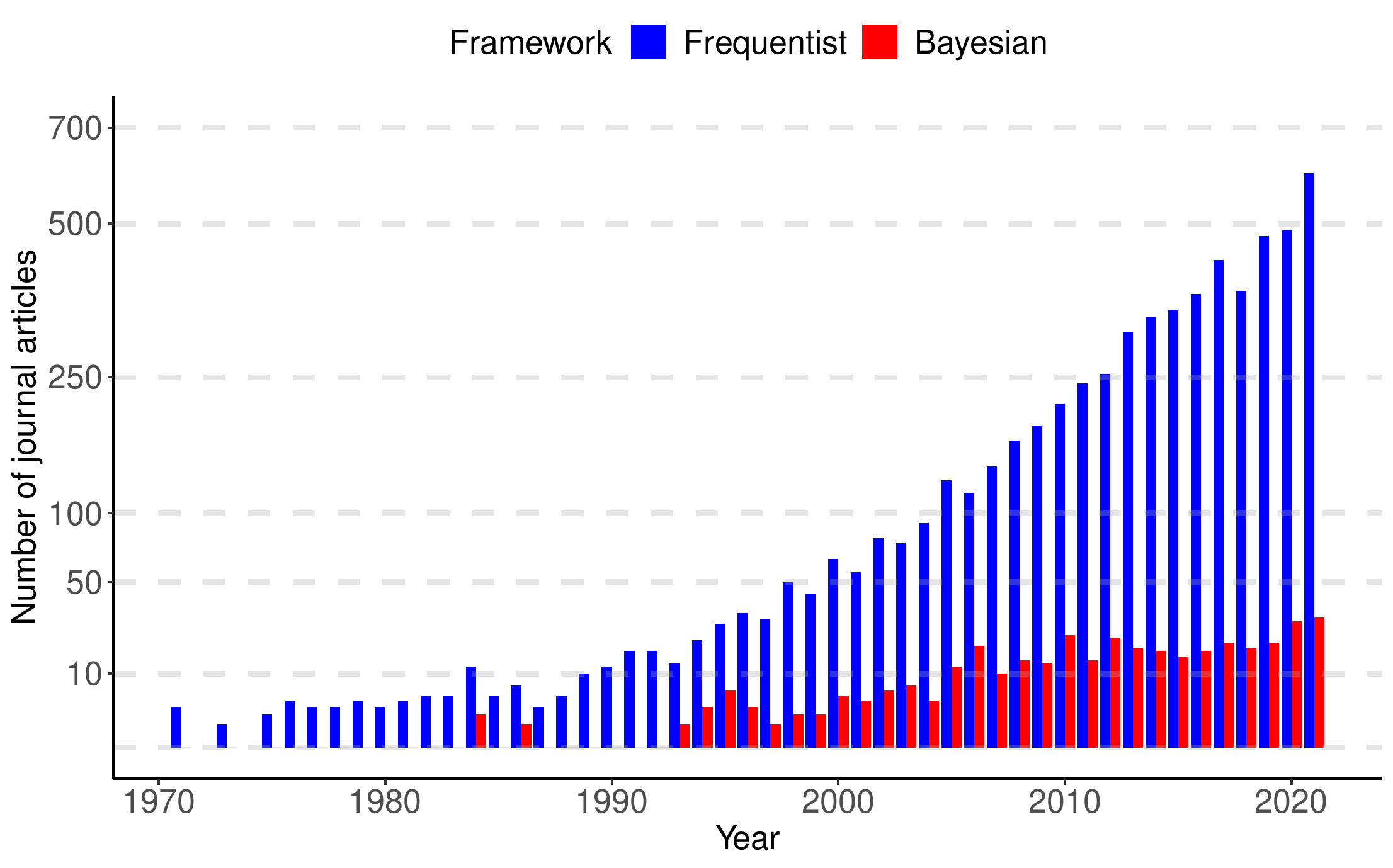}
\caption{
Publication trend of the nonlinear mixed-effect models categorized by frequentist and Bayesian frameworks. the x-axis represents the year from January 1st, 1970 to December 31st, 2021. The value on the y-axis is the number of published articles in each year. (Data sources: PubMed.gov)
}
\label{fig:Publication_trend_NLME}
\end{figure}

As of January 1970 to December 2021, PubMed.gov (\href{https://pubmed.ncbi.nlm.nih.gov/}{pubmed.ncbi.nlm.nih.gov/-}) searched of ``nonlinear mixed-effect models" yielded $6,288$ publications. Among the published articles, nearly 94\% of works used frequentist approaches ($5,929$ articles), while only 6\% of works adopted Bayesian approaches ($359$ articles). Figure \ref{fig:Publication_trend_NLME} displays a bar plot based on the published articles, categorized by the frequentist and Bayesian approaches over time. In the panel, it is observed that, until the late 1980s, the Bayesian research was nearly dormant, but since the early 1990s, Bayesian works begin to re-emerge, and the gap between frequentist and Bayesian works is becoming gradually narrower as time evolves. 

The dormancy of the Bayesian approaches until the late 1980s is mainly due to the time-consuming nature of the calculations based on sampling scheme which was previously impossible by the limitation of computing power. Fortunately, a breakthrough in computer processors (for e.g., Am386 in 1991, Pentium Processor in 1993, etc) took place in the early 1990s, driving the computing revolution to solve computationally intense problems, and this has given the statistical community the ability to solve statistical questions by using Bayesian methods. This timeline is also aligned with the widespread of Markov chain Monte Carlo (MCMC) sampling techniques in the Bayesian community \citep{hammersley2013monte,green2015bayesian}. Since then, the Bayesian community has been gradually gaining the momentum to leverage the rapidly growing developments of computing power, and now, assorted Bayesian software packages (e.g., \textsc{JAGS} \citep{plummer2003jags}, \textsc{BUGS} \citep{lunn2009bugs}, and \textsc{Stan} \citep{stan}) are available for researchers to answer scientific questions arising from industrial and academic research. 

To understand the rise of the Bayesian approaches, we want to first understand what will be some advantages of using Bayesian methods over frequentist methods in the context of nonlinear mixed effects models. As the primary focus of this review paper is to provide the readers with some insight on methodologies and practical implementation of using Bayesian approaches, our comparison and exposition below are described from an operational viewpoint. Table \ref{table:Comparison of modeling strategies} summarizes modeling strategies of using the frequentist and Bayesian approaches for the nonlinear mixed effects models. Broadly speaking, the usual estimation method of the frequentist computation is optimization, while that of the Bayesian computation is sampling. Normally, it is known that the former is much faster than the latter. This is not surprising because a sampling scheme, by its nature, needs to explore a wide range of the parameter space, whereas the optimization only needs to find the best point estimate, which is often described by the maximum likelihood estimate. In many practical problems, widely used frequentist optimization algorithms are the first-order approximation \citep{beal1982estimating}, Laplace approximation \citep{wolfinger1993laplace}, and stochastic approximation of expectation-maximization algorithm \citep{delyon1999convergence}. They will be briefly discussed in Subsection \ref{subsec:Likelihood from vector-form (b)}. As for the Bayesian sampling algorithms, combinations of Gibbs sampler \citep{lee2021gibbs}, Metropolis-Hastings algorithm
\citep{robert1999metropolis}, Hamiltonian Monte Carlo
\citep{neal2011mcmc}, and No-U-Turn sampler
\citep{hoffman2014no} are popularly used, among many others \citep{dwivedi2018log,ma2019sampling,neal2003slice}. We explain these in detail in Section \ref{sec:Bayesian inference and implementation}.

\begin{table}[H] 
\caption{ 
Comparison of modeling strategies using in Frequentist and Bayesian approaches for the nonlinear mixed effects models from implementational viewpoint.\label{table:Comparison of modeling strategies}}
\newcolumntype{C}{>{\centering\arraybackslash}X}
\begin{tabularx}{\textwidth}{CCC}
\toprule
\textbf{Characteristic}	& \textbf{Frequentist}	& \textbf{Bayesian}\\
\midrule
Estimation objective & Maximize a likelihood \citep{davidian2003nonlinear,davidian2017nonlinear,vonesh1996lin} & Sample from a posterior \citep{lee2007structural,wakefield1996bayesian,lee2020estimation}
\\
\\
Computation algorithm &  First-order approximation \citep{beal1982estimating}, Laplace approximation \citep{wolfinger1993laplace}, and stochastic approximation of EM algorithm \citep{delyon1999convergence}
& 
Gibbs sampler \citep{lee2021gibbs}, 
Metropolis-Hastings algorithm
\citep{robert1999metropolis}, 
Hamiltonian Monte Carlo
\citep{neal2011mcmc}, and
No-U-Turn sampler
\citep{hoffman2014no}
\\
\\
Software		& \textsc{SAS} \citep{sas1999sas}, \textsc{NONMEM} \citep{beal1992nonmem},\textsc{Monolix} \citep{lavielle2005monolix},\textsc{nlmixr} \citep{fidler2019nonlinear}  			& 
\textsc{JAGS} \citep{lunn2000winbugs}, \textsc{BUGS} \citep{lunn2009bugs}, \textsc{Stan} \citep{stan}, \textsc{brms} \citep{burkner2017brms}
\\
\\
Advantages & Relatively fast computation speed, the objectivity of inference results, and
widely available software packages to implement complex models & 
Inherent uncertainty quantification, better small sample performance, and  utility of prior knowledge
\\
\\
Disadvantages & 
Need large-sample theory for uncertainty quantification and cannot incorporate prior knowledge
 	& Need high computing power for big data and require Bayesian expertise in prior elicitation
\\
\bottomrule
\end{tabularx}
\end{table}

The stark difference of using frequentist and Bayesian approaches may be the procedure of describing an uncertainty underlying the parameter estimation for the nonlinear mixed effects models. Here, the parameter which is of primary interest is the population-level parameters (also called fixed effects), typical values for the individual-level parameters. In many cases, frequentist $95\%$ confidence intervals for the parameters of the models are constructed by assuming that asymptotic normality of maximum likelihood estimator holds in a finite sample study, which is actually the most accurate in large sample scenario \citep{chernoff1956large}. Most frequentist software packages, such as \textsc{NONMEM} \citep{beal1992nonmem,lindstrom1990nonlinear}, \textsc{Monolix} \citep{lavielle2005monolix}, and \textsc{nlmixr} \citep{fidler2019nonlinear}, by default, may print out a $95\%$ confidence interval of the form, ``\textsl{Estimate}  $\pm$ 1.96 $\times$ \textsl{Standard Error}", or some transformation of the lower and upper bounds, if necessary, such that the \textsl{Standard Error} is calculated by using (observed) Fisher information matrix \citep{wand2007fisher,breslow1993approximate,kang2012standard}. Using such a scheme in small sample studies is highly likely to overlook the gap between the reality of the data and the idealistic asymptotic situation.

In contrast, as for the Bayesian approaches, the large-sample theory is not needed for the uncertainty quantification, and the procedure to obtain $95\%$ posterior credible intervals is a lot easier than obtaining $95\%$ confidence intervals (See Chapter 4 of \citep{gelman1995bayesian}). Furthermore, Bayesian credible intervals based on percentiles of posterior samples allow for a strongly skewed distribution, wherein frequentist confidence intervals (based on large-sample theory) may induce a non-negligible approximating error due to the deviation from the asymptotic normality. Along with that, Bayesian methods are highly appreciated when researchers want to incorporate prior knowledge from previous studies into the model so that posterior inference provides the researchers with an updated view on the problem, possibly with a more accurate estimation. Using prior information would be particularly useful in small-sample contexts \citep{smid2020bayesian,rupp2004bayes}. For example, in medical device clinical trials, some opportunities and challenges in developing a new medical device are: (i) there is often a great deal of prior information for a medical device; (ii) a medical device evolves in relatively small increments from previous generations to a new generation; (iii) there are only a few numbers of patients for the trials; and (iv) companies need to make a rational decision promptly to reduce cost. In those settings, Bayesian methods have been demonstrated to be suitable, and their proper use is guided by Food and Drug Administration \citep{bonangelino2011bayesian,campbell2017bayesian}.


\subsection{Bayesian workflow}\label{subsec:Bayesian workflow}
\begin{figure}[h!]
\centering
\includegraphics[width=\textwidth]{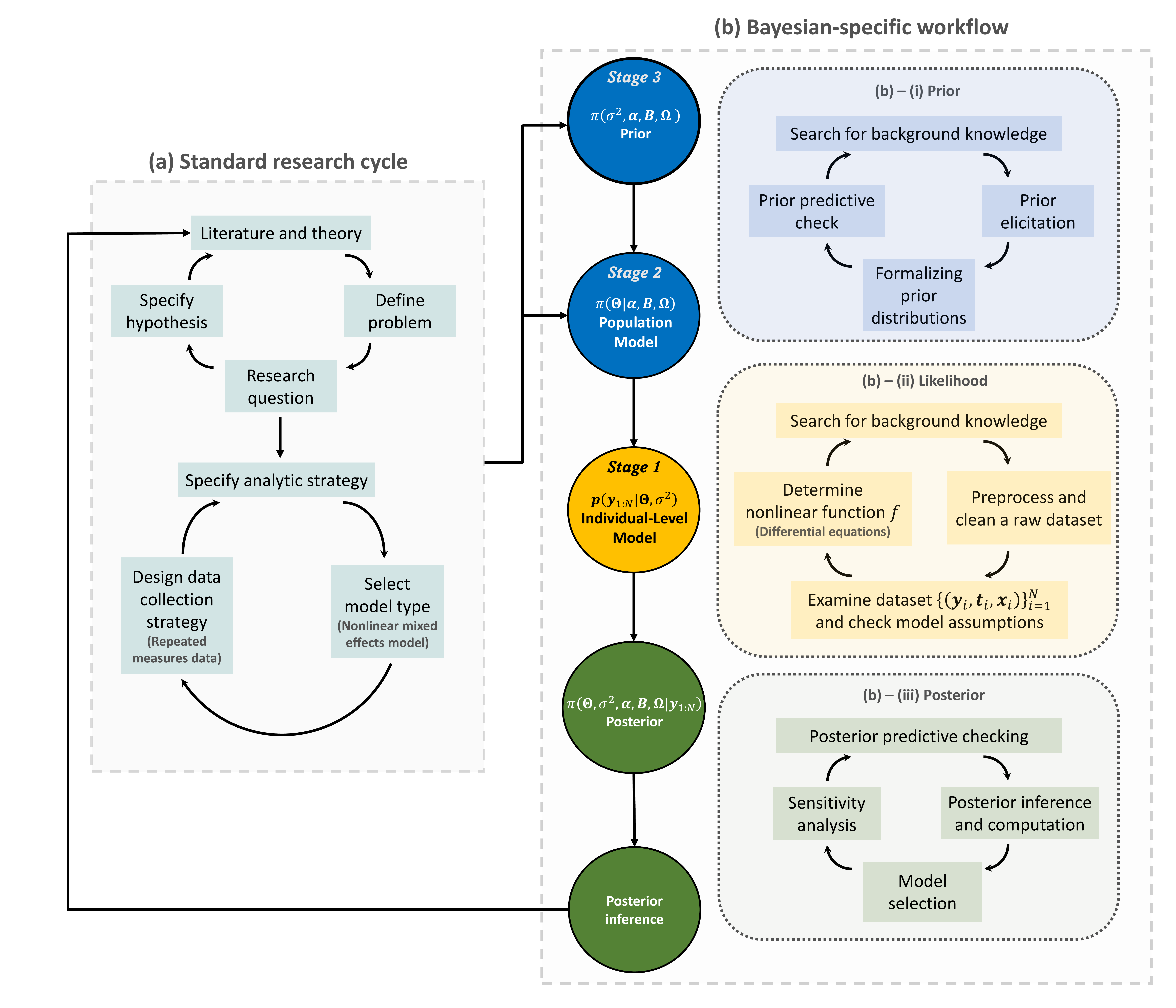}
\caption{
The Bayesian research cycle. A research cycle using Bayesian nonlinear mixed effects model comprises two steps: (a) standard research cycle and (b) Bayesian-specific workflow. Standard research cycle involves literature review, defining a problem and specifying the research question and hypothesis. Bayesian-specific workflow comprises three sub-steps: (b) - (i) formalizing prior distributions based on background knowledge and prior elicitation; (b) - (ii) determining the likelihood function based on a nonlinear function $f$; and (b) - (iii) making a posterior inference. The resulting posterior inference can be used to start a new research cycle. Distributions for prior, likelihood, and posterior are colored in blue, yellow, and violet, respectively. $\bm{\Theta}$, model matrix; $\sigma^{2}$, error variance parameter; $\bm{\alpha}$, intercepts; $\textbf{B}$, coefficient matrix; $\bm{\Omega}$, covariance matrix; $p(.)$, probability distribution; $\pi(.)$, prior or posterior probability distribution;  $\{(\textbf{y}_{i},\textbf{t}_{i},\textbf{x}_{i})\}_{i=1}^{N}$, data.
}
\label{fig:Bayesian_Workflow}
\end{figure}

We outline the first two steps in the Bayesian workflow of using Bayesian nonlinear mixed effects models described in Figure \ref{fig:Bayesian_Workflow}. The panel includes some mathematical notations that are consistently used throughout the paper. These notations will be clearly understood later. The aim of our explanation at this point is to provide readers with a blueprinted plan to implement Bayesian modeling strategies for the nonlinear mixed effects models. We assume that readers are familiar with basic concepts and generic workflow in Bayesian statistics; see \citep{hoff2009first,o2004bayesian,gelman1995bayesian} for those basic concepts and refer to the review paper by \citep{van2021bayesian} and references therein for detailed concepts and general terminologies used in workflow, such as prior and posterior predictive checks, and prior elicitation, etc. 

The first step of the Bayesian research cycle is (a) standard research cycle \citep{blaxter2010research,neuman2016understanding}. Some beginning activities at this step involve reviewing literature, defining a problem, and specifying a research question and a hypothesis. After that, researchers specify which analytic strategy would be taken to solve the research question and suggest possible model types, followed by data collection. The data type arising in this process may include a response variable and some covariates that are grouped longitudinally, which then formulates repeated measures data of a population of interest. Furthermore, if there appears to be some nonlinear temporal tendency at each subject, then a possible model type for the analysis is a nonlinear mixed effects model \citep{pinheiro2006mixed,davidian2017nonlinear}.

The second step of the Bayesian research cycle is (b) Bayesian-specific workflow. Logically, the first thing to do at this step is to determine prior distributions (see Step (b) - (i) in Figure \ref{fig:Bayesian_Workflow}). The selection of priors is often viewed as one of the most crucial choices that a researcher makes when implementing a Bayesian model as it can have a substantial impact on the final results \citep{gelman2017prior}. As exemplified earlier in the context of Bayesian medical device trials, using a prior in small sample studies may improve the estimation accuracy, but unthoughtful choice of priors would lead to a significant bias in estimation. Prior elicitation effort would require Bayesian expertise to formulate domain expert’s knowledge in a probabilistic form \citep{garthwaite2005statistical}. Strategies for prior elicitation include asking domain experts to provide suitable values for the hyperparameters of the prior \citep{o2006uncertain,howard2000proof}. After prior is specified, one can check the appropriateness of the priors through prior predictive checking process \citep{levy2011bayesian}. For almost all practical problems, prior distribution of Bayesian nonlinear mixed effect models can be hierarchically represented as follow: (1) a prior for the parameters used in likelihood, often called `population-level model' in the literature of mixed effects modeling; and (2) a prior for the parameters used in the population-level model and the likelihood. It is important to note that, the former type of prior distribution (that is, (1)) is also a requirement to implement frequentist approaches for the nonlinear mixed effects model, as a name of `distribution for random effects'. Essentially, the defining factor of the Bayesian framework is the latter type of prior distribution (that is, (2)), which is fixed in the frequentist framework, as a name of `fixed effects'. Some prior options of the latter type will be discussed in Section \ref{sec:Prior options}.

The second task is to determine the likelihood function (see Step (b) - (ii) in the panel). At this time, the raw dataset collected in (a) standard research cycle should be cleaned and preprocessed. Before embarking on more serious statistical modeling, it is a common practice to get some insight about the research question via exploratory data analysis and have a discussion with domain experts such as clinical pharmacologists, clinicians, physicians, engineers, etc. To some extent, eventually, all these efforts are to determine a nonlinear function (denoted as $f$ in this paper) that best represents the temporal profiles of all subjects. This nonlinear function is a \emph{known} function because it should be specified by researchers. In other words, the branch of the nonlinear mixed effects models belongs to parametric statistics. However, one technical challenge is that, in many problems, such a nonlinear function is represented as a solution of a differential equation system \citep{wang2014estimating,botha2021particle}, and therefore there is no guarantee that we can conveniently work with a closed-form expression of the nonlinear function. For example, if researchers wish to work with nonlinear differential equations \citep{fucik2014nonlinear,verhulst2006nonlinear}, then some approximation via differential equation solver \citep{cohen1996cvode,dormand1980family} may be needed to calculate the nonlinear function. As such, most software packages dedicated to implementing nonlinear mixed effect model, or more generally, Bayesian hierarchical model, are equipped with several built-in differential equation solvers \citep{beal1992nonmem,stan,margossiantorsten}. For instance, visit the website (\href{https://mc-stan.org/docs/2_29/stan-users-guide/ode-solver.html}{mc-stan.org/-}) to see the functionality supported in Stan \citep{stan}.

Finally, the likelihood is combined with the prior to form the posterior distribution (see Step (b) - (iii) in the panel). Given the important roles that the prior and the likelihood have in determining the posterior, this step must be conducted with care. The implementational challenge at this step is to construct an efficient MCMC sampling algorithm. The basic idea behind MCMC here is the construction of a sampler that simulates a Markov chain that is converging to the posterior distribution. One can use software packages if prior distributions to be implemented in Bayesian models exist in the list of prior options available in the packages. Otherwise, professional programmers and Bayesian statisticians are needed to make codes manually; this review paper will be useful for that purpose. Another activity important at this step is to compare multiple models with different priors and nonlinear functions, specified in Step (b) - (i) and (ii), and select the best model out of them. This topic is broadly called the model selection \citep{chipman2001practical}, which will be discussed in Section \ref{sec:Model selection}.

\section{Applications of Bayesian nonlinear mixed effects model in real-life problems}\label{sec:Bayesian Nonlinear mixed effects model}
\subsection{The setting}\label{subsec:The setting}
To exemplify circumstances for which the nonlinear mixed effects model is a suitable modeling framework, we review challenges from several diverse applications. Table \ref{table:Summary of examples} summarize four real-life problems that will be illustrated in the next subsections. 
\begin{table}[H] 
\caption{Summary of examples.\label{table:Summary of examples}}
\newcolumntype{C}{>{\centering\arraybackslash}X}
\begin{tabularx}{\textwidth}{CCCC}
\toprule
\textbf{Research field}	&
\textbf{Problem}	&
 \textbf{Objective}
 &
  \textbf{References}
 \\
\midrule
Pharmaceutical industry & Pharmacokinetics analysis & Estimation of typical values of pharmacokinetics parameters &
\citep{gibaldi1982pharmacokinetics,jambhekar2009basic,sheiner1992population,ette2004population}
\\
\\
Oil and gas industry & Decline curve analysis & 
Prediction of estimated ultimate recovery &
\citep{lewis1918some,lee2021bayesian,fetkovich1980decline,harris2014study}
\\
\\
Financial industry & Yield curve modeling  &
Estimation of the interest rate parameters over time
&
\citep{nelson1987parsimonious,diebold2006forecasting,svensson1994estimating,dahlquist1996estimating}
\\
\\
Epidemiology & Epidemic spread prediction&
Prediction of final epidemic size and finding risk factors
&
\citep{lee2020estimation,wang2020prediction,wilke2020predicting}
\\
\bottomrule
\end{tabularx}
\end{table}

\subsection{Example 1: Pharmacokinetics analysis}\label{subsec:Example 1: Pharmacokinetics analysis}
\begin{figure}[h!]
\centering
\includegraphics[width=\textwidth]{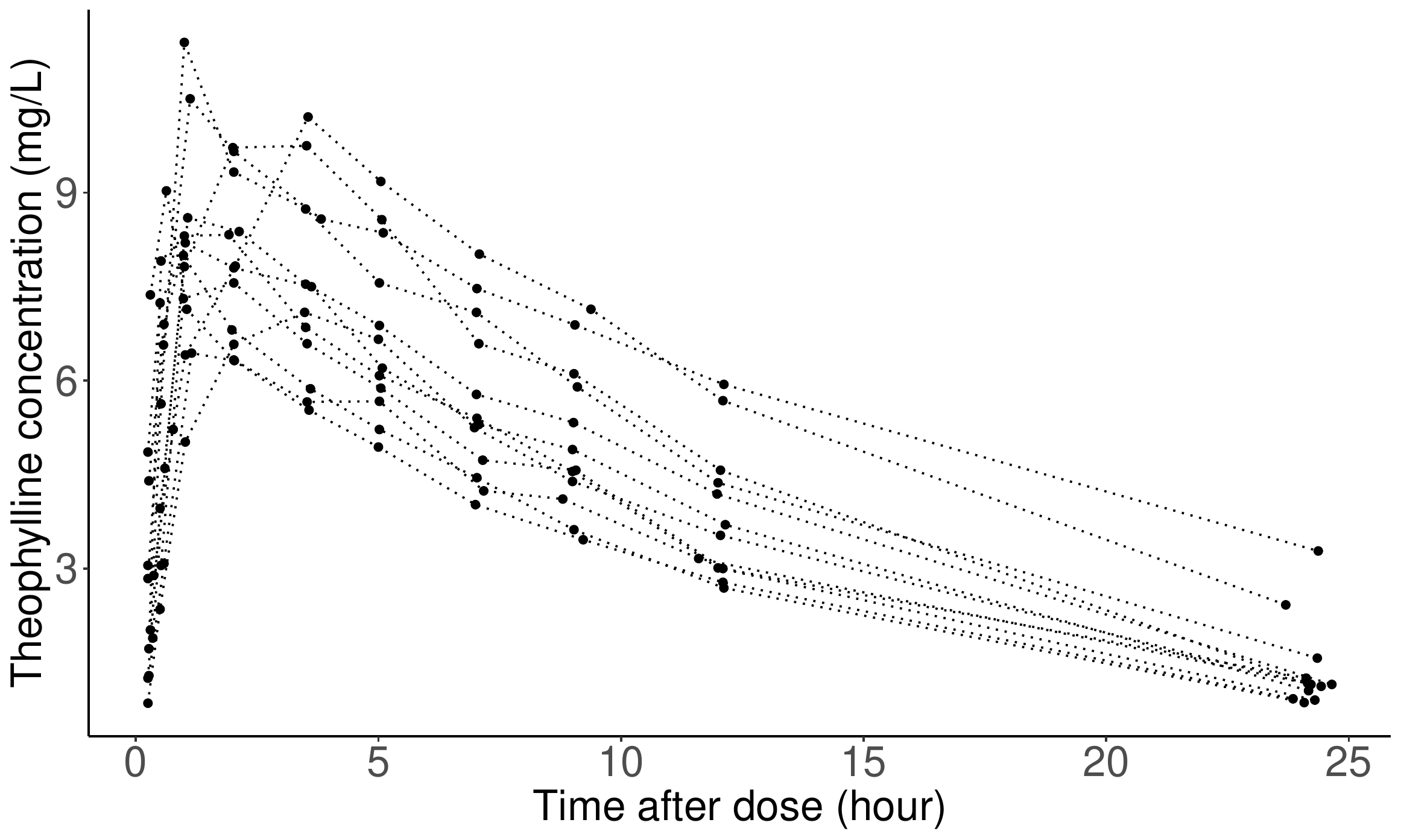}
\caption{
Theophylline concentrations for 12 subjects following an oral dose}
\label{fig:Theophylline_data}
\end{figure}

Studies of the pharmacokinetics of drugs help
us learn about the variability in drug disposition in a population \citep{bonate2011pharmacokinetic}. Figure \ref{fig:Theophylline_data} shows theophylline concentration in the plasma as a function of time after oral administration of the same amount of anti-asthmatic theophylline for 12 subjects. (The data considered here are courtesy of Dr. Robert A. Upton of the University of California, San Francisco.) As seen in the panel, concentration trajectories have a similar functional shape for all individuals. However, $C_{max}$ and $t_{max}$ (peak concentration and time when it is achieved), absorption, and elimination phases are substantially different across subjects. Clinical pharmacologists believe that these differences are attributable to between-subject variation in the underlying pharmacokinetic processes, explained by Absorption, Distribution, Metabolism, and Excretion (ADME), understanding of which is crucial in a new drug development in the pharmaceutical industry.

In pharmacokinetics analysis, often abbreviated by `PK analysis', it is routine to use compartmental modeling to describe the amount of drug in the body by dividing the whole body into one or more compartments \citep{rowland2005clinical}. For theophylline, a one-compartment model is normally used, which assumes that the entire body acts like a single, uniform compartment; see page 30 from \citep{gabrielsson2001pharmacokinetic} for a detailed explanation about the model:
\begin{align}
\label{eq:One_cpt}
C(t) &= \frac{  D F  k_{a}}{V(k_{a} - Cl/V)}
\left\{
\exp 
\left( 
-\frac{Cl}{V} t
\right)
-\exp (-k_{a} t) 
\right\},
\end{align}
where $C(t)$ is drug concentration at time $t$ for a single subject following oral dose $D$ at $t = 0$. Here, $F$ is the bioavailability which expresses the proportion of a drug that gains access to the systemic circulation. $k_{a}$ is the absorption rate constant describing how quickly drug is absorbed from the gut into the systemic circulation. $V$ is the volume of the central compartment. $Cl$ is the clearance rate representing the volume of plasma from which drug is eliminated per unit time. Eventually, the pharmacokinetic processes for a given subject is summarized by the $4$-dimensional vector with `PK parameters' $(F,k_{a},V,Cl)$. Obviously, it is the modeler's discretion to proceed with a more complex PK model such as a three compartment models with nonlinear clearance to fit the data, but in this case, over-parameterization should be carefully examined \citep{dua2015tutorial}.

Typically, the dataset collected in a drug development program includes demographic and clinical covariates obtained from each subject, for e.g., body weight, height, age, sex, creatinine clearance, albumin, etc; and furthermore, one can also involve genetic information in an individual’s response to drugs. Most covariates are measured at baseline, before assigning the drug, while some covariates can be measured at every sampling time. One of the crucial goals of PK analysis is to illustrate the effect of such covariates on the PK parameters \citep{xu2018full}. The causal relationship inferred by the covariate analysis can be used to support physicians in making the necessary judgments about the medicines that they prescribe, tailored to individual patients \citep{roses2000pharmacogenetics}.

In PK report for a new drug application to government authorities like U.S. Food and Drug Administration (FDA) or European Medicines Agency (EMA), PK parameters are summarized by mean or median, and very importantly, estimates of parameter precision. Estimates of parameter precision can provide valuable information regarding the adequacy of the data to support those parameters \citep{guidance1999population}. Parameter uncertainty can be estimated through several methods, including bootstrap procedures \citep{efron1982jackknife}, log-likelihood profiling \citep{broeker2020assessing}, or using the asymptotic standard errors of parameter estimates, and recently, Bayesian approaches draw a lot of attention from the pharmaceutical industry \citep{bauer2019nonmem}. Particularly, Bayesian approach for the population PK analysis can be very useful when there is prior knowledge about PK parameters learned from preclinical studies, published works, etc, and one wants to incorporate them into the prior specification for PK parameters \citep{wakefield1996bayesian}.

\subsection{Example 2: Decline curve analysis}\label{subsec:Example 2: Decline curve analysis}
The US shale boom–a product of technological advances in horizontal drilling and hydraulic fracturing that unlocked new stores of energy–has greatly benefited the growth in the US economy. Horizontal drilling is a directional drilling technology such that a well is drilled parallel to the reservoir bedding plane \citep{giger1984reservoir}. Well productivity of a horizontal well is known to be often 3 to 5 times greater than that of a vertical well \citep{al1991productivity,mukherjee1991parametric}, but also costs 1.5 to 2.5 times more than a vertical well \citep{joshi2003cost}. Therefore, the eventual success of the drilling project of unconventional shale wells relies on a large degree of well construction costs \citep{valdes2013uncertainty}. Because of very low permeability, and a flow mechanism very different from that of conventional reservoirs, estimates for the shale well construction cost often contain high levels of uncertainty. For this reason, one of the crucial tasks of petroleum engineers is to quantify the uncertainty associated with the process of oil or gas production to reduce the extra initial risk for the projects.


\begin{figure}[h!]
\centering
\includegraphics[width=\textwidth]{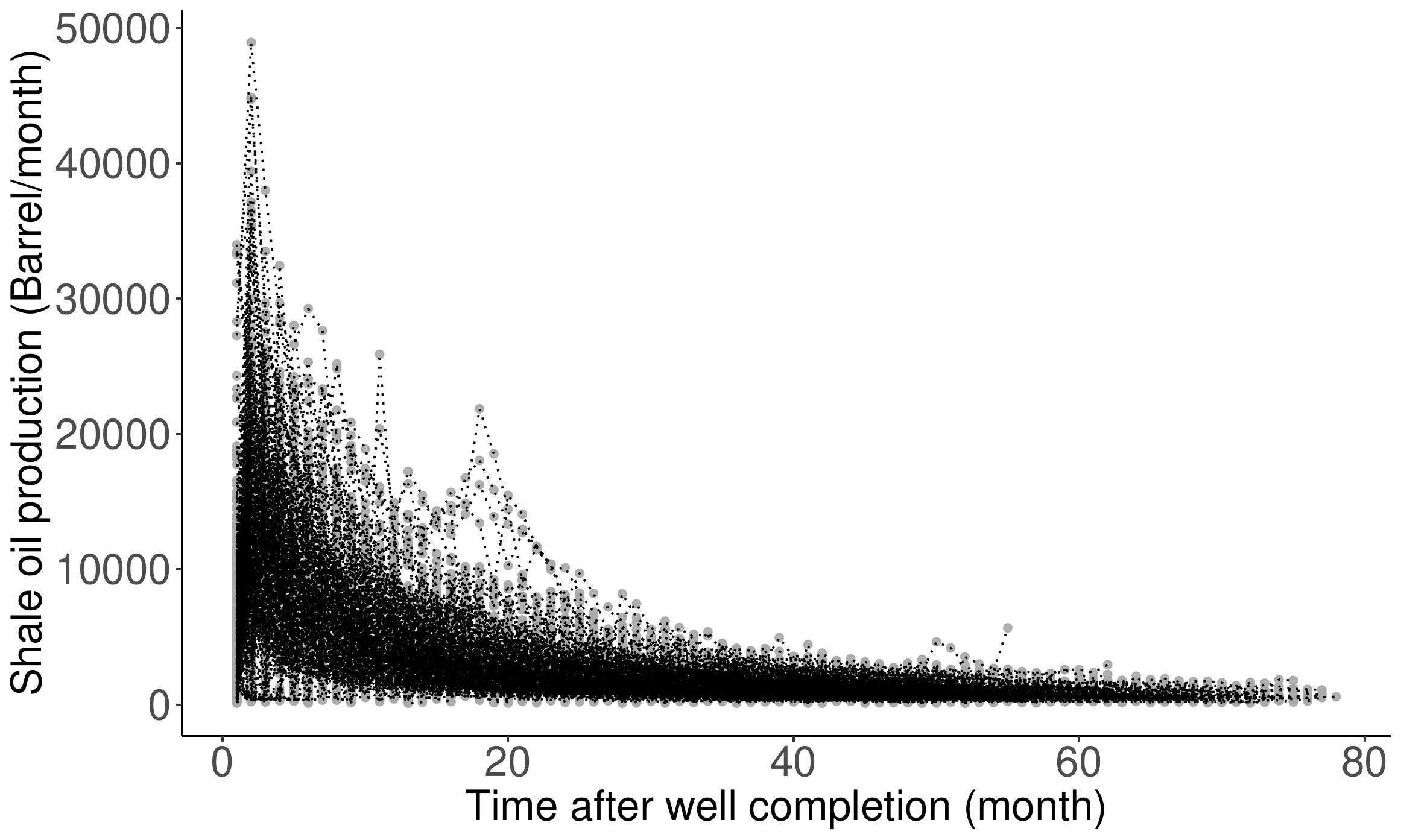}
\caption{
Production rates for 360 shale oil wells after completion.}
\label{fig:Shale_Oil_Production}
\end{figure}

Figure \ref{fig:Shale_Oil_Production} shows monthly production rate trajectories of 360 shale oil wells completed in the Eagle Ford Shale of South Texas, studied by \citep{lee2021bayesian}. The declining pattern manifested in the trajectories is commonly observed in almost all oil production rate time series data following well completion. (Here, the completion is terminology in petroleum engineering, meaning the process of transforming a well ready for the initial production \citep{bellarby2009well}.) Decline curve analysis (DCA), introduced by \citep{lewis1918some} around 100 years ago, is one of the most popularly utilized methods for petroleum engineers. Its purpose is to (i) theorize a curve describing the declining pattern, (ii) analyze the declining production rates, (iii) characterize the well-productivity, and (iv) forecast the future performance of oil and gas wells. Particularly, estimation and uncertainty quantification of estimated ultimate recovery (EUR) (here, EUR is a special jargon defined as an approximated quantity of oil from a well which is potentially recoverable by the end of its producing life \citep{currie2010continuous}) is the utmost important task and a starting point in the decision-making process for future drilling projects. Also, the oil and gas companies comply with financial regulations about EUR outlined by the U.S. Securities and Exchange Commission: see 
\href{https://www.sec.gov/rules/final/2008/33-8995.pdf}{www.sec.gov/-} for the regulations.

Most curves used in DCA are derived from solving certain differential
equations that describe a hidden dynamic from production rate trajectory \citep{arps1945analysis,weibull1951,ilk2008exponential,valko2010better,
clark2011decline,duong2011}. See  \cite{fetkovich1980decline,ali2015production,
harris2014study,miao2018comparison} for an overview of such curves. \citep{lee2021bayesian} studied Arps' hyperbolic, stretched exponentiated decline, Duong, and Weibull curves to fit the trajectories shown in the Figure \ref{fig:Shale_Oil_Production}. Particularly, the Duong model was developed for unconventional reservoirs with very low permeability:
\begin{align*}
P(t) = 
q_{1} t^{-m}
\exp 
\left\{
\frac{a (t^{1-m}  -1 )  }{1-m}
\right\},
\end{align*}
where $P(t)$ is the production rate at  time $t$  for a single well following completion. $q_{1}$ is the initial rate coefficient, and $m$ and $a$ are additional model parameters. We note that the parameters, $q_{1}$, $m$ and $a$, have their own meanings in terms of well-productivity: see \citep{duong2011} for the interpretation. That being said, the well-productivity for a given well is summarized by the $3$-dimensional parameter vector, $(q_{1},m,a)$. In modeling perspective, the variation of the well-productivity across different wells is attributable to the different values for $(q_{1},m,a)$.  To explain this variability, one can regress the values $(q_{1},m,a)$ on the well-design parameters such as true vertical depth, measure depth, etc. The causal relationship inferred by the covariate analysis will be used in a future drilling project. Geological information of wells can be also incorporated to make a spatial prediction for the EUR at a new location, as researched by \citep{lee2021bayesian}. 
\subsection{Example 3: Yield curve modeling}\label{subsec:Example 3: Yield curve modeling}
\begin{figure}[h!]
\centering
\includegraphics[width=\textwidth]{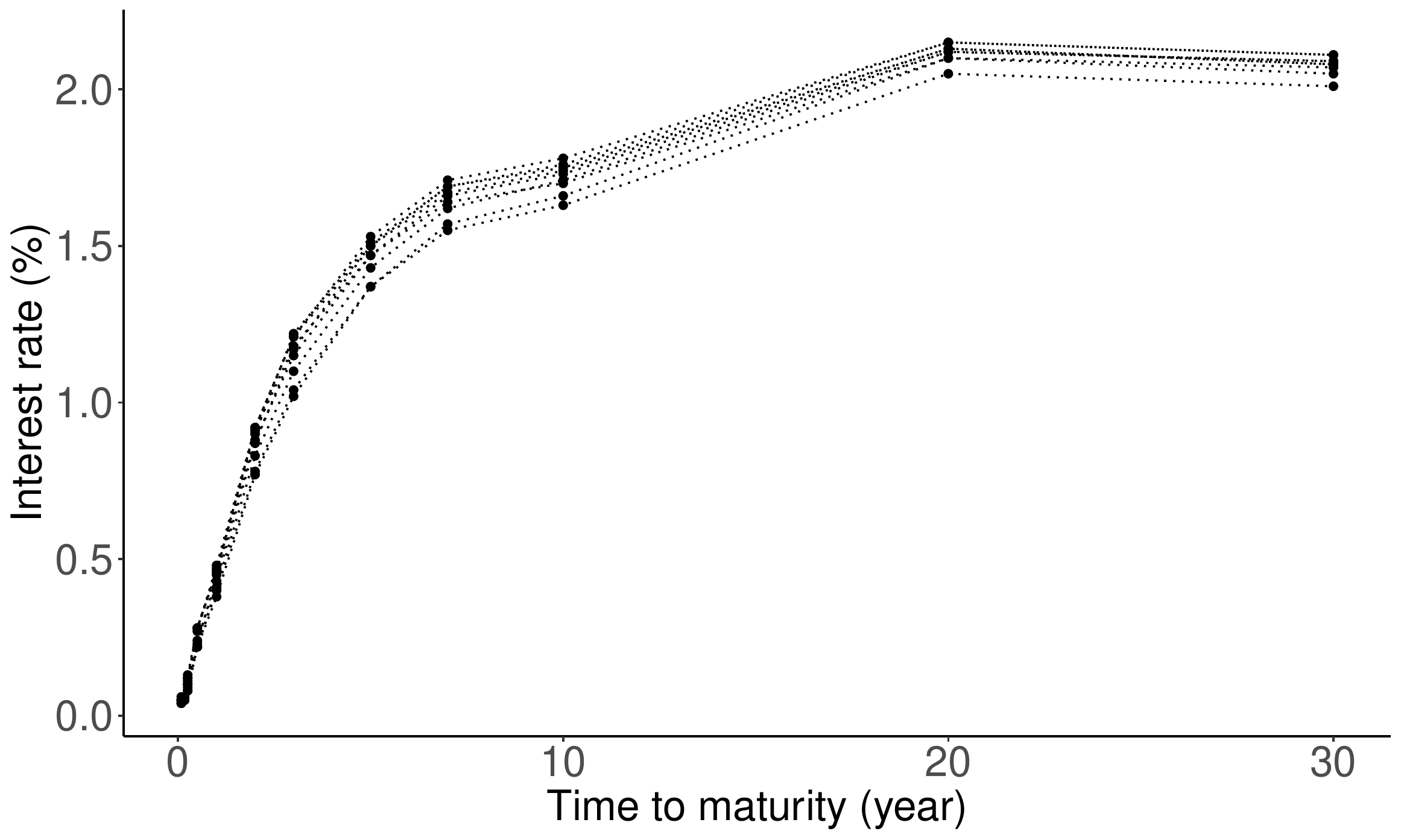}
\caption{
Daily Treasury par yield curve rates from January 3rd to January 13th, 2022.
}
\label{fig:Term_structure}
\end{figure}

Macroeconomists, financial economists, and market participants all attempt to build good models of the `yield curve' \citep{duffee2013forecasting}. The yield curve on a given day is a curve showing the interest rates across different maturity spans (1 month, one year, five years, etc.) for a similar debt contract at a particular date. It determines the interest rate pattern (i.e., cost of borrowing), which can be used to calculate a bond's price \citep{gurkaynak2007us}. Figure \ref{fig:Term_structure} shows daily treasury par yield curve rates spanning from January 3rd to January 13th, 2022, with maturities up to 30 years. The data source is from the U.S. Department of the treasury (\href{https://www.treasury.gov/resource-center/data-chart-center/interest-rates/Pages/TextView.aspx?data=yield}{www.treasury.gov/-}). As seen from the panel, the shape of the yield curve displays a slightly delayed humped shape. Economists believe that such a shape of the yield curve has an important implication on the economic growth \citep{zaloom2009read}.

The Nelson-Siegel model \citep{nelson1987parsimonious} is a very popular model in the literature to fit the term structure:
\begin{align*}
Y(\tau)  = 
\beta_{0} 
+
\beta_{1}
\left\{
\frac{1 - \exp (-\lambda \tau)}{\lambda \tau}
\right\}
+
\beta_{2}
\left\{
\frac{1 - \exp (-\lambda \tau)}{\lambda \tau}
-
\exp (-\lambda \tau)
\right\},
\end{align*}
where $Y(\tau)$ denotes the (zero-coupon) yield at evaluated at $\tau$, and $\tau$ denotes the time to maturity. The model parameters have a specific financial meaning: $\beta_{0}$, $\beta_{1}$, and $\beta_{2}$ are related long-term, short-term, and mid-term effects on the interest rate, respectively, and $\lambda$ is referred to as a decay factor \citep{diebold2006forecasting}. Each of the yield curves is summarized by the $4$-dimensional parameter $(\beta_{0},\beta_{1},\beta_{2},\lambda)$, and it is known that model can capture a wide range of possible shapes of the yield curve \citep{nelson1987parsimonious,diebold2006forecasting,hays2012functional,chen2014adaptive}.
Therefore, the Nelson‐Siegel model is extensively used by central banks and monetary policymakers \citep{ac049878412005zero}. For example, The Federal Reserve updates estimates of $(\beta_{0},\beta_{1},\beta_{2},\lambda)$ once per week: visit the website (\href{https://www.federalreserve.gov/data/yield-curve-tables/feds200628_1.html}{www.federalreserve.gov/-}). In recent years, there has been a great deal of interest in the uncertainty quantification of the Nelson-Siegel parameters over time, and their relationship with macroeconomic variables such as inflation and real activity, etc, in financial applications: refer to  \citep{chen2014adaptive,hautsch2012bayesian,diebold2008global,cruz2011estimating} for some of those works.
\subsection{Example 4: Early stage of epidemic}\label{subsec:Example 4: Early stage of epidemic}
\begin{figure}[h!]
\centering
\includegraphics[width=\textwidth]{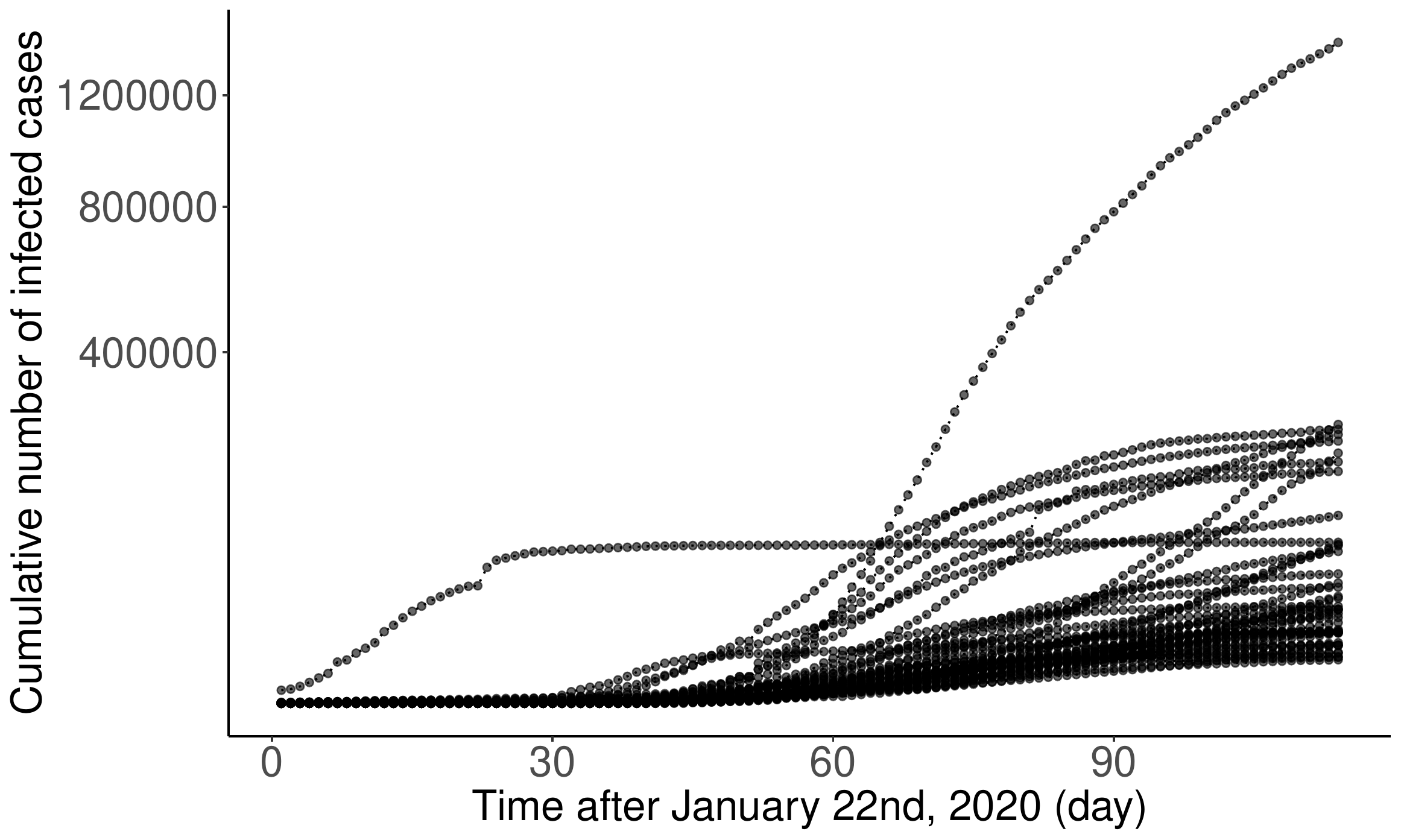}
\caption{
Daily trajectories for cumulative numbers of COVID-19 infections for 40 countries from January 22nd to May 14th, 2020.}
\label{fig:COVID_19_infection}
\end{figure}

Novel coronavirus disease 2019 (COVID-19) is a big threat to global health. The rapid spread of the virus has created a pandemic, and countries all over the world are struggling with a surge in COVID-19 infected cases. Figure \ref{fig:COVID_19_infection} displays the daily infection trajectories describing the cumulative numbers of infected cases for 40 countries, spanning from January 22nd to May 14th, 2020, studied by \citep{lee2020estimation}. The data source is from COVID-19 Data Repository by the Center for Systems Science and Engineering at Johns Hopkins University (\href{https://coronavirus.jhu.edu/map.html}{coronavirus.jhu.edu/-}). Refer to Table S.1 in \citep{lee2020estimation} for the list of 40 countries. The time frame of the authors' research was the early stage of the pandemic when there was no drug or other therapeutics approved by the US FDA. 

In general, during an early phase of a pandemic, information regarding the disease is very limited and scattered even if it exists. In spite of that, it is crucial to predict future cases of infection or death. In such a situation, one consideration is to use data integration (also called `borrowing information'), combining data from diverse sources and eliciting useful information with a unified view of them. Additionally, it is very important to find risk factors relevant to the disease. Reliable and early risk assessment of a developing infectious disease outbreak allow policymakers to make swift and well-informed decisions that would be needed to ensure epidemic control. Quantifying uncertainty about the final epidemic size is also very important.

Richards growth curve \cite{richards1959flexible}, so-called the generalized logistic curve \cite{nelder1962182}, is a popularly used growth curve for population studies in situations where growth is not symmetrical about the point of inflection \cite{seber2003nonlinear,anton1988calculus}. There are variant reparamerized forms of the Richards curve in the literature \cite{causton1969computer,birch1999new,kahm2010grofit,cao2019new}, and one of the frequently used form is
\begin{align*}
I(t) = \frac{a}{[1 + \xi \exp \{ - b(t - c)\}]^{1/\xi}},
\end{align*}
where $I(t)$ is the cumulative number of infected cases at time $t$. Here, epidemiological meanings of the parameters, $a$, $b$, and $c$, are the final epidemic size, infection rate, and lag phase of the trajectory, respectively. The parameter $\xi$ is the shape parameter, and there seems no clear epidemiological meaning \cite{wang2012richards}. Each infection trajectory in Figure \ref{fig:COVID_19_infection} can be characterized by the $4$-dimensional parameters $(a,b,c,\xi)$ if the Richards curve is used. Due to its flexibility originating from the shape parameter $\xi$, Richards curve has been widely used in epidemiology for real-time prediction of outbreak of diseases, possibly at an early phase of the pandemic when there is no second wave. Examples include SARS \cite{hsieh2004sars,hsieh2009richards}, dengue fever \cite{hsieh2009intervention,hsieh2009turning}, pandemic influenza H1N1 \cite{hsieh2010pandemic}, and COVID-19 outbreak \cite{lee2020estimation,wu2020generalized}. 
\subsection{Statistical problem}\label{subsec:Statistical problem}
In the previous subsections, we presented a range of examples in which nonlinear mixed effects models can be exploited. They have their own challenges to solve the problems that are representative of issues many researchers have to deal with in other areas: for example, (1) how to describe a possible nonlinear clearance with a limited number of patients; (2) how to handle an enormously large number of shale oil wells and make a spatial prediction of EUR at a new location; (3) how to describe the dynamic of the financial parameters over time; and (4) how to integrate data from different sources to produce more accurate forecast on the epidemic size. 

An emerging issue accompanied by these problems, requested from researchers, government entities, domain experts, etc, is how to quantify the uncertainty associated with parameter estimation and prediction. Although the traditional nonlinear mixed effects models, based on the maximum likelihood method, can provide confidence intervals and statistical tests, calculations of those generally involve approximations that are most accurate for large sample sizes, as discussed in Subsection \ref{subsec:A rise of Bayesian approaches}. On the other hand, in the Bayesian approach - in which the prior automatically imposes the parameter constraints - inferences about parameter values based on the posterior distribution usually require integration rather than maximization, and no further approximation is involved. For that reason, the Bayesian approach is often suggested as a viable alternative to the frequentist approach to solving the problems.

We now formulate these problems as a statistical problem. First, we summarize common features of the dataset for the analysis.
\begin{itemize}
\baselineskip=15pt
\item[(1)] There exists repeated measures of a continuous response over time for each subject; 
\item[(2)] There exists a variation of individual observations over time;
\item[(3)] There exists a variation from subject-to-subject in trajectories;
\item[(4)] There exist covariates measured at baseline for each subject.
\end{itemize}


The subject of sampling units considered in the statistical analysis is quite comprehensive. We have seen that it can be a patient, a shale oil well, a particular date, and a country. As the unique identifier, we assign the index $i$ to each individual. By denoting $N$ as the number of individuals (i.e., the sample size), the index $i$ will take an integer from $1$ to $N$. The sample size $N$ available for the data analysis substantially varies across different industrial problems as well as subfields within the same industry. For example, the number of shale oil wells on Eagle Ford Shale Play can be as large as $6,000$ \citep{lee2021bayesian}. As for the pharmaceutical industry, in phase I cancer clinical trials, the number of cancer patients $N$ may be strictly confined to $25$ \citep{lee2021optimization}, but for phase III trials for non-oncology drug studies, $N$ can be as large as $2,000$ \citep{dugel2020hawk}. 

Here the term `time' is meant in the broadest sense. It can be a calendar time, a nominal time, a time after some event (for e.g., the time after dose from Figure \ref{fig:Theophylline_data} and the time after well completion from Figure \ref{fig:Shale_Oil_Production}), or a time to some event (for e.g., the time to maturity from Figure \ref{fig:Term_structure}). Essentially, time can be defined as a physical quantity that can be indexed with consecutive integers to produce a temporal record. Another important characteristic of the time is that each subject may have different time points where observations are measured. In this article, we use $t_{ij}$ to represent the time point, where the integer $j=1,2,\cdots,M_{i}$ indexes the time point from the earliest to the last observations. Thus, $M_{i}$ represents the number of repeated observations for the $i$-th individual. When $M_{i}$ is relatively small (or large), we say the repeated measures are sparsely observed (or densely observed). For example, the theophylline and yield curve data shown in Figure \ref{fig:Theophylline_data} and \ref{fig:Term_structure} are sparse data, while the oil production and COVID-19 data shown in Figure \ref{fig:Shale_Oil_Production} and \ref{fig:COVID_19_infection} are dense data.

As for the repeated measures, $y_{ij}$ denotes the continuous response of the $i$-th subject at the time point $t_{ij}$. We assume that $y_{ij}$ has been already pre-processed so that it is ready to be used for statistical modeling. For most applications, it may be necessary first to transform the data into some new representation before training the model. For example, as seen from Figure \ref{fig:Shale_Oil_Production}, oil productions vary substantially across different wells. For that reason, the authors \citep{lee2021bayesian} take a logarithm on the productions to derive the response $y_{ij}$, followed by appropriate statistical modeling on the log-scale. To some extent, data pre-processing may enhance the performance of the model. 

Suppose that researchers collected $P$ number of covariates at the baseline from each subject $i$ ($i=1,\cdots,N$). Here, the baseline refers to the time point $t_{i1}$ (or possibly right before the time point $t_{i1}$), where at the first response $y_{i1}$ has not been observed yet. Let $x_{ib}$ denote the $b$-th covariate of the $i$-th subject ($b=1,\cdots,P$). In general, there are two types of covariates: time-invariant and time-varying covariates. This article mainly concerns the former type. As similar to $N$, the number of covariates $P$ substantially varies across industries and specific problems. For instance, in pharmacogenetics analysis, the number of protein-coding genes $P$ would be around $20,000$ \citep{willyard2018new}. In the oil and gas industry, if we consider most of the covariates obtained from the well completion procedure, $P$ could be at least $100$
\citep{lee2021bayesian}.

In conclusion, the dataset for the statistical analysis can be represented by the collection of the $N$ triplets $\{(\textbf{y}_{i},\textbf{t}_{i},\textbf{x}_{i})\}_{i=1}^{N}$. Here, for each subject $i$ ($i=1,\cdots,N$), we formulated two $M_{i}$-dimensional vectors $\textbf{y}_{i}=(y_{i1},\cdots, y_{ij}, \cdots,y_{iM_{i}})^{\top}$ and $\textbf{t}_{i}=(t_{i1},\cdots, t_{ij}, \cdots,t_{iM_{i}})^{\top}$, and a $P$-dimensional vector $\textbf{x}_{i} = (x_{i1},\cdots,  x_{ib} , \cdots, x_{iP})^{\top}$.

\begin{figure}[h!]
\centering
\includegraphics[width=\textwidth]{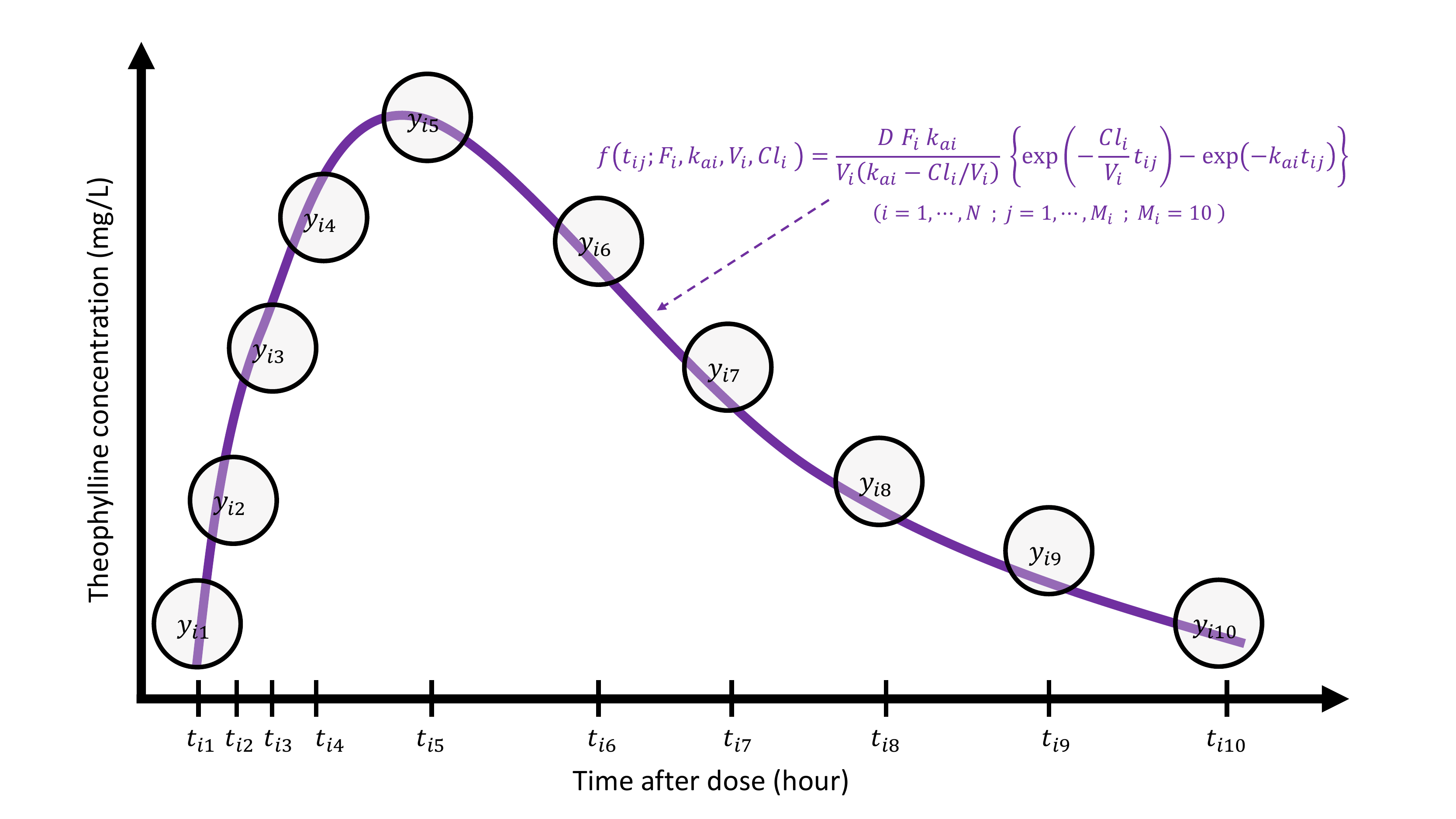}
\caption{
Pictorial illustration of PK modeling for the theophylline data.}
\label{fig:Concept_of_NONMEM}
\end{figure}

The data structures described up to this point are commonly encountered in longitudinal data studies \citep{demidenko2013mixed}. Essentially, the feature of dataset motivating the use of nonlinear mixed effects models is that, for each subject $i$, the response vector $\textbf{y}_{i}$ displays some nonlinear tendency over time $\textbf{t}_{i}$, as seen in Figure \ref{fig:Theophylline_data}, \ref{fig:Shale_Oil_Production}, \ref{fig:Term_structure}, and \ref{fig:COVID_19_infection}. To explain this nonlinearity, a researcher needs to theorize some nonlinear function, denoted as $f$, such as one compartment, Duong, Nelson-Siegel, and Richards models, depending on the contexts. The construction of such functions relies on human modelers’ abstraction of data into a suitable dynamical system, which is often represented by a differential equation. Such a differential equation has a finite number of parameters that control the dynamic of the solution of the system, understanding of which is vital for causal inference for the nature of the system by associating with covariates $\textbf{x}_{i}$. 

Figure \ref{fig:Concept_of_NONMEM} displays a pictorial description about how a PK modeler would see the theophylline concentration trajectory from the modeling perspective, where she theorized that the one compartment model (\ref{eq:One_cpt}) would be suitable to describe the trajectories $\textbf{y}_{i}$ over time $\textbf{t}_{i}$ for each subject $i$ ($i=1,\cdots,N$). Then the $10$-dimensional vector $\textbf{y}_{i}$ is summarized by a $4$-dimensional PK parameter vector $(F_{i},k_{ai},V_{i},Cl_{i})$; the dimension reduction is intrinsically embedded in this process. As each of the parameters $F_{i}$, $k_{ai}$, $V_{i}$, and $Cl_{i}$ has an important clinical meaning, it is very natural to ask how they are related with $P$ covariates $\textbf{x}_{i}$ to induce a causal relationship. For the purpose of modeling, it may be necessary to transform the original PK parameters $(F_{i},k_{ai},V_{i},Cl_{i})\in [0,1] \times (0,\infty)^{3}$ to model parameters $(\theta_{1i},\theta_{2i},\theta_{3i},\theta_{4i}) \in \mathbb{R}^{K}$ ($K=4$) so that elements $\theta_{li}$ ($l=1,\cdots,K$) are supported on the real number by taking transformations $\theta_{1i} = \log \{F_{i}/(1 - F_{i})\}$, $\theta_{2i} = \log k_{ai}$, $\theta_{3i} = \log V_{i}$, and $\theta_{4i} = \log Cl_{i}$. As these transformations were taken only for the modeling purpose, interpretations on the PK parameter for the PK report should be carried out after transforming back to the original scale.

\section{The model}\label{sec:The model}
\subsection{Basic model}\label{subsec:Basic model}
Assume that we have dataset for a statistical analysis $\{(\textbf{y}_{i},\textbf{t}_{i},\textbf{x}_{i})\}_{i=1}^{N}$ from $N$ subjects, as explained in Subsection \ref{subsec:Statistical problem}. We consider a basic version of the model here. Extensions are discussed in Subsection \ref{subsec:Priors for covariance} and Section \ref{sec:Extensions and recent developments}. The usual Bayesian nonlinear hierarchical model may then be written as a three-stage hierarchical model as follows:
\begin{itemize}
\baselineskip=15pt
\item[]$\bullet$ \emph{\textbf{Stage 1: Individual-Level Model}}
\begin{align}
\label{eq:Stage 1: Individual-Level Model}
y_{ij} &=
f(t_{ij} ; \bm{\theta}^{i}) + \epsilon_{ij}, \quad \epsilon_{ij}\sim \mathcal{N}(0,\sigma^{2}), \quad (i=1,\cdots,N; j  = 1,\cdots,M_{i}).
\end{align}
In (\ref{eq:Stage 1: Individual-Level Model}), the conditional mean $\mathbb{E}[y_{ij}| \bm{\theta}^{i},\sigma^{2}] = f(t_{ij} ; \bm{\theta}^{i})$ is a known function governing within-individual temporal behavior dictated by a $K$-dimensional parameter $\bm{\theta}^{i} =  (\theta_{1i},\theta_{2i},$ $\cdots,$ $\theta_{li},\cdots,\theta_{Ki})^{\top}\in \mathbb{R}^{K}$ specific to the subject $i$. We assume that the residuals, $\epsilon_{ij}$, are normally distributed with mean zero and with an unknown variance, $\sigma^{2}$.
\item[]$\bullet$ \emph{\textbf{Stage 2: Population Model}}
\begin{align}
\label{eq:Stage 2: Population Model}
\theta_{li} &= \alpha_{l} + \textbf{x}_{i}^{\top} \bm{\beta}_{l} + \eta_{li}, \quad
\eta_{li}\sim \mathcal{N}(0,\omega_{l}^{2}), \quad (i=1,\cdots,N; l=1,\cdots,K).
\end{align}
In (\ref{eq:Stage 2: Population Model}), the $l$-th model parameter $\theta_{li}$ is used as the response of an ordinary linear regression with predictor $\textbf{x}_{i}$, with intercept $\alpha_{l} \in \mathbb{R}$ and coefficient vector $\bm{\beta}_{l}=(\beta_{l1},\beta_{l2},\cdots,\beta_{lP})\in \mathbb{R}^{P}$. By letting $\bm{\eta}_{i}=(\eta_{1i},\eta_{2i},\cdots,$ $\eta_{li},\cdots,\eta_{Ki})\in \mathbb{R}^{K}$, we assume that the $\bm{\eta}_{i}$ is distributed according a $K$-dimensional Gaussian distribution $\mathcal{N}(\textbf{0}, \bm{\Omega})$ with covariance matrix $\bm{\Omega} = \text{diag}(\omega_{1}^{2},\omega_{2}^{2},\cdots,\omega_{l}^{2},\cdots, \omega_{K}^{2})\in \mathbb{R}^{K \times K}$. The diagonality in $\bm{\Omega}$ implies that each model parameter are uncorrelated across $l$.
\item[] $\bullet$ \emph{\textbf{Stage 3: Prior}}
\begin{align}
\label{eq:Stage 3: Prior}
\sigma^{2} &\sim \pi(\sigma^{2}),\quad \alpha_{l}\sim \pi(\alpha_{l}), \quad
\bm{\beta}_{l}\sim \pi(\bm{\beta}_{l}), \quad \omega_{l}^{2} \sim \pi(\omega_{l}^{2}), \quad (l=1,\cdots,K).
\end{align}
Distributions in (\ref{eq:Stage 3: Prior}) are chosen to encapsulate any information or belief which have been formulated about the parameters. We suggest some popularly used prior options in Section \ref{sec:Prior options}. 
\end{itemize}

\begin{figure}[h!]
\centering
\includegraphics[scale=0.6]{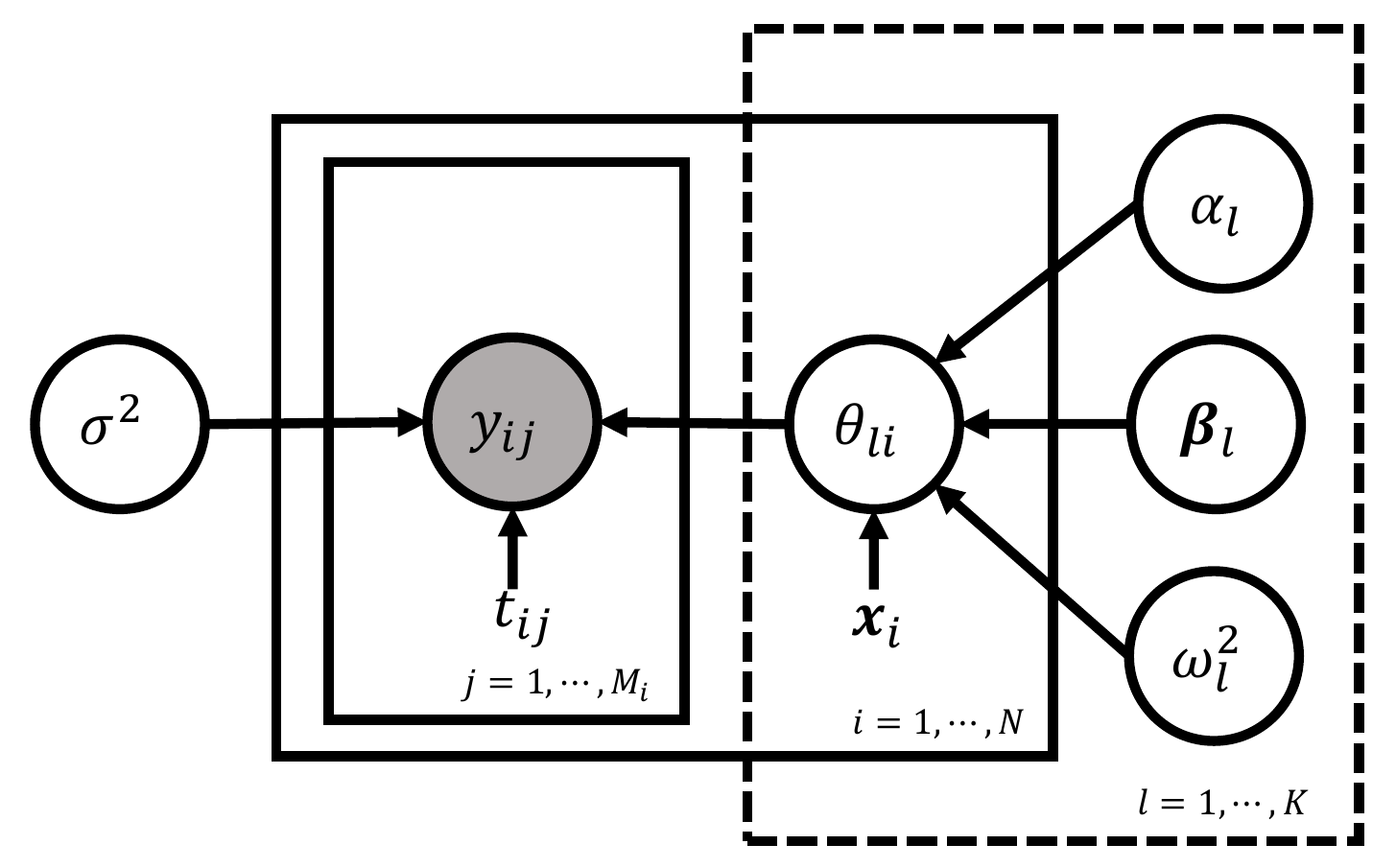}
\caption{
The basic model (\ref{eq:Stage 1: Individual-Level Model}) -- (\ref{eq:Stage 3: Prior}) as a graphical model.}
\label{fig:framework_BNONMEM}
\end{figure}

Directed asymmetric graphical (DAG) model representation of the basic model (\ref{eq:Stage 1: Individual-Level Model}) -- (\ref{eq:Stage 3: Prior}) is depicted in Figure \ref{fig:framework_BNONMEM}. Following the grammar of the graphical model (Chapter 8 of \citep{bishop2006pattern}), the circled variables indicate stochastic variables, while observed ones are additionally colored in grey. Non-stochastic quantities are uncircled. The arrows indicate the conditional dependency between the variables.

\subsection{Vectorized form of the basic model}\label{subsec:Vector form of the basic model}
We will often wish to write the hierarchy (\ref{eq:Stage 1: Individual-Level Model}) -- (\ref{eq:Stage 3: Prior}) for the $i$-th individual's entire response vector and represent it with an equivalent vector-form. This turns out to be useful to develop relevant computational algorithms. We first introduce a $K\times N$ dimensional matrix frequently used throughout this article:
\begin{align}
\label{eq:model_matrix}
\bm{\Theta}= 
\begin{bmatrix}
\theta_{11} & \cdots &\theta_{1i}& \cdots &\theta_{1N}\\
\vdots & &\vdots &&\vdots \\
\theta_{l1}  &\cdots& \theta_{li} &\cdots& \theta_{lN}\\
\vdots & &\vdots &&\vdots \\
\theta_{K1}   &\cdots& \theta_{Ki} &\cdots& \theta_{KN}
\end{bmatrix}\in \mathbb{R}^{K \times N}.
\end{align}
The matrix (\ref{eq:model_matrix}) is referred to as \emph{model matrix} because it comprises of scalar model parameters $\{ \theta_{li} \}_{l=1,i=1}^{K,N}$ from all subjects. Indeed, most of computational techniques either via frequentist or Bayesian setting in the literature have been developed to overcome an obstacle of a nonlinear association of the model matrix $\bm{\Theta}$ into the mean function $f$. 

In (\ref{eq:model_matrix}), the subject index $i$ is stacked column-wisely, while model parameter index $l$ is stacked row-wisely, different from the conventional way adopted in most statistics. The column indexing for the subjects (i.e., stacking individual-based vector horizontally) shown in (\ref{eq:model_matrix}) is often adopted in modern computation theory of deep learning \citep{goodfellow2016deep}, and one of the main advantages of using this indexing is that it may give some pedagogical insights on the use of vectorization toward the entries $\{ \theta_{li} \}_{l=1,i=1}^{K,N}$ to exploit parallel computations, stochastic updating, etc, in optimization or sampling techniques.

The model matrix $\bm{\Theta}$ (\ref{eq:model_matrix}) 
can be re-expressed as $\bm{\Theta} = [
\bm{\theta}^{1}  \cdots  \bm{\theta}^{i}  \cdots  \bm{\theta}^{N}
]\in \mathbb{R}^{K \times N}$, obtained by stacking the individual model parameter vector in Stage 1 (\ref{eq:Stage 1: Individual-Level Model}). Alternatively, we can represent the matrix with $\bm{\Theta} = [
\bm{\theta}_{1}  \cdots  \bm{\theta}_{l}  \cdots  \bm{\theta}_{K}
]^{\top}\in \mathbb{R}^{K \times N}$ by defining a $N$-dimensional vector corresponding the $l$-th model parameter across all subjects $\bm{\theta}_{l} = (\theta_{l1},\theta_{l2},\cdots,\theta_{lN})^{\top}\in \mathbb{R}^{N}$ ($l=1,\cdots,K$). Former and latter indexing method are referred to as $i$-indexing and $l$-indexing, respectively.

We are now in a position to re-write the hierarchy (\ref{eq:Stage 1: Individual-Level Model}) -- (\ref{eq:Stage 3: Prior}) using the vector notations:
\begin{itemize}
\baselineskip=15pt
\item[] $\bullet$ \emph{\textbf{Stage 1: Individual-Level Model}}
\begin{align}
\label{eq:Stage 1: Individual-Level Model_Vector_Form}
\textbf{y}_{i} &= \bm{f}_{i}(\textbf{t}_{i}, \bm{\theta}^{i}) +\bm{\epsilon}_{i},
\quad
\bm{\epsilon}_{i} \sim \mathcal{N}_{M_{i}}(\textbf{0}, \sigma^{2}  \textbf{I}), \quad (i=1,\cdots,N).
\end{align}
In (\ref{eq:Stage 1: Individual-Level Model_Vector_Form}), $\bm{f}_{i}(\textbf{t}_{i}, \bm{\theta}^{i})$ is a $M_{i}$-dimensional vector whose elements are temporally stacked: $\bm{f}_{i}(\textbf{t}_{i}, \bm{\theta}^{i}) =  (f(t_{i1} ; \bm{\theta}^{i}),f(t_{i2} ; \bm{\theta}^{i}),\cdots,f(t_{iM_{i}} ; \bm{\theta}^{i}))^{\top}$ for the subject $i$. The vector $\bm{\epsilon}_{i}$ is distributed according to the $M_{i}$-dimensional Gaussian distribution with mean $\textbf{0}$ and covariance matrix $\sigma^{2}  \textbf{I}$.
\item[]$\bullet$ \emph{\textbf{Stage 2: Population Model ($l$-indexing)}}
\begin{align}
\label{eq:Stage 2: Population Model_Vector_Form}
\bm{\theta}_{l} &= \textbf{1}\alpha_{l} + \textbf{X} \bm{\beta}_{l} + \bm{\eta}_{l}, \quad
\bm{\eta}_{l} \sim \mathcal{N}_{N}(0,\omega_{l}^{2} \textbf{I}), \quad (l=1,\cdots,K).
\end{align}
In (\ref{eq:Stage 2: Population Model_Vector_Form}), for each $l$, the $N$-dimensional model parameter vector $\bm{\theta}_{l}$ is used as the response vector of an ordinary linear regression: (i) $N$-by-$P$ design matrix $\textbf{X} = [\textbf{x}_{1} \textbf{x}_{2} \cdots \textbf{x}_{N}]^{\top}$; (ii) intercept $\alpha_{l}$; (iii) coefficient vector $\bm{\beta}_{l}$, and (iv) isotropic Gaussian error vector $\bm{\eta}_{l} = (\eta_{l1},\eta_{l2},\cdots,\eta_{lN})^{\top}$ with variance $\omega_{l}^{2}$. (Notation $\textbf{1}$ in (\ref{eq:Stage 2: Population Model_Vector_Form}) represents an all-ones vector.). 

\item[] $\bullet$ \emph{\textbf{Stage 2': Population Model ($i$-indexing)}}
\begin{align}
\label{eq:Stage 2: Population Model_Vector_Form_theory}
\bm{\theta}^{i} & = \bm{\alpha} + \bm{B} \textbf{x}_{i} +  \bm{\eta}^{i}, \quad \bm{\eta}^{i} \sim \mathcal{N}_{K}(\textbf{0}, \bm{\Omega}),\quad (i=1,\cdots,N).
\end{align}
Equation in (\ref{eq:Stage 2: Population Model_Vector_Form_theory}) is derived by incorporating each of the $N$ columns of the model matrix (\ref{eq:model_matrix}). Here, $\bm{\alpha}$ represents a $K$-dimensional vector $\bm{\alpha} = (\alpha_{1},\alpha_{2},\cdots,\alpha_{K})^{\top}$, and $\bm{B}$ represents a $K$-by-$P$ matrix with rows $\bm{\beta}_{l}$ ($l=1,\cdots,K$). Here, the $K$-dimensional vector $\bm{B} \textbf{x}_{i}$ in the right-hand side of (\ref{eq:Stage 2: Population Model_Vector_Form_theory}) is the mathematically identical to $\textbf{X}_{i} \bm{\beta}$, where 
$\textbf{X}_{i} = \textbf{I}_{K} \otimes \textbf{x}_{i}^{\top} \in \mathbb{R}^{K \times KP}$ and $\bm{\beta} = (\bm{\beta}_{1},\bm{\beta}_{2},\cdots,\bm{\beta}_{K}) \in \mathbb{R}^{KP}$ ($\textbf{I}_{K}$ is the $K$-by-$K$ identity matrix and $\otimes$ represents the Kronecker matrix product.). The error vector $\bm{\eta}^{i} = (\eta_{1i},\eta_{2i},\cdots,\eta_{Ki})^{\top}$ is distributed according a $K$-dimensional Gaussian distribution with mean $\textbf{0}$ and covariance matrix $\bm{\Omega} = \text{diag}(\omega_{1}^{2},\omega_{2}^{2},\cdots,\omega_{K}^{2})$.
\item[] $\bullet$\emph{\textbf{Stage 3: Prior}}
\begin{align}
\label{eq:Stage 3: Prior_Vector_Form}
\sigma^{2} &\sim \pi(\sigma^{2}),\quad \bm{\alpha} \sim \pi(\bm{\alpha}), \quad
\bm{B} \sim \pi(\bm{B} ), \quad \bm{\Omega} \sim \pi(\bm{\Omega}).
\end{align}
Each of the parameter blocks in $(\sigma^{2},\bm{\alpha} , \bm{B}, \bm{\Omega})$ is assumed to be independent a \emph{priori}.
\end{itemize}

\begin{figure}[h!]
\centering
\includegraphics[width=\textwidth]{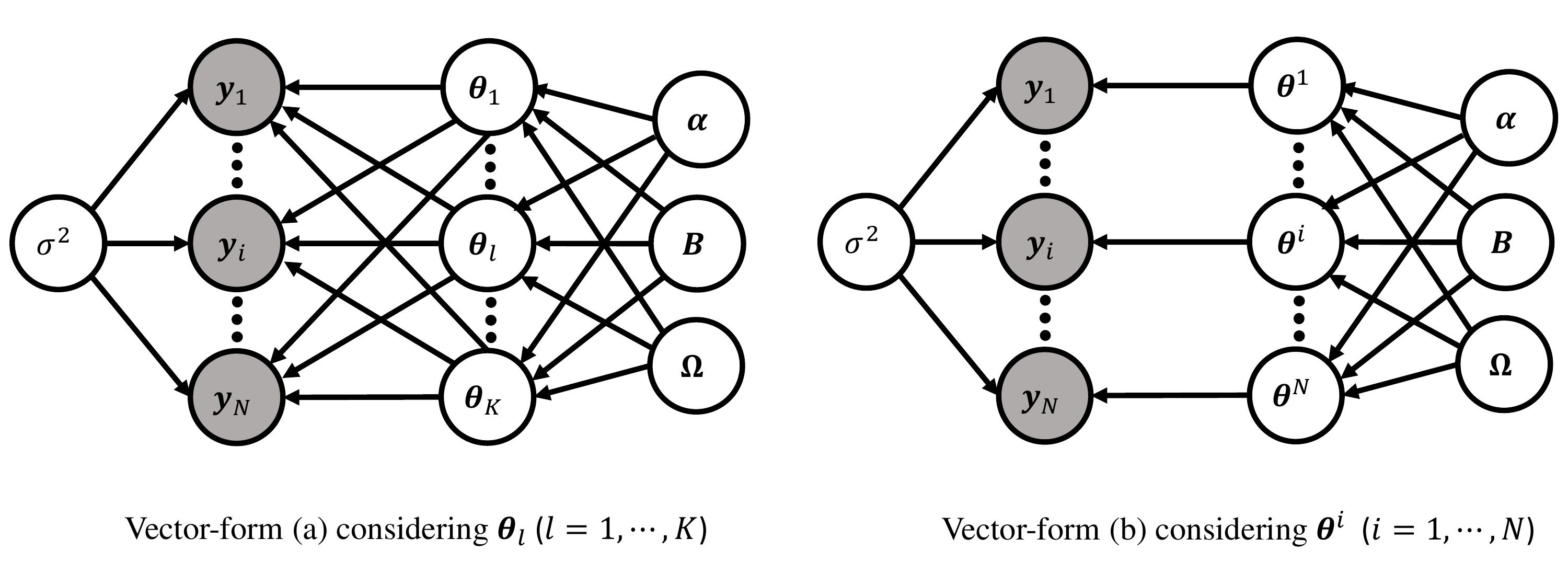}
\caption{
DAG representations of the basic model (\ref{eq:Stage 1: Individual-Level Model}) -- (\ref{eq:Stage 3: Prior}) in vector-form (a) [Stage 1-(\ref{eq:Stage 1: Individual-Level Model_Vector_Form}), Stage 2-(\ref{eq:Stage 2: Population Model_Vector_Form}) ($l$-indexing), and Stage 3-(\ref{eq:Stage 3: Prior_Vector_Form})] (left) and vector-form (b) [Stage 1-(\ref{eq:Stage 1: Individual-Level Model_Vector_Form}), Stage 2'-(\ref{eq:Stage 2: Population Model_Vector_Form_theory}) ($i$-indexing), and Stage 3-(\ref{eq:Stage 3: Prior_Vector_Form})] (right). Two vector forms are equivalent except for the way how the model matrix $\bm{\Theta}$ (\ref{eq:model_matrix}) is vectorized.}
\label{fig:framework_BNONMEM_two_vector_forms}
\end{figure}

To summarize, we derived two equivalent vectorized formulations representing the basic model (\ref{eq:Stage 1: Individual-Level Model}) -- (\ref{eq:Stage 3: Prior}) according to how the model matrix $\bm{\Theta}$ (\ref{eq:model_matrix}) is vectorized:
\begin{itemize}
\baselineskip=15pt
\item[] $\bullet$ \textbf{Vector-form (a)}: Stage 1-(\ref{eq:Stage 1: Individual-Level Model_Vector_Form}), Stage 2-(\ref{eq:Stage 2: Population Model_Vector_Form}) ($l$-indexing), and Stage 3-(\ref{eq:Stage 3: Prior_Vector_Form});
\item[] $\bullet$ \textbf{Vector-form (b)}: Stage 1-(\ref{eq:Stage 1: Individual-Level Model_Vector_Form}), Stage 2'-(\ref{eq:Stage 2: Population Model_Vector_Form_theory}) ($i$-indexing), and Stage 3-(\ref{eq:Stage 3: Prior_Vector_Form}).
\end{itemize}
Figure \ref{fig:framework_BNONMEM_two_vector_forms} displays DAG representations of the two vector forms of the basic model. In vector-form (a), $K$ latent nodes $\{\bm{\theta}_{l}\}_{l=1}^{K}$ are fully connected toward the $N$ response vectors $\{\textbf{y}_{i} \}_{i=1}^{N}$. On the other hand, in vector-form (b), $N$ latent nodes $\{\bm{\theta}^{i}\}_{i=1}^{N}$ are bijectively connected to the $N$ response vectors $\{\textbf{y}_{i} \}_{i=1}^{N}$ for each subject $i$. These two ways of looking at the framework of the Bayesian nonlinear hierarchical models complement a more proper understanding of the framework and provide modelers with a statistical insight. For example, vector-form (a) is useful to understand the mathematics for $P$ regressors in the population regression, while vector-form (b) makes us easy to comprehend the role of the population covariance matrix $\bm{\Omega}$.

\section{Likelihood}\label{sec:Likelihood}
\subsection{Outline}\label{subsec:Outline}
In this section, we investigate a likelihood function based on the basic model (\ref{eq:Stage 1: Individual-Level Model}) -- (\ref{eq:Stage 3: Prior}). As illustrated in Bayesian workflow in Subsection \ref{subsec:Bayesian workflow}, likelihood theory is fundamental of Bayesian inference (see Figure \ref{fig:Bayesian_Workflow}). Therefore, it is worth spending time to re-study the formulation of the likelihood function. Here, one caveat is that, due to the hierarchical nature of the nonlinear mixed effects model, a notion of the likelihood function depends on what part of the model specification is considered to be part of the likelihood, and what is not. Most papers directly consider the marginal likelihood that will be discussed in Subsection \ref{subsec:Likelihood from vector-form (b)}. In this paper, before marching there, we study other two formulations of the likelihood in Subsection \ref{subsec:Likelihood based on Stage 1} and \ref{subsec:Likelihood from vector-form (a)} to get some pedagogical insights. We will briefly discuss popularly used frequentist computing strategies in Subsection \ref{subsec:Likelihood from vector-form (b)}. 
\subsection{Likelihood based on Stage 1}\label{subsec:Likelihood based on Stage 1}
As in most of the statistical models, a natural starting point for inference is maximum likelihood estimation. We start with considering only Stage 1 from the basic model in Section \ref{subsec:Basic model} and ignore Stage 2 and 3 for now. Then the likelihood function for the $i$-th subject is 
\begin{align}
\nonumber
\mathcal{L}(\bm{\theta}^{i},\sigma^{2}|\textbf{y}_{i})
& 
=
p(\textbf{y}_{i}|\bm{\theta}^{i},\sigma^{2})
=
\mathcal{N}_{M_{i}}(\textbf{y}_{i}|
\bm{f}_{i}(\textbf{t}_{i}, \bm{\theta}^{i})
,
\sigma^{2}  \textbf{I} 
), \quad (i = 1, \cdots, N).
\end{align}
Therefore, the likelihood function based on the $N$ subjects $\textbf{y}_{1:N} = \{\textbf{y}_{i}\}_{i=1}^{N}$ is 
\begin{align}
\label{eq: likelihood for y_1:N}
\mathcal{L}(\bm{\Theta},\sigma^{2}|\textbf{y}_{1:N})
& =
\prod_{i=1}^{N}
\mathcal{N}_{M_{i}}(\textbf{y}_{i}|
\bm{f}_{i}(\textbf{t}_{i}, \bm{\theta}^{i})
,
\sigma^{2}  \textbf{I} 
).
\end{align}
Now, we maximize the likelihood (\ref{eq: likelihood for y_1:N}) with respect to the model matrix $\bm{\Theta}$ (\ref{eq:model_matrix}) given $\sigma^{2}$ fixed:
\begin{align}
\nonumber
\widehat{\bm{\Theta}}
 &= 
 \text{argmax}_{\bm{\Theta} \in \mathbb{R}^{K \times N}}
\log\
\mathcal{L}(
\bm{\Theta},\sigma^{2}|
\textbf{y}_{1:N})
\\
\nonumber
&=
 \text{argmin}_{
\bm{\Theta} \in \mathbb{R}^{K \times N} 
 }
\sum_{i=1}^{N}
\|
\textbf{y}_{i}
-
\bm{f}_{i}(\textbf{t}_{i}, \bm{\theta}^{i})
\|_2^{2}\\
\label{eq:non_info_borrowing}
&=
 \text{argmin}_{
\bm{\Theta} \in \mathbb{R}^{K \times N} 
 }
\sum_{i=1}^{N}
\sum_{j = 1}^{M_{i}}
\Big(
y_{it}
-
f(t_{ij} ; \bm{\theta}^{i})
\Big)^{2}
,
\end{align}
where $\|\cdot \|_{2}$ is the Euclidean norm.

The estimator $\widehat{\bm{\Theta}} 
=
[
\widehat{\bm{\theta}}^{1} 
\cdots
\widehat{\bm{\theta}}^{i }
\cdots
\widehat{\bm{\theta}}^{N}
]
\in \mathbb{R}^{K \times N}$ of the model matrix $\bm{\Theta}$ (\ref{eq:model_matrix}) can be obtained by various optimization techniques such as Newton-Raphson method or Gradient descent method \citep{boyd2004convex}. Noting from the summation across $i$ in (\ref{eq:non_info_borrowing}), $N$ estimators $\{\widehat{\bm{\theta}}^{i}\}_{i=1}^{N}$ are independent. We can obtain an estimator of the variance $\sigma^{2}$ by plugging $\widehat{\bm{\Theta}}$ into the likelihood (\ref{eq: likelihood for y_1:N}), and then maximize with respect to the $\sigma^{2}$. To investigate a denoised temporal tendency for the trajectory $\textbf{y}_{i}$, we can simply plug $\widehat{\bm{\theta}}^{i}$ into the function $f(t_{ij} ; \bm{\theta}^{i})$ $(j=1,\cdots,M_{i})$. To see a future pattern, we can extrapolate the function by extending the time index beyond the last time point $t_{iM_{i}} $. Eventually, the illustrated approach is based on traditional least squares estimation. 

Unfortunately, there are three major drawbacks in this approach. First, it forfeits the opportunity to use `information borrowing' \citep{lee2020estimation} to improve a predictive accuracy due to the ignorance of Stage 2. What happens in Stage 2  (\ref{eq:Stage 2: Population Model}) is to borrow strength across $N$ individuals to produce a better estimator for $\bm{\Theta}$ (\ref{eq:model_matrix}) than an estimator simply based on individual data. A similar issue can be found in the Clemente problem from \citep{efron2010future} where the James-Stein estimator \citep{james1992estimation} better predicts than an individual hitter-based estimator. Another example applied to epidemic data can be found in \citep{lee2020estimation}. Second, it does not well-aligned with the generic motivation to use the mixed effects models whose primary purpose is to understand “typical” values for the model parameters in $f$, representing whole subjects, which should be addressed by making an inference about the parameters $\bm{\alpha}$, $\bm{B}$, and $\bm{\Omega}$. Third, it only produces point estimates for the parameters, failing to describe the underlying uncertainty. 

A remedy of the first two drawbacks is the consideration of Stage 1 (\ref{eq:Stage 1: Individual-Level Model}) and 2 (\ref{eq:Stage 2: Population Model}) hierarchically in a single model, leading to a frequentist version of nonlinear mixed effects models, which will be discussed in Subsection \ref{subsec:Likelihood from vector-form (a)} and \ref{subsec:Likelihood from vector-form (b)}. To describe relevant uncertainty within the frequentist framework one may resort to bootstrap methods \citep{efron1982jackknife}. Another solution resolving all the three drawbacks at once is to incorporate Stage 1 (\ref{eq:Stage 1: Individual-Level Model}), 2 (\ref{eq:Stage 2: Population Model}), and 3 (\ref{eq:Stage 3: Prior}) in a fully Bayesian way, resulting in a Bayesian version of nonlinear mixed effects models, which is the main topic in this paper. 
\subsection{Likelihood based on Stage 1 and 2 from vector-form (a)}\label{subsec:Likelihood from vector-form (a)}
A likehood function based on vector-form (a) is derived.  More specifically, we examine the frequentist setting where the assumptions of Stage 1-(\ref{eq:Stage 1: Individual-Level Model_Vector_Form}) and Stage 2-(\ref{eq:Stage 2: Population Model_Vector_Form}) are considered, while the parameters introduced in Stage 3-(\ref{eq:Stage 3: Prior_Vector_Form}) are fixed (i.e., no prior assumptions). 

The individual model on Stage 1 (\ref{eq:Stage 1: Individual-Level Model_Vector_Form}) yields a conditional density $p(\textbf{y}_{i} | \bm{\theta}^{i},\sigma^{2}) = \mathcal{N}_{M_{i}} (\textbf{y}_{i} | \bm{f}_{i}(\textbf{t}_{i}, \bm{\theta}^{i}) , \sigma^{2}I )  $ for each subject $i = 1,\cdots,N$. Under the population assumption on Stage 2 (\ref{eq:Stage 2: Population Model_Vector_Form}), we have the density $p(\bm{\theta}_{l} |\alpha_{l},\bm{\beta}_{l},\omega_{l}^{2}) = \mathcal{N}_{N} (\bm{\theta}_{l}|\textbf{1} \alpha_{l} + \textbf{X}\bm{\beta}_{l},\omega_{l}^{2}\textbf{I}) $ for each model parameter index $l=1,\cdots,K$. The joint density of $(\textbf{y}_{1:N},\bm{\theta}_{1:K})$ given parameters $\sigma^{2},\bm{\alpha} ,\bm{B}$ and $\bm{\Omega}$ is a product-form distribution:
\begin{align*}
p(\textbf{y}_{1:N},\bm{\theta}_{1:K}
|
\sigma^{2},\bm{\alpha} ,\bm{B},\bm{\Omega} 
) &=
\bigg\{
\prod_{i =1}^{N}
 \mathcal{N}_{M_{i}}(\textbf{y}_{i}|
\bm{f}_{i}(\textbf{t}_{i}, \bm{\theta}^{i})
,\sigma^{2} \mathbf{I})
\bigg\}
\cdot
\bigg\{
\prod_{l =1}^{K}
\mathcal{N}_{N} (\bm{\theta}_{l}|\textbf{1} \alpha_{l} + \textbf{X}\bm{\beta}_{l},\omega_{l}^{2}\textbf{I}) 
\bigg\},
\end{align*}
where $\textbf{y}_{1:N} = \{\textbf{y}_{1},\textbf{y}_{2},\cdots,\textbf{y}_{N}\}$ and $\bm{\theta}_{1:K} = \{\bm{\theta}_{1},\bm{\theta}_{2},\cdots,\bm{\theta}_{K}\}$. Now, the next step is to integrate out the latent model parameters $\bm{\theta}_{1:K}$ (or equivalently, the model matrix $\bm{\Theta}$ (\ref{eq:model_matrix})) from the density above to get a likelihood for the $(\sigma^{2},\bm{\alpha} ,\bm{B},\bm{\Omega})$:
\begin{align}
\label{eq:marginal_density_Y_vector_form_a}
\mathcal{L}(
\sigma^{2},\bm{\alpha} ,\bm{B},\bm{\Omega} 
|
\textbf{y}_{1:N})
=
\int 
p(\textbf{y}_{1:N},\bm{\theta}_{1:K}
|
\sigma^{2},\bm{\alpha} ,\bm{B},\bm{\Omega} 
)
d \bm{\theta}_{1:K}.
\end{align}
In most cases, the integral (\ref{eq:marginal_density_Y_vector_form_a}) is not tractable due to the non-linearity of the function $\bm{f}_{i}(\textbf{t}_{i}, \bm{\theta}^{i})$ with respect to the $\bm{\theta}^{i}$. Although it may be possible to use numerical techniques for the evaluation of the integral (\ref{eq:marginal_density_Y_vector_form_a}), this might require enormous computational effort, which is not really appreciated in the literature due to the high-dimensionality of the integral involving the $KN$ dimensional model parameters $\bm{\theta}_{1:K}$.

\subsection{Likelihood based on Stage 1 and 2' from vector-form (b)}\label{subsec:Likelihood from vector-form (b)}
A likehood function based on vector-form (b) adopting $i$-indexing is derived here. As similar to Subsection \ref{subsec:Likelihood from vector-form (a)}, we preserve the assumption of Stage 1-(\ref{eq:Stage 1: Individual-Level Model_Vector_Form}) and Stage 2'-(\ref{eq:Stage 2: Population Model_Vector_Form_theory}), but work with fixing the parameters in Stage 3-(\ref{eq:Stage 3: Prior_Vector_Form}). In these specification, for each index $i=1,\cdots,N$, the individual model on Stage 1 (\ref{eq:Stage 1: Individual-Level Model_Vector_Form}) and population model on Stage 2 (\ref{eq:Stage 2: Population Model_Vector_Form_theory}) lead to densities $p(\textbf{y}_{i} | \bm{\theta}^{i},\sigma^{2}) = \mathcal{N}_{M_{i}} (\textbf{y}_{i} | \bm{f}_{i}(\textbf{t}_{i}, \bm{\theta}^{i}) , \sigma^{2}I )  $ and $p(\bm{\theta}^{i} |\bm{\alpha} ,\bm{B},\bm{\Omega}) = \mathcal{N}_{K} (\bm{\theta}^{i}|\bm{\alpha} + \bm{B} \textbf{x}_{i} , \bm{\Omega}) $, respectively. Thus, the joint density of $(\textbf{y}_{i} , \bm{\theta}^{i})$ given parameters $\sigma^{2},\bm{\alpha} ,\bm{B}$ and $\bm{\Omega}$ is 
\begin{align*}
p(\textbf{y}_{i} , \bm{\theta}^{i} |\sigma^{2},\bm{\alpha} ,\bm{B},\bm{\Omega} ) &= p(\textbf{y}_{i} | \bm{\theta}^{i},\sigma^{2}) 
\cdot p(\bm{\theta}^{i} |\bm{\alpha} ,\bm{B},\bm{\Omega})\\
&=
\mathcal{N}_{M_{i}} (\textbf{y}_{i} | \bm{f}_{i}(\textbf{t}_{i}, \bm{\theta}^{i}) , \sigma^{2}\textbf{I} ) 
\cdot
\mathcal{N}_{K} (\bm{\theta}^{i}|\bm{\alpha} + \bm{B} \textbf{x}_{i} , \bm{\Omega}).
\end{align*}
Given the parameters $(\sigma^{2},\bm{\alpha} ,\bm{B},\bm{\Omega})$, the ordered pairs in the collection $\{ (\textbf{y}_{i} , \bm{\theta}^{i}) \}_{i=1}^{N}$ are conditionally independent across individuals. Therefore, a likelihood for the $(\sigma^{2},\bm{\alpha} ,\bm{B},\bm{\Omega})$ is based on the marginal density of $\textbf{y}_{1:N}=\{\textbf{y}_{1},\textbf{y}_{2},\cdots,\textbf{y}_{N} \}$:
\begin{align}
\label{eq:likelihood_of_NONMEM}
\mathcal{L}(
\sigma^{2},\bm{\alpha} ,\bm{B},\bm{\Omega} 
|
\textbf{y}_{1:N})
&=
\prod_{i=1}^{N}
\int 
\mathcal{N}_{M_{i}} (\textbf{y}_{i} | \bm{f}_{i}(\textbf{t}_{i}, \bm{\theta}^{i}) , \sigma^{2}\textbf{I} ) 
\cdot
\mathcal{N}_{K} (\bm{\theta}^{i}|\bm{\alpha} + \bm{B} \textbf{x}_{i} , \bm{\Omega})
d \bm{\theta}^{i}\\
\label{eq:likelihood_of_NONMEM_eta}
&=
\prod_{i=1}^{N}
\int 
\mathcal{N}_{M_{i}} (\textbf{y}_{i} | \bm{f}_{i}(\textbf{t}_{i}, 
\bm{\alpha} + \bm{B} \textbf{x}_{i} + \bm{\eta}^{i}
) , \sigma^{2}\textbf{I} ) 
\cdot
\mathcal{N}_{K} (\bm{\eta}^{i}|\textbf{0} , \bm{\Omega})
d \bm{\eta}^{i},
\end{align}
where the last equality is derived by using the change of variable (\ref{eq:Stage 2: Population Model_Vector_Form_theory}). The last expression (\ref{eq:likelihood_of_NONMEM_eta}) is a standard mathematical formulation that many frequentist computing strategies are constructed: see Equation (3.2) from \citep{davidian2003nonlinear}.

As the model parameter $\bm{\theta}^{i}$ in (\ref{eq:likelihood_of_NONMEM}) (or similarly, $\bm{\eta}^{i}$ in (\ref{eq:likelihood_of_NONMEM_eta}) which is often called random effect in the frequentist framework) participates to the function $f$ in a non-linear fashion, the integral generally cannot be obtained in a closed-form. Benefiting from a conditional independence \citep{dawid1979conditional}, dimensionality of the $N$ integrals (\ref{eq:likelihood_of_NONMEM}) is much lower than that of the integral (\ref{eq:marginal_density_Y_vector_form_a}) based on vector-form (a). Analytically, the likelihood functions of the basic model (\ref{eq:Stage 1: Individual-Level Model}) -- (\ref{eq:Stage 3: Prior}) based on vector-form (a) (\ref{eq:marginal_density_Y_vector_form_a}) and vector-form (b) (\ref{eq:likelihood_of_NONMEM}) may be equivalent. That being said, minimization of the two functions with respect to the parameters $(\sigma^{2},\bm{\alpha},\bm{B},\bm{\Omega})$ yields the same solution, $(\widehat{\sigma^{2}},\widehat{\bm{\alpha}},\widehat{\bm{B}},\widehat{\bm{\Omega}})$, so-called maximum likelihood estimators (MLE). 

We shall briefly discuss on MLE computations. One approach would be to perform a multivariate numerical integration (for e.g., Gauss-Hermite quadrature \citep{liu1994note}) to each of the $N$ integrals (\ref{eq:likelihood_of_NONMEM}), and then obtain the MLE by maximizing the product of the $N$ numerical integrals with respect to the parameters $(\sigma^{2},\bm{\alpha},\bm{B},\bm{\Omega})$ \citep{hedeker1994random}. This approach turns out to be computationally so expensive and may have poor converge properties due to the following two reasons \citep{vonesh2002conditional}. First, the numerical integration necessitates increasingly expensive iterative procedures within a MLE algorithm as the correlation of the model parameters (or equivalently, random effects) increases. Second, convergence property may be highly deteriorated when the number of model parameters $K$ is large (i.e., high-dimensional integral) and the number of sampling times $M_{i}$ is small (i.e., sparse data) due to the `curse of dimensionality' \citep{hinrichs2014curse}.

A class of common approaches for the  MLE computations is based on analytical approximation to each of the $N$ integrals (\ref{eq:likelihood_of_NONMEM_eta}) \citep{beal1982estimating,lindstrom1990nonlinear,vonesh1992mixed,goldstein1991nonlinear,vonesh1996note},  and some of them have been successfully adopted to industrial software like \textsc{NONMEM} \citep{beal1992nonmem,bauer2019nonmem} and \textsc{SAS} \citep{sas1999sas}. Here, we illustrate a key idea of the first-order method attributed to \citep{beal1982estimating}. Let us define a mapping $\bm{g}_{i}(\bm{\eta}^{i})=\bm{f}_{i}(\textbf{t}_{i}, \bm{\alpha} + \bm{B} \textbf{x}_{i} + \bm{\eta}^{i}) : A \subset \mathbb{R}^{K} \rightarrow  \mathbb{R}^{M_{i}}$ for each subject $i$, where $A$ is an open set with $\textbf{0} \in A$. Suppose that $\bm{g}_{i}(\bm{\eta}^{i})$ is smooth on the set $A$: then, by Taylor's theorem (page 375 of \citep{marsden1993elementary}), we have the best linear approximation of the mapping $\bm{g}_{i}(\bm{\eta}^{i})$ at the origin $\textbf{0}$ given by $\bm{g}_{i}(\bm{\eta}^{i}) \approx  \bm{g}_{i}(\textbf{0}) + \textbf{D}\bm{g}_{i}(\textbf{0}) \bm{\eta}^{i}$, where $\textbf{D}\bm{g}_{i}(\textbf{0})\in \mathbb{R}^{M_{i} \times K}$ is the Jacobian matrix of $\bm{g}_{i}(\bm{\eta}^{i})$ at $\textbf{0}$. Now, we shall replace the function $\bm{g}_{i}(\bm{\eta}^{i}) = \bm{f}_{i}(\textbf{t}_{i}, \bm{\alpha} + \bm{B} \textbf{x}_{i} + \bm{\eta}^{i})$ in integral (\ref{eq:likelihood_of_NONMEM_eta}) with the resulting approximation $\bm{g}_{i}(\textbf{0}) + \textbf{D}\bm{g}_{i}(\textbf{0}) \bm{\eta}^{i}$ for each $i$ ($i=1,\cdots,N$), leading to a closed-form expression

\begin{align}
\label{eq:likelihood_of_NONMEM_FOCE}
\widetilde{\mathcal{L}}(
\sigma^{2},\bm{\alpha} ,\bm{B},\bm{\Omega} 
|
\textbf{y}_{1:N})
&=
\prod_{i=1}^{N}
\int 
\mathcal{N}_{M_{i}} (\textbf{y}_{i} | \bm{g}_{i}(\textbf{0}) + \textbf{D}\bm{g}_{i}(\textbf{0}) \bm{\eta}^{i}
 , \sigma^{2} \textbf{I} ) 
\cdot
\mathcal{N}_{K} (\bm{\eta}^{i}|\textbf{0} , \bm{\Omega})
d \bm{\eta}^{i}\\
\nonumber
&=
\prod_{i=1}^{N}
\mathcal{N}_{M_{i}}
(\textbf{y}_{i} | 
\bm{g}_{i}(\textbf{0}),
 \textbf{D}\bm{g}_{i}(\textbf{0})
 \bm{\Omega}
  \textbf{D}\bm{g}_{i}(\textbf{0})^{\top}
  +
   \sigma^{2} \textbf{I} 
).
\end{align}
To summarize, a linearization was used to convert the nonlinear mixed effects model to  a linear mixed effects model, in some sense, equivalent to Lindley-Smith form \citep{lindley1972bayes}. This enables us to integrate out the random vector $\bm{\eta}^{i}$ from the $N$ integrals (\ref{eq:likelihood_of_NONMEM_FOCE}), deriving a marginal likelihood (\ref{eq:likelihood_of_NONMEM_FOCE}) to approximate the exact marginal likelihood (\ref{eq:likelihood_of_NONMEM}). MLE $(\widehat{\sigma^{2}},\widehat{\bm{\alpha}},\widehat{\bm{B}},\widehat{\bm{\Omega}})$ can be obtained by jointly maximizing $\widetilde{\mathcal{L}}(
\sigma^{2},\bm{\alpha} ,\bm{B},\bm{\Omega} 
|
\textbf{y}_{1:N})$ (\ref{eq:likelihood_of_NONMEM_FOCE}) assuming the approximation is exact. 

An another way to compute the MLE is through the use of expectation-maximization (EM) algorithm \citep{dempster1977maximum}. Borrowing terms from EM updating process \citep{meng1991using}, $\textbf{y}_{i}$, $\bm{\theta}^{i}$, $(\textbf{y}_{i}, \bm{\theta}^{i})$, $(\sigma^{2},\bm{\alpha} ,\bm{B},\bm{\Omega})$ and $p(\textbf{y}_{i} , \bm{\theta}^{i} |\sigma^{2},\bm{\alpha} ,\bm{B},\bm{\Omega} )$ (i.e., the integrand in (\ref{eq:likelihood_of_NONMEM})) can be viewed as observable incomplete data, missing data, complete data, unknown parameters, and density of complete data, respectively, for the $i$-th subject. The goal is to maximize the exact marginal likelihood $\mathcal{L}(
\sigma^{2},\bm{\alpha} ,\bm{B},\bm{\Omega} 
|
\textbf{y}_{1:N})$ (\ref{eq:likelihood_of_NONMEM}) by iterating E-step and M-step, leading to the MLE $(\widehat{\sigma^{2}},\widehat{\bm{\alpha}},\widehat{\bm{B}},\widehat{\bm{\Omega}})$. The E-step computes a conditional expected log-likelihood of $(\sigma^{2},\bm{\alpha} ,\bm{B},\bm{\Omega})$ based on the hierarchy (\ref{eq:Stage 1: Individual-Level Model}) -- (\ref{eq:Stage 2: Population Model}), followed by the M-step that maximizes the function with respect to $(\sigma^{2},\bm{\alpha} ,\bm{B},\bm{\Omega})$. The nonlinearity associated with the model matrix $\bm{\Theta}$ (\ref{eq:model_matrix}) makes the E-step intractable. As a remedy, variant versions of the EM algorithm are proposed; see \citep{walker1996algorithm,allassonniere2021new,kuhn2005maximum,samson2007saem} for a technical detail applied to a hierarchy similar to the basic model. Among them, the scheme of stochastic approximation EM algorithm proposed by \citep{delyon1999convergence}, splitting the E-step into two steps, namely a simulation step and a stochastic approximation step, is widely used in many applications for its numerical stability, fast computation, and theoretical soundness \citep{kuhn2004coupling,allassonniere2010construction}, which has been successfully deployed as industrial software including \textsc{Monolix} \citep{lavielle2005monolix} as well as open source software such as \textsc{R} package \textsc{nlmixr} \citep{fidler2019nonlinear}.
\section{Bayesian inference and implementation}\label{sec:Bayesian inference and implementation}
\subsection{Bayesian inference}\label{subsec:Bayesian inference}
We briefly overview two contrasting workflows of Bayesian and frequentist approaches for nonlinear mixed effects models before moving to a technical detail. Both settings allow the randomness in the model matrix $\bm{\Theta}$ (\ref{eq:model_matrix}), but then, they diverge when it comes to how parameters $(\sigma^{2},\bm{\alpha},\bm{B},\bm{\Omega})$ are treated. Bayesians treat $(\sigma^{2},\bm{\alpha},\bm{B},\bm{\Omega})$ as random, while frequentists regard it as fixed. To conceptualize a subtlety arising from this difference, let us recap frequentist computing strategies discussed in Subsection \ref{subsec:Likelihood from vector-form (b)}. There, the model matrix $\bm{\Theta}$ was eventually integrated out from the joint density of $(\textbf{y}_{1:N},\bm{\Theta})$, either approximately or exactly, to derive a marginal likelihood of $(\sigma^{2},\bm{\alpha},\bm{B},\bm{\Omega})$ from which the MLE $(\widehat{\sigma^{2}},\widehat{\bm{\alpha}},\widehat{\bm{B}},\widehat{\bm{\Omega}})$ is computed via various optimization methods. After that, frequentists apply standard Bayesian formulas, such as posterior density, posterior mean, and so on, to estimate $\bm{\Theta}$ \citep{demidenko2013mixed}. 

In contrast, to drive the Bayesian engine, one would need an appropriate prior $\pi(\sigma^{2},\bm{\alpha},\bm{B},\bm{\Omega})$. After that, the entire collection of parameters $(\bm{\Theta}, \sigma^{2},\bm{\alpha},\bm{B},\bm{\Omega})$ will be updated through the Bayes’ theorem post observing the data $\textbf{y}_{1:N}$, leading to the posterior density $\pi(\bm{\Theta},\sigma^{2},\bm{\alpha},\bm{B},\bm{\Omega}|\textbf{y}_{1:N})$ \citep{bernardo2009bayesian}. The essence of the Bayesian viewpoint is that there is no logical distinction between $\bm{\Theta}$ and $(\sigma^{2},\bm{\alpha},\bm{B},\bm{\Omega})$, which are associated with the random and fixed effects, respectively, from the frequentist perspective. In Bayesian framework, both $\bm{\Theta}$ and $(\sigma^{2},\bm{\alpha},\bm{B},\bm{\Omega})$ are random quantities. It is important to point out that the likelihood principle is naturally incorporated in the Bayes’ theorem \citep{lindley1972bayesian}. Clearly, modern data complications such as enormous volume, large dimensionality, and multi-level structures may necessitate a sophistication on the prior specifications.

We are now in a position to describe the Bayesian analysis for the basic model (\ref{eq:Stage 1: Individual-Level Model})--(\ref{eq:Stage 3: Prior}), that assumed independence, a \emph{priori}, for each parameter blocks $\sigma^{2}$, $\bm{\alpha},$ $\bm{B}$, and $\bm{\Omega}$, $\pi(\sigma^{2},\bm{\alpha},\bm{B},\bm{\Omega}) = \pi(\sigma^2)
\cdot
\pi(\bm{\alpha})
\cdot
\pi(\bm{B})
\cdot
\pi(\bm{\Omega})$. (Our logic below can be generalized to a more complex prior setting.) As discussed in Section \ref{sec:The model}, it is the discretion of modeler how she would treat the model matrix $\bm{\Theta}$ (\ref{eq:model_matrix}) with $l$-indexing $\bm{\theta}_{1:K} = \{\bm{\theta}_{l} \}_{l=1}^{K}$ or $i$-indexing $\bm{\theta}^{1:N} = \{\bm{\theta}^{i} \}_{i=1}^{N}$, leading to vector forms (a) and (b), respectively. For the sake of readability, we illustrate the Bayesian inference by using the vector-form (a), but we will sometimes use the vector-form (b) when this seems more understandable.


A central task in the application of the Bayesian nonlinear mixed effects models is to evaluate the posterior density, or indeed to compute expectation with respect to the density:
\begin{align*}
\pi(\bm{\Theta},\sigma^2,  \bm{\alpha}, \bm{B}, \bm{\Omega} |\textbf{y}_{1:N})
&=
\frac{\pi(\textbf{y}_{1:N},\bm{\Theta},\sigma^2,  \bm{\alpha}, \bm{B}, \bm{\Omega})}{p(\textbf{y}_{1:N})}
\\
&\propto 
\pi(\textbf{y}_{1:N},\bm{\Theta},\sigma^2,  \bm{\alpha}, \bm{B}, \bm{\Omega})
\\
&=
\underbrace{p(\textbf{y}_{1:N} | \bm{\Theta},\sigma^2)}_{\text{Stage 1}}
\cdot
\underbrace{\pi(\bm{\Theta}| \bm{\alpha}, \bm{B}, \bm{\Omega})}_{\text{Stage 2}}
\cdot
\underbrace{\pi(\sigma^2)
\cdot
\pi(\bm{\alpha})
\cdot
\pi(\bm{B})
\cdot
\pi(\bm{\Omega})}_{\text{Stage 3}},
\end{align*}
where the last equation can be detailed as follows
\begin{align}
\label{eq:joint_posterior}
\bigg\{
\prod_{i =1}^{N}
\mathcal{N}_{M_{i}} (\textbf{y}_{i} | \bm{f}_{i}(\textbf{t}_{i}, \bm{\theta}^{i})
,\sigma^{2} \mathbf{I})
\bigg\}
\cdot
\bigg\{
\prod_{l =1}^{K}
\mathcal{N}_{N}
(
\bm{\theta}_{l}
|
\textbf{1}
\alpha_{l}
+\textbf{X}\bm{\beta}_{l}
,\omega_{l}^2 \mathbf{I})
\pi(\alpha_{l})
\pi( \omega_{l}^{2})
\pi(\bm{\beta}_{l})
\bigg\}\cdot
\pi(\sigma^{2}).
\end{align}
From a Bayesian perspective, all inferential problems regarding the parameter $(\bm{\Theta},\sigma^2,  \bm{\alpha}, \bm{B}, \bm{\Omega} )$ may be addressed in terms of the posterior distribution $\pi(\bm{\Theta},\sigma^2,  \bm{\alpha}, \bm{B}, \bm{\Omega} |\textbf{y}_{1:N})$ (\ref{eq:joint_posterior}). Unfortunately, for almost all problems, the distribution is intractable. In such situations, we need to resort to approximation techniques, and these fall broadly into two classes, according to whether they rely on stochastic \citep{casella1992explaining,neal2011mcmc,murray2010elliptical,neal2003slice} or deterministic \citep{ranganath2014black,wang2013variational,minka2013expectation,blei2017variational} approximations. See \citep{andrieu2003introduction,zhang2018advances} for review papers for these techniques. In this article, we are mainly focused on the stochastic approximation. The basic idea behind the methodology is to construct a Markov chain whose stationary distribution is the posterior distribution $\pi(\bm{\Theta},\sigma^2,  \bm{\alpha}, \bm{B}, \bm{\Omega} |\textbf{y}_{1:N})$ (\ref{eq:joint_posterior}).
\subsection{Gibbs sampling algorithm}\label{subsec:Gibbs sampling algorithm}
We resort to MCMC technique \citep{andrieu2003introduction} to sample from the full joint density $\pi(\bm{\Theta},\sigma^2,  \bm{\alpha}, \bm{B}, \bm{\Omega} |\textbf{y}_{1:N})$ (\ref{eq:joint_posterior}). Among many MCMC techniques, we use the Gibbs sampling algorithm \citep{casella1992explaining,lee2021gibbs} to exploit the conditional independence \citep{dawid1979conditional} induced by the hierarchical formulation. A generic Gibbs sampler would cycle in turn through each of the conditional distributions for the parameter blocks $\bm{\Theta},\sigma^2,  \bm{\alpha}, \bm{B}$, and $\bm{\Omega}$ as follows:
\begin{itemize}
\baselineskip=15pt
\item[] \textbf{\emph{Step 1.}} Sample $\bm{\Theta}$ from its full conditional distribution 
\begin{align}
\label{eq:full_condi_theta}
\pi(\bm{\Theta}|\sigma^2,  \bm{\alpha}, \bm{B}, \bm{\Omega},\textbf{y}_{1:N})&\propto
\bigg\{
\prod_{i =1}^{N}
\mathcal{N}_{M_{i}} (\textbf{y}_{i} | \bm{f}_{i}(\textbf{t}_{i}, \bm{\theta}^{i})
,\sigma^{2} \mathbf{I})
\bigg\}
\cdot
\bigg\{
\prod_{l =1}^{K}
\mathcal{N}_{N}
(
\bm{\theta}_{l}
|
\textbf{1}
\alpha_{l}
+\textbf{X}\bm{\beta}_{l}
,\omega_{l}^2 \mathbf{I})
\bigg\};
\end{align}
\item[] \textbf{\emph{Step 2.}} Sample $\sigma^2$ from its full conditional distribution
\begin{align}
\label{eq:full_condi_sigma_sq}
\pi(\sigma^2|\bm{\Theta}, \bm{\alpha}, \bm{B}, \bm{\Omega},\textbf{y}_{1:N})&\propto
\bigg\{
\prod_{i =1}^{N}
\mathcal{N}_{M_{i}} (\textbf{y}_{i} | \bm{f}_{i}(\textbf{t}_{i}, \bm{\theta}^{i})
,\sigma^{2} \mathbf{I})
\bigg\}
\cdot
\pi(\sigma^{2});
\end{align}
\item[] \textbf{\emph{Step 3.}} Sample $\bm{\alpha}$ from its full conditional distribution 
\begin{align}
\label{eq:full_condi_alpha}
\pi(\bm{\alpha}|\sigma^2, \bm{\Theta}, \bm{B}, \bm{\Omega},\textbf{y}_{1:N})&\propto
\prod_{l =1}^{K}
\mathcal{N}_{N}
(
\bm{\theta}_{l}
|
\textbf{1}
\alpha_{l}
+\textbf{X}\bm{\beta}_{l}
,\omega_{l}^2 \mathbf{I}) \cdot
\pi(\alpha_{l});
\end{align}
\item[] \textbf{\emph{Step 4.}} Sample $\bm{B}$ from its full conditional distribution 
\begin{align}
\label{eq:full_condi_B}
\pi(\bm{B} | \bm{\Theta}, \sigma^2,  \bm{\alpha}, \bm{\Omega},\textbf{y}_{1:N})&\propto
\prod_{l =1}^{K}
\mathcal{N}_{N}
(
\bm{\theta}_{l}
|
\textbf{1}
\alpha_{l}
+\textbf{X}\bm{\beta}_{l}
,\omega_{l}^2 \mathbf{I}) \cdot
\pi(\bm{\beta}_{l});
\end{align}
\item[] \textbf{\emph{Step 5.}} Sample $\bm{\Omega}$ from its full conditional distribution 
\begin{align}
\label{eq:full_condi_Omega}
\pi(\bm{\Omega} | \bm{\Theta},\sigma^2,  \bm{\alpha}, \bm{B},\textbf{y}_{1:N})&\propto
\prod_{l =1}^{K}
\mathcal{N}_{N}
(
\bm{\theta}_{l}
|
\textbf{1}
\alpha_{l}
+\textbf{X}\bm{\beta}_{l}
,\omega_{l}^2 \mathbf{I}) \cdot
\pi( \omega_{l}^{2}).
\end{align}
\end{itemize}
Sampling the model matrix $\bm{\Theta}$ (\ref{eq:model_matrix}) at \textbf{\emph{Step 1}} is independent of the choice of the priors, which we discuss shortly. On the other hand, sampling the parameters $(\sigma^2,  \bm{\alpha}, \bm{B}, \bm{\Omega})$ at \textbf{\emph{Step 2}}, \textbf{\emph{3}}, \textbf{\emph{4}}, and \textbf{\emph{5}} depends on the prior choices for the parameters in Stage 3 (\ref{eq:Stage 3: Prior_Vector_Form}); we discuss this topic in Section \ref{sec:Prior options}. 
\subsection{Parallel computation for model matrix}\label{subsec:Parallel computation for model matrix}
One of the most computer-intensive steps to implement the Gibbs sampler in Subsection \ref{subsec:Gibbs sampling algorithm} is \textbf{\emph{Step 1}} to sample the model matrix $\bm{\Theta} \in \mathbb{R}^{K \times N}$ (\ref{eq:model_matrix}), or equivalently its entries $\{\theta_{li} \}_{l=1,i=1}^{K,N}$, from the full conditional distribution $\pi(\bm{\Theta}|\sigma^2,  \bm{\alpha}, \bm{B}, \bm{\Omega},\textbf{y}_{1:N})$ (\ref{eq:full_condi_theta}). Clearly, the nonlinear participation of the model parameters to the function $f$ makes the conditional distribution intractable, hence, non-conjugate sampling is unavoidable, which may suffer from a slow convergence. At the same times, due to the Markovian nature of the Gibbs algorithm, it is difficult to parallelize the whole steps of the Gibbs sampler, which creates difficulties in slower languages like \textsc{R} \citep{team2013language}. Nevertheless, the increasing number of parallel cores that are available at a very low price drives more and more interest in `parallel sampling algorithms' that can benefit from the available parallel processing units on computers \citep{lee2010utility,suchard2010understanding}.

We suggest a framework of parallel computations to efficiently update the model matrix $\bm{\Theta} \in \mathbb{R}^{K \times N}$. This framework can be particularly appreciated under the setting of Bayesian nonlinear mixed effects models when the number of subjects $N$ is a lot larger than the number of model parameters $K$ ($N \gg K$). 

The first version of parallel sampling algorithms is based on scalar updating. For the derivation, we start with analyzing the full conditional posterior distribution of a single element $\theta_{li}$ ($l=1,\cdots,K; i=1,\cdots,N$):
\begin{align}
\nonumber
\pi(\theta_{li}|-) &= \pi(\theta_{li}|\theta_{-li},\sigma^2,  \bm{\alpha}, \bm{B}, \bm{\Omega},\textbf{y}_{1:N})
\propto
\pi(\bm{\Theta}|\sigma^2,  \bm{\alpha}, \bm{B}, \bm{\Omega},\textbf{y}_{1:N})\\
\nonumber
&\propto
\bigg\{
\prod_{i =1}^{N}
\mathcal{N}_{M_{i}} (\textbf{y}_{i} | \bm{f}_{i}(\textbf{t}_{i}, \theta_{1i},\cdots,\theta_{li},\cdots,\theta_{Ki}  )
,\sigma^{2} \mathbf{I})
\bigg\}
\cdot
\bigg\{
\prod_{l =1}^{K}
\prod_{i =1}^{N}
\mathcal{N}
(\theta_{li}
|
\alpha_{l}
+\textbf{x}_{i}^{\top}\bm{\beta}_{l}
,\omega_{l}^2)
\bigg\}
\\
\label{eq:full_condi_theta_ind}
&\propto
\mathcal{N}_{M_{i}} (\textbf{y}_{i} | \bm{f}_{i}(\textbf{t}_{i}, \theta_{1i},\cdots,\theta_{li},\cdots,\theta_{Ki}  )
,\sigma^{2} \mathbf{I})
\cdot
\mathcal{N}
(\theta_{li}
|
\alpha_{l}
+\textbf{x}_{i}^{\top}\bm{\beta}_{l}
,\omega_{l}^2),
\end{align}
where the notation $\theta_{-li}$ represents the all the entries of $\bm{\Theta}$ except for $\theta_{li}$, that is, $\theta_{-li} = \{ \theta_{li} \}_{l=1,i=1}^{K,N} - \{\theta_{li} \}$. Here, we used a conventional notation in Bayesian computation: `$\pi(\theta_{li}|-)$' indicates the density $\pi(\theta_{li}|\theta_{-li},\sigma^2,  \bm{\alpha}, \bm{B}, \bm{\Omega},\textbf{y}_{1:N})$, where the notation `$-$' in $\pi(\theta_{li}|-)$ implies all the parameters except for the $\theta_{li}$ in the basic model (\ref{eq:Stage 1: Individual-Level Model}) -- (\ref{eq:Stage 3: Prior}) along with $N$ observations.

\begin{figure}[h!]
\centering
\includegraphics[scale=0.5]{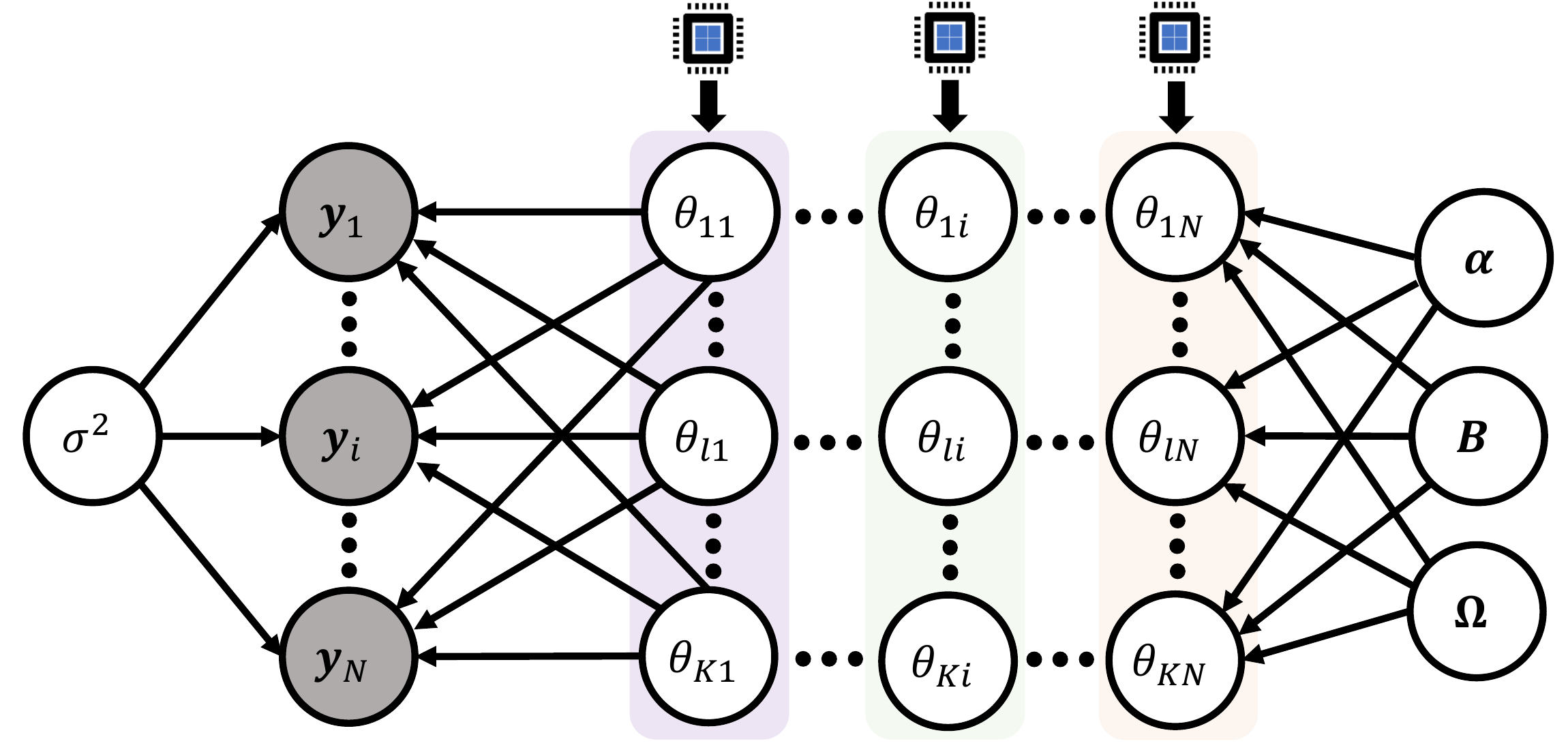}
\caption{
A pictorial description on the use of parallel computation to the basic model (\ref{eq:Stage 1: Individual-Level Model}) -- (\ref{eq:Stage 3: Prior}) to update the model matrix $\bm{\Theta} = [
\bm{\theta}^{1}  \cdots  \bm{\theta}^{i}  \cdots  \bm{\theta}^{N}
]\in \mathbb{R}^{K \times N}$ (\ref{eq:model_matrix}). In the parallel computation, a single CPU is assigned to an individual subject $i=1,\cdots,N$. 
}
\label{fig:Par_Com}
\end{figure}

Note that the proportional part of the full conditional $\pi(\theta_{li}|-)$ (\ref{eq:full_condi_theta_ind}) only involves the $i$-th column vector of the model matrix $\bm{\Theta}$ (\ref{eq:model_matrix}), that is, $\bm{\theta}^{i} = (\theta_{1i},\cdots,\theta_{Ki})^{\top} \in \mathbb{R}^{K}$ in its analytic expression. This implies that we can update the $K$ entries of the vector $\bm{\theta}^{i}$ ($i=1,\cdots,N$) independently across subjects. Parallel sampling algorithm can be completed by assigning a single CPU process to each of the subjects $i$. Within the step to sample the $K$ entries from the vector $\bm{\theta}^{i}$, it is required to use Gibbs iterative procedure to update the scalar components. Authors \citep{lee2021bayesian,lee2020estimation} applied this technique to update the model matrix for a Bayesian nonlinear mixed effects model to train the dataset explained in Subsection \ref{subsec:Example 2: Decline curve analysis} and \ref{subsec:Example 4: Early stage of epidemic}. Figure \ref{fig:Par_Com} displays the schematic idea of the parallel sampling algorithm.

The second version of parallel sampling algorithms is based on vector updating. We analyze the full conditional posterior distribution of the vector $\bm{\theta}^{i}$ ($i=1,\cdots,N$):
\begin{align}
\nonumber
\pi(\bm{\theta}^{i}|-) &= \pi(\bm{\theta}^{i}|\bm{\theta}^{-i},\sigma^2,  \bm{\alpha}, \bm{B}, \bm{\Omega},\textbf{y}_{1:N})
\propto
\pi(\bm{\Theta}|\sigma^2,  \bm{\alpha}, \bm{B}, \bm{\Omega},\textbf{y}_{1:N})\\
\nonumber
&\propto
\prod_{i =1}^{N}
\mathcal{N}_{M_{i}} (\textbf{y}_{i} | \bm{f}_{i}(\textbf{t}_{i}, \bm{\theta}^{i})
,\sigma^{2} \mathbf{I})
\cdot
\mathcal{N}_{K} (\bm{\theta}^{i}|\bm{\alpha} + \bm{B} \textbf{x}_{i} , \bm{\Omega})
\\
\label{eq:full_condi_theta_vec}
&\propto
\mathcal{N}_{M_{i}} (\textbf{y}_{i} | \bm{f}_{i}(\textbf{t}_{i}, \bm{\theta}^{i})
,\sigma^{2} \mathbf{I})
\cdot
\mathcal{N}_{K} (\bm{\theta}^{i}|\bm{\alpha} + \bm{B} \textbf{x}_{i} , \bm{\Omega}),
\end{align}
where the notation $\bm{\theta}^{-i}$ represents the all the column vectors of $\bm{\Theta}$ except for $\bm{\theta}^{i}$. As similar to the first version, we can use the parallel computation to update the model matrix $\bm{\Theta}$ by simultaneously sampling from the full-conditional density $\pi(\bm{\theta}^{i}|-)$ (\ref{eq:full_condi_theta_vec}) across subjects by assigning one CPU to each individual.


\subsection{Elliptical slice sampler}\label{subsec:Elliptical slice sampler}
Due to the issue of non-conjugacy to sample from the univariate density $\pi(\theta_{li}|-)$ (\ref{eq:full_condi_theta_ind}) or $K$-dimensional density $\pi(\bm{\theta}^{i}|-)$ (\ref{eq:full_condi_theta_vec}), the choice of a suitable MCMC method and further the choice of a proposal distribution is crucial for the fast convergence of the Markov chain simulated from the \textbf{\emph{Step 1}} within the Gibbs sampler in Section \ref{subsec:Gibbs sampling algorithm}. The Metropolis-Hastings (MH) algorithm \citep{hastings1970monte,metropolis1953equation} is the first solution to consider in such intractable situations: see the Algorithm 1 from \citep{robert2015metropolis}. In practice, the performances of the MH algorithm are highly dependent on the choice of the proposal density \citep{chib1995understanding}. In the past decades, numerous MH-type algorithms to improve computational efficiency have been developed, and these fall broadly into two classes, according to whether the proposal density reflects a gradient information \citep{dwivedi2018log,ma2019sampling,duane1987hybrid,neal2011mcmc} or not \citep{murray2010elliptical,mengersen1996rates}. In specific, the gradient information, here, refers to the first-order derivative of the minus of the log of the target density (i.e., $
\nabla U(\theta_{li}) =-\nabla\log\ \pi(\theta_{li}|-) \in \mathbb{R} $ or $
\nabla U(\bm{\theta}^{i}) = -\nabla \log\ \pi(\bm{\theta}^{i}|-) \in \mathbb{R} ^{K}$, where the notation $\nabla$ represents the gradient operator). Typically, gradient-based samplers are attractive in terms of rapid exploration of the state space, but the cost of the gradient computation can be prohibitive when the sample size $N$ or model dimension $K$ is extremely large \citep{chen2014stochastic,aicher2019stochastic}. Fortunately, this requirement can be made less burdensome by using automatic differentiation \citep{griewank2008evaluating}.

In the present subsection, we introduce an efficient gradient-free sampling technique, the elliptical slice sampler (ESS) proposed by \citep{murray2010elliptical}, to simulate a Markov chain from the density $\pi(\theta_{li}|-)$ (\ref{eq:full_condi_theta_ind}). (The sampling logic can be directly applied to the situation to sample from the density $\pi(\bm{\theta}^{i}|-)$ (\ref{eq:full_condi_theta_vec}) by simply replacing $\theta_{li}$ by $\bm{\theta}^{i}$ and changing the dimension of relevant distributions, stochastic processes, etc, from $1$ to $K$.) Conceptually, MH and ESS algorithms are similar in that both comprise two steps: a proposal step and a criterion step. A difference between the two algorithms arises in the criterion step. If a new candidate does not pass the criterion, then MH algorithm takes the current state as the next state: whereas, ESS re-proposes a new candidate until rejection does not take place, rendering the algorithm rejection-free. The utility of ESS can be appreciated when an analytic expression of a target density can be factored to have a Gaussian prior distribution. Unlike the traditional MH algorithm that requires the proposal variance or density, ESS is fully automated, no tuning is required.

To adapt the ESS to our example, we re-write the density $\pi(\theta_{li}|-)$ (\ref{eq:full_condi_theta_ind}) as the following form:
\begin{align}
\label{eq:ESS_theta_ind}
\pi(\theta_{li}|-)
&= 
\frac{1}{Z}
\mathcal{L}(\theta_{li})
\cdot
\mathcal{N}
(\theta_{li}
|
\alpha_{l}
+\textbf{x}_{i}^{\top}\bm{\beta}_{l}
,\omega_{l}^2),
\end{align}
where $
\mathcal{L}(\theta_{li}) = 
\exp\ \{-\|
\textbf{y}_{i} - \bm{f}_{i}(\textbf{t}_{i}, \theta_{1i},\cdots,\theta_{li},\cdots,\theta_{Ki})
 \|_{2}^{2}/(2\sigma^{2}) \}$, and $Z$ is the normalization constant. Introducing the notation $\mathcal{L}(\theta_{li})$ is our intention that we shall treat the function as a likelihood part temporarily only when sampling from the density (\ref{eq:ESS_theta_ind}). 
Alternatively, one can proceed with the choice $\mathcal{L}_{2}(\theta_{li}) = 
\mathcal{N}_{M_{i}} (\textbf{y}_{i} | \bm{f}_{i}(\textbf{t}_{i}, \theta_{1i},\cdots,\theta_{li},\cdots,\theta_{Ki})
,\sigma^{2} \mathbf{I})$ as a likelihood part to operate ESS, which then change the normalization constant accordingly. We recommend to use the simplest functional form for the likelihood part, if possible, to reduce computation cost. 

Algorithm \ref{alg:ESS} summarizes the ESS in an algorithmic form where the situation is at the $(s+1)$-th iteration of the Gibbs sampler. Therefore, the goal is to draw $\theta_{li}^{(s+1)}$ from the target density $\pi(\theta_{li}|-)$ ($l=1,\cdots,K; i = 1,\cdots,N$) (\ref{eq:ESS_theta_ind}), where we already have $\theta_{li}^{(s)}$ as the current state for the target variable $\theta_{li}$ realized from the $s$-th iteration:

\begin{algorithm}[h!]
\caption{ESS to sample from $\pi(\theta_{li}|-)$ (\ref{eq:full_condi_theta_ind})}\label{alg:ESS}
\SetAlgoLined
\baselineskip=8pt
\textbf{Goal : }  Sampling from the full conditional posterior distribution
$$\pi(\theta_{li}|-)
\propto
\mathcal{L}(\theta_{li})
\cdot
\mathcal{N}
(\theta_{li}
|
\mu_{li}
,\omega_{l}^2),$$
where 
$\mathcal{L}(\theta_{li}) = 
\exp\ \{-\|
\textbf{y}_{i} - \bm{f}_{i}(\textbf{t}_{i}, \theta_{1i},\cdots,\theta_{li},\cdots,\theta_{Ki})
 \|_{2}^{2}/(2\sigma^{2}) \}$ and $\mu_{li} =\alpha_{l}
+\textbf{x}_{i}^{\top}\bm{\beta}_{l}$.
\\
\textbf{Input : } Current state $\theta_{li}^{(s)}$.\\
\textbf{Output : } A new state $\theta_{li}^{(s+1)}$.\\
\begin{itemize}
\item[a. ] Choose an ellipse centered at $\mu_{li}$: $\nu \sim \mathcal{N}( \mu_{li} , \omega_{l}^2 )$.
\item[b. ] Define a criterion function: 
\begin{align*}
\alpha(\theta_{li}, \theta_{li}^{(s)}) = \text{min}\{\mathcal{L}(\theta_{li}) / \mathcal{L}(\theta_{li}^{(s)}),1\}: \mathbb{R} \rightarrow [0,1].
\end{align*}
\item[c. ] Choose a threshold and fix: $u \sim \mathcal{U}nif[0,1]$.
\item[d. ] Draw an initial proposal $\theta_{li}^{*}$: 
\begin{align*}
\phi &\sim \mathcal{U}nif(-\pi,\pi];\\
\theta_{li}^{*}&=(\theta_{li}^{(s)} - \mu_{li}) \cos  \phi
+
(\nu -\mu_{li}) \sin\ \phi
+
\mu_{li}.
\end{align*}
\item[e. ] \textbf{if}  \textbf{(} $u < \alpha(\theta_{li}^{*}, \theta_{li}^{(s)})$  \textbf{)} $\{$ $\theta_{li}^{(s+1)} = \theta_{li}^{*}$  $\}$ \textbf{else} $\{$ \\
\indent$\quad$ Define a bracket : $(\phi_{\text{min}},\phi_{\text{max}} ]= (-\pi, \pi ]$.\\
\indent$\quad$ \textbf{while} \textbf{(} $u \geq \alpha(\theta_{li}^{*}, \theta_{li}^{(s)})$ \textbf{)} $\{$\\
\indent$\quad\quad$ Shrink the bracket and try a new point :
\\
\indent$\quad\quad$ \textbf{if} \textbf{(} $\phi > 0$ \textbf{)} $\phi_{\text{max}} = \phi$ \textbf{else} $\phi_{\text{min}} = \phi$ \\
\indent$\quad\quad$ $\phi \sim \mathcal{U}nif(\phi_{\text{min}},\phi_{\text{max}}]$\\
\indent$\quad\quad$
$\theta_{li}^{*}=(\theta_{li}^{(s)} - \mu_{li})\cos
 \phi
+
(\nu -\mu_{li}) \sin\ \phi
+
\mu_{li}.$
\\
\indent$\quad\quad$ $\}$\\
\indent$\quad\quad$ $\theta_{li}^{(s+1)} = \theta_{li}^{*}$\\
\indent$\quad$ $\}$
\end{itemize}
\end{algorithm}

\subsection{Metropolis adjusted Langevin algorithm}\label{subsec:MALA}
We introduce Metropolis adjusted Langevin algorithm (MALA) \citep{dwivedi2018log,ma2019sampling} which is popular for its use of problem-specific proposal distribution based on the gradient information of the target density. The main idea of MALA is to use Langevin dynamics to
construct the Markov chain. To adapt the sampling technique to our example, we re-write the density $\pi(\theta_{li}|-)$ (\ref{eq:full_condi_theta_ind}) as the following form:
\begin{align}
\label{eq:targer_density_step1}
\pi(
\theta_{li}
|
-)
 = \frac{ \exp (- U(\theta_{li}))}{\int_{-\infty}^{\infty} \exp (- U(\theta_{li})) d\theta_{li}} \quad \text{for all } \theta_{li}\in \mathbb{R},
\end{align}
where $ 
U(\theta_{li})
=
(1/\{2 \sigma^{2}\})
\cdot
\|
\textbf{y}_{i}
-
\bm{f}_{i}(\textbf{t}_{i} ; \theta_{1i}, \cdots, \theta_{li}, \cdots, \theta_{Ki})
\|_2^{2}
+
(1/\{2 \omega_{l}^{2}\})
\cdot
(\theta_{li} -\alpha_{l} -  \textbf{x}_{i}^{\top}\bm{\beta}_{l} )^{2}$. Now, we consider a stochastic differential equation \citep{oksendal2003stochastic} that characterizes the evolution of the Langevin diffusion with the drift term set by the gradient of the log of the density (\ref{eq:targer_density_step1}):
\begin{align}
\label{eq:Langevin dynamics}
d\theta_{li} (t) &=  \nabla \log \pi(\theta_{li} (t)|-) dt + \sqrt{2} dW(t) = - \nabla U(\theta_{li} (t) ) dt + \sqrt{2} dW (t),
\end{align}
where $\{W(t) \mid t\geq0\}$ is a standard $1$-dimensional Wiener process, or Brownian motion \citep{uhlenbeck1930theory}. In (\ref{eq:Langevin dynamics}), $t$ indexes a fictitious continuous time. $\nabla$ represents the gradient operator with respect to $\theta_{li}$. Under faily mild conditions on the function $U(\theta_{li})$, the equation (\ref{eq:Langevin dynamics}) has a strong solution $\{\theta_{li} (t) \mid t \geq 0\}$ that is a Markov process \citep{roberts1996geometric}. Furthermore, it can be shown that the distribution of $\{\theta_{li} (t) \mid t \geq 0\}$ converges to the invariant distribution $\pi(
\theta_{li}
|
-)$ (\ref{eq:targer_density_step1}) as $t \rightarrow \infty$.

Since solving the equation (\ref{eq:Langevin dynamics}) is very difficult, a first-order Euler-Maruyama discretization \citep{asai2013numerical} is used to approximate the solution to the equation:
\begin{align}
\label{eq:discretization_Langevin_Dynamics}
\theta_{li}^{[s+1]}&\leftarrow
\theta_{li}^{[s]}
 - \delta \cdot \nabla U(\theta_{li}^{[s]}) + \sqrt{2\delta} Z,\quad Z\sim \mathcal{N}(0,1),
\end{align}
where $\delta$ is the step size of discretization, and $[s]$ indexes the discrete time steps. This recursive update defines the Langevin Monte Carlo algorithm. Typically, to handle the discretization error and satisfy the detailed balance \citep{casella1999monte} to make Markov chain converge to the target distribution $\pi(
\theta_{li}
|
-)$ (\ref{eq:targer_density_step1}), the MH correction is needed. Algorithm \ref{alg:MALA} details MALA to sample from the $\pi(
\theta_{li}
|
-)$ ($l=1,\cdots,K; i = 1,\cdots,N$) (\ref{eq:targer_density_step1}):

\begin{algorithm}[h!]
 \caption{MALA to sample from $\pi(\theta_{li}|-)$ (\ref{eq:full_condi_theta_ind})}
\label{alg:MALA}
\baselineskip=8pt
\SetAlgoLined
\textbf{Goal : } Sampling from the full conditional posterior distribution
$$
\pi(
\theta_{li}
|
-)
 \propto 
 \exp\ \left(-U(\theta_{1i})\right),
$$
where 
$ 
U(\theta_{li})
=
\|
\textbf{y}_{i}
-
\bm{f}_{i}(\textbf{t}_{i} ; \theta_{1i}, \cdots, \theta_{li}, \cdots, \theta_{Ki})
\|_2^{2}/(2 \sigma^{2})
+
(\theta_{li} -\alpha_{l} -  \textbf{x}_{i}^{\top}\bm{\beta}_{l} )^{2}/(2 \omega_{l}^{2})
.$
\\
\textbf{Input : } Current state $\theta_{li}^{(s)}$ and step size $\delta$.\\
\textbf{Output : } A new state $\theta_{li}^{(s+1)}$.\\
\begin{itemize}
\item[a. ] Define a criterion function: 
$$
\alpha(
\theta_{li},
\theta_{li}^{(s)}) = \text{min}\left\{
\frac{
 \exp\ (-U(\theta_{li}))}{ \exp\ (-U(\theta_{li}^{(s)}))}
\cdot
\frac{
\mathcal{J}(\theta_{li}^{(s)}|\theta_{li})
}{
\mathcal{J}(\theta_{li}|\theta_{li}^{(s)})
}
,1\right \}: \mathbb{R} \rightarrow [0,1].
$$
\item[b. ] Choose a threshold $u$: $u \sim \mathcal{U}nif[0,1].$
\item[c. ] Draw a proposal $\theta_{li}^{*}$: $$\theta_{li}^{*} \sim
\mathcal{J}(\theta_{li}|\theta_{li}^{(s)}) = \mathcal{N}
(\theta_{li}
|
\theta_{li}^{(s)} 
- \delta \cdot \nabla U(\theta_{li}^{(s)})
, 2\delta).$$
\item[e. ] \textbf{if}  \textbf{(} $u < 
\alpha(
\theta_{li}^{*},
\theta_{li}^{(s)})
$  \textbf{)} $\{$ $\theta_{li}^{(s+1)} = \theta_{li}^{*}$  $\}$ \textbf{else} 
$\{$ $ \theta_{li}^{(s+1)} = \theta_{li}^{(s)}$ $\}$
\end{itemize}
\end{algorithm} 

\subsection{Hamiltonian Monte Carlo}\label{subsec:Hamiltonian Monte Carlo}
We introduce Hamiltonian Monte Carlo (HMC) algorithm that employs Hamiltonian dynamics to efficiently explore the parameter space \citep{duane1987hybrid,neal2011mcmc}. Among many MH-type sampling algorithms, HMC has been recognized as one of the most effective algorithms due to its rapid mixing rate and small discretization error. By that reason, HMC has been deployed as the default sampler in many open packages such as \textsc{Stan} \citep{carpenter2017stan} and \textsc{Tensorflow} \citep{abadi2016tensorflow}. A key idea of HMC distinctive from ESS and MALA is the introduction of an auxiliary momentum variable, which is typically assumed to follow as a Gaussian distribution and independent of the target variable. By doing so, the HMC can produce distant proposals for the target variable, thereby avoiding the slow exploration of the state space that results from the diffusive behavior of simple random-walk proposals. 

We adapt the HMC to our example. We shall first take a look at a joint density:
\begin{align}
\nonumber
\pi(\theta_{li},\phi_{li}
|
-) &=
\pi(
\theta_{li}
|
-)
\cdot
\pi(
\phi_{li}
)\\
\nonumber
&=
\frac{ \exp (- U(\theta_{li}))}{\int_{-\infty}^{\infty} \exp (- U(\theta_{li})) d\theta_{li}}
\cdot
\frac{1}{\sqrt{2\pi m_{li}}}
\exp\ \left(- \frac{\phi_{li}^{2}}{2 m_{li}} \right)
\\
\label{eq:targer_joint_density_HMC}
&\propto 
\exp\ (
-U(\theta_{li})
-K(\phi_{li})
)
\quad \text{for all } (\theta_{li},\phi_{li})\in \mathbb{R} \times \mathbb{R},
\end{align}
where $ 
U(\theta_{li})
=
(1/\{2 \sigma^{2}\})
\cdot
\|
\textbf{y}_{i}
-
\bm{f}_{i}(\textbf{t}_{i} ; \theta_{1i}, \cdots, \theta_{li}, \cdots, \theta_{Ki})
\|_2^{2}
+
(1/\{2 \omega_{l}^{2}\})
\cdot
(\theta_{li} -\alpha_{l} -  \textbf{x}_{i}^{\top}\bm{\beta}_{l} )^{2}$ and 
$K(\phi_{li}) = \phi_{li}^{2}/(2 m_{li})$. The auxiliary variable $\phi_{li}$ is distributed according to the univariate Gaussian distribution $\mathcal{N}(\phi_{li}|0,m_{li})$ with variance $m_{li}$. Note that it holds $\int \pi(\theta_{li},\phi_{li}
|
-) d \phi_{li} = \pi(\theta_{li}|-)$ due to the independence between $\theta_{li}$ and $\phi_{li}$. Therefore, our ultimate goal is to sample from the joint density $\pi(\theta_{li},\phi_{li}
|
-)$ (\ref{eq:targer_joint_density_HMC}), and take only $\theta_{li}$ by marginalization.

Noting from (\ref{eq:targer_joint_density_HMC}), the negative of joint log-posterior is
\begin{align}
\label{eq:Hamiltonian_function}
H(\theta_{li},\phi_{li})
&=
U(\theta_{li}) + 
K(\phi_{li}),\quad \text{for all } (\theta_{li},\phi_{li})\in \mathbb{R} \times \mathbb{R}.
\end{align}
The physical analogy of the bivariate function $H(\theta_{li},\phi_{li}): 
\mathbb{R} \times \mathbb{R} \rightarrow \mathbb{R}$ (\ref{eq:Hamiltonian_function}) is a Hamiltonian \citep{duane1987hybrid,leimkuhler2004simulating}, which describe the sum of a potential energy $U(\theta_{li})$, defined at the position $\theta_{li}$, and a kinetic energy $K(\phi_{li}) = \phi_{li}^{2}/(2 m_{li})$, where the auxiliary variable $\phi_{li}$ can be interpreted as a momentum variable and the variance $m_{li}$ denotes a mass. 

Now, we construct a Hamiltonian system by taking a derivative of $H$ (\ref{eq:Hamiltonian_function}) with respect to $\theta_{li}$ and $\phi_{li}$, and introducing a continuous fictitious time $t$:
\begin{align}
\label{eq:Hamiltonian_Eq_1}
\frac{d\theta_{li}(t)}{dt}
&=
\frac{\partial H (\theta_{li},\phi_{li})}{\partial \phi_{li}} = 
\frac{\partial K(\phi_{li})}{\partial \phi_{li}} = \frac{\phi_{li}(t)}{m_{li}},\\
\label{eq:Hamiltonian_Eq_2}
\frac{d\phi_{li}(t)}{dt}
&=
-
\frac{\partial H (\theta_{li},\phi_{li})}{\partial \theta_{li}} = 
-
\frac{\partial U(\theta_{li})}{\partial \theta_{li}} = 
-\nabla  U(\theta_{li}(t)),
\end{align}
where $\nabla$ represents the gradient operator with respect to $\theta_{li}$.

The Hamiltonian system (\ref{eq:Hamiltonian_Eq_1}) - (\ref{eq:Hamiltonian_Eq_2}) has three nice properties. Assume that $(\theta_{li}(t),\phi_{li}(t)): [a,b]\rightarrow \mathbb{R} \times \mathbb{R}$ is a solution curve of the system, where $a,b \in \mathbb{R} \cup \{\pm \infty \}$. Then, the following relationships hold:
\begin{itemize}
\baselineskip=15pt
\item[] (a) \emph{preservation of total energy}: $H(\theta_{li}(t),\phi_{li}(t))=H(\theta_{li}(0),\phi_{li}(0))$ for all $t \in [a,b]$;
\item[] (b) \emph{preservation of volume}: $d\theta_{li}(t) d\phi_{li}(t) =d\theta_{li}(0) d\phi_{li}(0)$ for all $t \in [a,b]$;
\item[] (c) and \emph{time reversibility}: The mapping $T_{s}$ from state at $t$, $(\theta_{li}(t),\phi_{li}(t))$, to the state at time $t+s$, $(\theta_{li}(t+s),\phi_{li}(t+s))$, is one-to-one, and hence has an inverse $T_{-s}$.
\end{itemize}
Three properties are eventually related with the following nice properties of the HMC: (a) a high probability of acceptance of proposals; (b) a simple analytic form of acceptance ratio (no need to consider a hard-to-compute Jacobian factor); and (c) a detailed balance with respect to the target density $\pi(
\theta_{li}
|
-)$. For a detailed description and extensive review see \citep{neal2011mcmc}. 

For practical applications, the differential equation system (\ref{eq:Hamiltonian_Eq_1}) - (\ref{eq:Hamiltonian_Eq_2}) cannot be solved analytically and numerical methods are required. As Hamiltonian $H$ in the system is separable (or equivalently, the joint density $\pi(\theta_{li},\phi_{li}
|
-)$ is factorizable), to traverse the state space more efficiently, the leapfrog integrator method is typically used, which involves a discretized step of the dynamics. As similar to the construction of MALA, the discretization errors arising from the leapfrog integration are addressed by MH correction step. Algorithm \ref{alg:HMC} details the HMC to sample from the target density $\pi(\theta_{li}|-) = \int \pi(\theta_{li},\phi_{li}
|
-) d \phi_{li}$ (\ref{eq:full_condi_theta_ind}). In the algorithm, $(s)$ indexes the  sampling iteration within the Gibbs sampler, while $[d]$ represents the index introduced due to the discretization. 

\begin{algorithm}[h!]
 \caption{HMC to sample from $\pi(\theta_{li}|-) = \int \pi(\theta_{li},\phi_{li}
|
-) d \phi_{li}$ (\ref{eq:full_condi_theta_ind})}
\label{alg:HMC}
\baselineskip=8pt
\SetAlgoLined
\textbf{Goal : } Sampling from the full conditional posterior distribution
$$
\pi(
\theta_{li}
|
-)
= \int \pi(\theta_{li},\phi_{li}
|
-) d \phi_{li}
 \propto 
 \exp\ \left(-U(\theta_{li})\right),
$$
where $\pi(\theta_{li},\phi_{li}
|
-)
\propto
\exp\
(
-
H(\theta_{li},\phi_{li})
)
=
\exp\ (
-U(\theta_{li})
-K(\phi_{li})
)
$ with $ 
U(\theta_{li})
=
\|
\textbf{y}_{i}$ $
-
\bm{f}_{i}( $ $\textbf{t}_{i} ; $ $\theta_{1i}, \cdots,  $ $\theta_{li}, \cdots, $ $\theta_{Ki})
\|_2^{2}/(2 \sigma^{2})
+
(\theta_{li} -\alpha_{l} -  \textbf{x}_{i}^{\top}\bm{\beta}_{l} )^{2}/(2 \omega_{l}^{2})$ and $K(\phi_{li}) = \phi_{li}^{2}/(2 m_{li})$.
\\
\textbf{Input : } Current state $\theta_{li}^{(s)}$, step size $\delta$, number of steps $L$, and mass $m_{li}$.\\
\textbf{Output : } A new state $\theta_{li}^{(s+1)}$.\\
\begin{itemize}
\item[a. ] Generate the initial momentum with mass $m_{li}$: $\phi_{li}^{[0]} \sim \mathcal{N}(0,m_{li})$.
\item[b. ] Define a criterion function: 
$$
\alpha(
(\theta_{li},\phi_{li})
,
(\theta_{li}^{(s)}
\phi_{li}^{[0]})
) = \text{min}\left\{
\frac{
 \exp\ (
-H(\theta_{li},\phi_{li})
  )}{ \exp\  (
-H(\theta_{li}^{(s)},
\phi_{li}^{[0]}
))}
,1\right \}: \mathbb{R} \times \mathbb{R} \rightarrow [0,1].
$$
\item[c. ] Simulate discretization of Hamiltonian dynamics (\ref{eq:Hamiltonian_Eq_1}) -- (\ref{eq:Hamiltonian_Eq_2}):
\begin{enumerate}
\item[i.]Set the initial pair of the solution curve: $(\theta_{li}^{[0]},
\phi_{li}^{[0]})=(\theta_{li}^{(s)}
\phi_{li}^{[0]}).$
\item[ii.]Make a half step for the momentum: 
$
\phi_{li}^{[0]} \leftarrow
\phi_{li}^{[0]} - (\delta/2) \nabla U(\theta_{li}^{[0]}).
$
\item[iii.] Alternate full steps for position and momentum: 
\\ \textbf{for}  \textbf{(} $ d = 1,\cdots, L $\textbf{)} \textbf{\{}
\\
Update position: 
$\theta_{li}^{[d]} \leftarrow \theta_{li}^{[d-1]}  +  (\delta / m_{li}) \phi_{li}^{[d-1]}.$
\\
Update momentum:
\textbf{if} \textbf{(} $d !=L$ \textbf{)} \textbf{\{}
$\phi_{li}^{[d]} \leftarrow \phi_{li}^{[d-1]} - \delta  \nabla U(\theta_{li}^{[d]}).$
\textbf{\}}
\\
\textbf{\}}
\item[iv.] Make a half step for momentum:
$
\phi_{li}^{[L]} \leftarrow
\phi_{li}^{[L-1]} - (\delta/2) \nabla U(\theta_{li}^{[L]}).
$
\item[v. ] Negate the last momentum: $\phi_{li}^{[L]} \leftarrow -\phi_{li}^{[L]}.$
\item[vi. ] Set the last pair of the solution curve as the proposal: $(\theta_{li}^{*},
\phi_{li}^{*})=(\theta_{li}^{[L]}
\phi_{li}^{[L]}).$
\end{enumerate}
\item[d. ] Choose a threshold $u$: $u \sim \mathcal{U}nif[0,1].$
\item[e. ] \textbf{if}  \textbf{(} $u < 
\alpha(
(\theta_{li}^{*},\phi_{li}^{*})
,
(\theta_{li}^{(s)}
\phi_{li}^{[0]})
)
$  \textbf{)} $\{$ $\theta_{li}^{(s+1)} = \theta_{li}^{*}$  $\}$ \textbf{else} 
$\{$ $ \theta_{li}^{(s+1)} = \theta_{li}^{(s)}$ $\}$
\end{itemize}
\end{algorithm} 

One caveat in the HMC is that no matter what we accept or reject the proposal, we draw a new momentum from the kinetic energy at every iteration. To check this, see the Step a in Algorithm \ref{alg:HMC}, where $\phi_{li}^{[0]}$ is drawn from the kinetic density $K(\phi_{li}) \propto \mathcal{N}(0,m_{li})$. The momentum $\phi_{li}^{[0]}$ is only used to formulate the initial pair $(\theta_{li}^{[0]},
\phi_{li}^{[0]})$, where $\theta_{li}^{[0]}$ is the current target state $\theta_{li}^{(s)}$, that will be guided by Hamiltonian dynamics (\ref{eq:Hamiltonian_Eq_1}) -- (\ref{eq:Hamiltonian_Eq_2}) via leapfrog integrator, eventually reaching to the last pair $(\theta_{li}^{[L]}
\phi_{li}^{[L]})$ which is used as the proposal $(\theta_{li}^{*},\phi_{li}^{*}).$ The momentum $\phi_{li}^{[0]}$ is deleted and we will draw a new momentum in the next iteration. This independent drawing of the momentum is the engine that enables HMC to produce distant proposals, but nevertheless maintains a high probability of acceptance.

The naive HMC (Algorithm \ref{alg:HMC}) requires the users to specify at least three parameters: a step size $\delta$, a number of steps $L$, and a mass $m_{li}$, for which to run a simulated Hamiltonian system. A poor choice of either of these parameters will result in a dramatic drop in the efficiency HMC. No-U-Turn Sampler (NUTS) developed by \citep{hoffman2014no} is an extension of HMC, which is designed to automatically turn the parameters $(\delta,L)$ while fixing $m_{li}=1$, making it possible to run NUTS with no hand-tuning at all. HMC and NUTS are general-purpose inference engines deployed in \textsc{Stan}.

We would like to highlight a difference between MALA (Algorithm \ref{alg:MALA}) and HMC (Algorithm \ref{alg:HMC}). Although both algorithms utilizes the gradient information (that is, $\nabla U(\theta_{li}) =-\nabla\log\ \pi(\theta_{li}|-)$), the former is based on stochastic differential equation (\ref{eq:Langevin dynamics}) and the latter is based on ordinary differential equation (\ref{eq:Hamiltonian_Eq_1}) -- (\ref{eq:Hamiltonian_Eq_2}). From the algorithmic perspective, MALA exhibits a single loop structure and is designed to directly employ the discretization of the underlying Langevin dynamics: see that the index $[s]$ resulting from the discretization in (\ref{eq:discretization_Langevin_Dynamics}) is directly used as the sampling index $(s)$ in Algorithm \ref{alg:MALA}. On the other hand, HMC has a double loop structure: the inner loop (i.e., Step c in Algorithm \ref{alg:HMC}) solves the Hamiltonian dynamics (\ref{eq:Hamiltonian_Eq_1}) -- (\ref{eq:Hamiltonian_Eq_2}) to make a proposal, while the outer loop judges the proposals. The index $[d]$ in the inner loop, resulting from the leapfrog integrator, and the sampling index $(s)$ of the outer loop in Algorithm \ref{alg:HMC} are not related \citep{zou2021convergence}. Therefore, one can set the number of steps $L$ for the leapfrog integrator by an arbitrary integer.
\section{Prior options}\label{sec:Prior options}
\subsection{Priors for variance}\label{subsec:Priors for variance}
Provided the assumption of the basic form (\ref{eq:Stage 1: Individual-Level Model}) -- (\ref{eq:Stage 3: Prior}), the random error terms $\{ \epsilon_{ij}\}_{i=1,j=1}^{N,M_{i}}$ in Stage 1 and $\{ \eta_{li}\}_{l=1,i=1}^{K,N}$ in Stage 2 are the stochastic sources of (remaining) intra-individual and inter-individual variabilities, respectively \citep{snijders2011multilevel}. Both terms are assumed to follow univariate Gaussian distributions in the basic model. This assumption can be generalized to multivariate Gaussian distribution, $t$-distribution, mixture of Gaussian distributions, etc, depending on the exhibition of the data or prior guess of perturbation associated with model matrix $\bm{\Theta}$ (\ref{eq:model_matrix}) \citep{meza2012estimation}.

Recall that the basic model assumes that data-level errors $\epsilon_{ij}$ are distributed according to $\mathcal{N}(0,\sigma^{2})$ with variance $\sigma^{2}$, independently across times $j=1,\cdots,M_{i}$ and subjects $i=1,\cdots,N$. (We discuss about the $\eta$-term in Subsection \ref{subsec:Priors for covariance}.) Therefore, the standard deviation $\sigma$ describes a vertical difference (i.e., measurement error) between the observation $y_{ij}$ and theory $f(t_{ij} ; \bm{\theta}^{i})$ across time and individuals. We can generalize the basic setting by replacing $\sigma^{2}$ with (a) $\sigma_{i}^{2}$ ($i=1,\cdots,N$) or (b) $\sigma_{ij}^{2}$ ($i=1,\cdots,N ; j=1,\cdots,M_{i}$) to accommodate the heterogeneity the measurement error (a) across subjects and (b) across subjects and time, respectively, provided sufficiently large sampling times $M_{i}$ \citep{wakefield1996bayesian}. 

For any prior $\pi(\sigma^{2})$, the full conditional posterior distribution of $\sigma^2$ (\ref{eq:full_condi_sigma_sq}) is given as
\begin{align}
\label{eq:full_condi_sigma_sq_2}
\pi(\sigma^2|\bm{\Theta}, \bm{\alpha}, \bm{B}, \bm{\Omega},\textbf{y}_{1:N})
&\propto
(\sigma^{2})^{-\sum_{i=1}^{N} M_{i}/2}
\exp
\left(
-\frac{1}{2 \sigma^2}
\sum_{i=1}^{N}
\|
\textbf{y}_{i} -\bm{f}_{i}(\textbf{t}_{i}, \bm{\theta}^{i})
 \|_{2}^{2}
\right)
\cdot
\pi(\sigma^{2}),
\end{align}
where $\|\textbf{a} \|_{2}^{2}$ represents the Euclidean norm of the vector $\textbf{a}$.

Popularly used priors $\pi(\sigma^{2})$ (or $\pi(\sigma)$) are (i) the Jeffreys prior $\pi(\sigma^{2})\propto 1/\sigma^{2}$ \citep{jeffreys1946invariant}; (ii) inverse-gamma prior $\pi(\sigma^{2}) = \mathcal{IG}(a_{\sigma^{2}}, b_{\sigma^{2}} )$ with shape $a_{\omega_{l}^{2}}>0$ and scale $b_{\omega_{l}^{2}}>0$; and (iii) half-Cauchy prior $\pi(\sigma)=\mathcal{C}^{+}(0,b_{\sigma}) = \{2/(\pi b_{\sigma})\} \cdot 1/\{ 1 + (\sigma/b_{\sigma})^{2}\}$ with scale $b_{\sigma}>0$. Note that half-Cauchy distribution should be given to the standard deviation $\sigma$, not variance $\sigma^{2}$. The first two prior options lead to the conjugate update to sample from the density $\pi(\sigma^2|-)$ (\ref{eq:full_condi_sigma_sq_2}). Although the third one induces non-conjugate update to sample from the density $\pi(\sigma|-)$, computationally efficient sampling can be constructed by using parameter expansion technique \citep{makalic2015simple} or slice sampler \citep{neal2003slice}.
\subsection{Priors for intercept and coefficient vector}\label{subsec:Priors for intercept and coefficient vector}
One of the central goals of using nonlinear mixed effects models is to identify significant covariates among the $P$ covariates $\textbf{x} = (x_{1},\cdots, x_{P})^{\top}$, explaining each of the model parameter $\theta_{l}$ $(l=1,\cdots,K)$. This is because the function $f$ in Stage 1 (\ref{eq:Stage 1: Individual-Level Model}) is typically derived from a differential equation system. Such a differential equation has model parameters $\{\theta_{l}\}_{l=1}^{K}$ which controls the dynamic of the solution of the system, and how the parameters are related with covariates is vital to understand causality. For example, in PK analysis, understanding whether and to what extent weight, renal status, disease status, etc, are associated with drug clearance may dictate how these factors can be considered in a dosing schedule.

We explain popularly used priors for the intercept and coefficient vector by taking the vector-form (a) [Stage 1-(\ref{eq:Stage 1: Individual-Level Model_Vector_Form}), Stage 2-(\ref{eq:Stage 2: Population Model_Vector_Form}), and Stage 3-(\ref{eq:Stage 3: Prior_Vector_Form})] because it directly embeds the framework of linear regression. For each model parameter index $l = 1,\cdots,K$, we re-write the equation (\ref{eq:Stage 2: Population Model_Vector_Form}) for the purpose of illustration:
\begin{align}
\label{eq:Stage 2: Regression}
\begin{bmatrix}
\theta_{l1} \\
\vdots \\
\theta_{li} \\
\vdots \\
\theta_{lN}
\end{bmatrix}
 &= 
\begin{bmatrix}
 \alpha_{l}\\
 \vdots\\
  \alpha_{l}\\
  \vdots \\
   \alpha_{l}
\end{bmatrix} 
 +
\begin{bmatrix}
x_{11} &\cdots & x_{1b} & \cdots & x_{1P} \\
\vdots &&\vdots & &\vdots\\
x_{i1} &\cdots & x_{ib} & \cdots & x_{iP} \\
\vdots &&\vdots & &\vdots\\
x_{N1} &\cdots & x_{Nb} & \cdots & x_{NP} 
\end{bmatrix} 
\begin{bmatrix}
 \beta_{l1}\\
 \vdots\\
 \beta_{lb}\\
  \vdots \\
 \beta_{lP}
\end{bmatrix} 
 +
\begin{bmatrix}
\eta_{l1} \\
\vdots \\
\eta_{li} \\
\vdots \\
\eta_{lN}
\end{bmatrix} 
\end{align}
where $\bm{\eta}_{l} = (\eta_{l1},\cdots,\eta_{li},\cdots,\eta_{lN})^{\top}\sim \mathcal{N}_{N}(\textbf{0},\omega_{l}^{2} \textbf{I})$. By the assumption (\ref{eq:Stage 3: Prior_Vector_Form}), we have priors $\alpha_{l}\sim	\pi(\alpha_{l})$ and $\beta_{lb}\sim	\pi(\beta_{lb})$ ($b=1,\cdots,P$). Note that the equation (\ref{eq:Stage 2: Regression}) is a Bayesian multivariate linear regression (page 149 of \cite{hoff2009first}), and the only difference from the usual context is that the response vector in (\ref{eq:Stage 2: Regression}) is latent. Therefore, almost all Bayesian regression techniques \citep{castillo2015bayesian,lee2020tail} can be used to the latent regression (\ref{eq:Stage 2: Regression}) provided that the model matrix $\bm{\Theta}$ (\ref{eq:model_matrix}) is efficiently realized in \textbf{\emph{Step 1}} within the Gibbs sampler.

Default choice for the prior $\pi(\alpha_{l})$ is the flat prior $\pi(\alpha_{l})\propto 1$, also called a uniform prior. Alternatively, a diffuse Gaussian prior $\pi(\alpha_{l}) = \mathcal{N}(a_{l} , b_{l}^{2})$ is also often used by fixing $b_{l}$ to be sufficiently large (saying $b_{l} = 10$ or $100$) and $a_{l}=0$. In either case, the full conditional density (\ref{eq:full_condi_alpha}) enjoys the conjugate update, hence, \textbf{\emph{Step 3}} in the Gibbs sampler in Subsection \ref{subsec:Gibbs sampling algorithm} seldom imposes computational burden. The idea behind the use of (nearly) non-informative priors for the intercept is that such priors induce almost minimal degree of Bayesian shrinkage, hence, allow the data to have (nearly) maximum effect on the posterior estimate for the intercept  \citep{makalic2016simple}. 

Now, we discuss on the Bayesian analysis for the coefficients. There are numerous choices for the prior of the coefficient vector $\bm{\beta}_{l} = (\beta_{l1},\cdots,\beta_{lb},\cdots,\beta_{lP}) \sim \pi(\bm{\beta}_{l} )$, which is not surprising because the linear regression is arguably one of the most researched topics in statistics. Here, we suggest some popular priors whose main utility fall broadly into two settings, according to whether the design matrix $\textbf{X}\in \mathbb{R}^{N \times P}$ in (\ref{eq:Stage 2: Regression}) is tall ($N \geq P$) or fat ($N < P$). In regression theory, the former setting is referred to as low-dimensional regression, and latter one is called high-dimensional regression \citep{tibshirani1996regression,zou2005regularization,fan2009ultrahigh}.

Under the \emph{tall design} ($N \geq P$), particularly when the number of subjects $N$ is much larger than the number of covariates $P$ ($N \gg P$), one important theoretical consideration is that, it is expected to see Bernstein–von Mises type results \citep{lu2017gaussian,wang2019frequentist,johnstone2010high} on the posterior inference for coefficients $\bm{\beta}_{l}$. Roughly speaking, the theorem, sometimes called the ``Bayesian Central Limit Theorem", states that the posterior distribution of $\bm{\beta}_{l}$ is approximately a normal distribution following a likelihood theory as sample size $N$ goes to infinity for any prior choice $\pi(\bm{\beta}_{l} )$ under certain regularity conditions. In reality, it is possible that due to an ill-conditioned design matrix $\textbf{X}$, a misspecification of error distribution for $\eta_{li}$, a small sample size $N$, an inappropriate choice of prior $\pi(\bm{\beta}_{l} )$, etc, the Bernstein-von Mises Phenomenon (page 151 from \citep{le2000asymptotics}) may not be empirically observed. But even such cases, it is known the influence of the prior distribution diminishes as $N$ grows, which means that using different priors for $\bm{\beta}_{l}$ may not sensitively change the resulting Bayesian inferences about $\bm{\beta}_{l}$, and furthermore, the inference outcomes obtained by Bayesian and frequentist methods agree in most instances under the tall design. For example, see results of \citep{davidian1992smooth} and \citep{wakefield1996bayesian} for pharmacokinetic applications. Some possible options for prior $\pi(\bm{\beta}_{l} )$ are (i) flat prior $\pi(\bm{\beta}_{l})\propto 1$; (ii) Gaussian diffuse prior $\pi(\bm{\beta}_{l}) = \mathcal{N}_{P}(\textbf{0}, \sigma_{\beta_{l}}^{2}  \textbf{I})$ with a large variance $\sigma_{\beta_{l}}^{2}$ \citep{wei2013bayesian}, and (iii) $g$-prior $\pi(\bm{\beta}_{l}) = \mathcal{N}_{P}(\textbf{0}, g \cdot \omega_{l}^{2} [\textbf{X}^{\top} \textbf{X}]^{-1} )$ for some positive value $g$ \citep{zellner1986assessing}. The suggested priors yield the conjugate update for \textbf{\emph{Step 4}} within the Gibbs sampler.

Now, we discuss some priors for $\pi(\bm{\beta}_{l})$ in the linear regression (\ref{eq:Stage 2: Regression}) under the \emph{fat design setting} ($N \ll P$). This setting can be applied to pharmacogenetics where one of the main interests is to find important genes that may influence pharmacokinetics or pharmacodynamics
\citep{pirmohamed2001pharmacogenetics,weinshilboum2006pharmacogenetics,arab2009pharmacogenetics}, where the number of genes $P$ is allowed to be a few thousand, while number of patients $N$ is confined to a few hundred. A fundamental assumption in this setting is \emph{sparsity assumption} on the coefficients $\bm{\beta}_{l}$. This means that many of the coefficients of $\bm{\beta}_{l} = (\beta_{l1},\cdots, \beta_{lP})^{\top}$ are (close to) zero. The true non-zero coefficients in the $\bm{\beta}_{l}$ are referred to as signal coefficients, while the remaining are called noise coefficients.  

Statisticians have devised a number of penalized regression techniques for estimating $\bm{\beta}_{l}$ under the sparsity assumption \citep{hastie2015statistical}.  From a Bayesian point of view, sparsity favoring mixture priors with separate control on the signal and noise coefficients have been proposed \citep{mitchell1988bayesian,george1995stochastic,johnson2010use,yang2016computational}, which is called the `spike-and-slab priors'. Although these priors often lead to attractive theoretical properties \citep{castillo2012needles,castillo2015bayesian}, computational issues and considerations that many of the $\beta_{lb}$'s ($b=1,\cdots,P$) may be small but not
exactly zero has led to a wide variety of `continuous shrinkage priors' \citep{park2008bayesian,tipping2001sparse,griffin2010inference,carvalho2010horseshoe,carvalho2009handling}, which can be unified through a global-local scale mixture representation \citep{polson2010shrink}.  The following hierarchies describe the sparse favoring priors: 
\begin{itemize}
\baselineskip=15pt
\item[] $\bullet$ \textbf{Spike-and-slab priors}. Each component of the coefficients $\bm{\beta}_{l}$ is assumed to be drawn from 
$$
\beta_{lb}|\tau_{l} \sim (1 - \tau_{l})\cdot \delta_{0}(\beta_{lb}) + \tau_{l} \cdot f(\beta_{lb}), \quad (l=1,\cdots,K; b = 1,\cdots,P),
$$
where $\tau_{l} = \textbf{Pr}[\beta_{lb} \neq  0 ]$. The function $\delta_{0}(\beta_{lb})$ is the Direc-delta function and $f(\beta_{lb})$ is a density supported on $\mathbb{R}$, called the spike and slab densities, respectively. The spike density shrinks noise coefficients to the exact zero, while the slab density captures signal coefficients by allowing a positive mass on the tail region \citep{george1997approaches,johnstone2004needles,castillo2015bayesian}.
\item[] $\bullet$ \textbf{Continuous shrinkage priors}. Each component of the coefficients $\bm{\beta}_{l}$ is assumed to be drawn from 
$$
\beta_{lb}|\lambda_{lb},\tau_{l}, \omega_{l}^{2} \sim \mathcal{N}(0,\lambda_{lb}^{2} \tau_{l}^{2} \omega_{l}^{2}), \quad (l=1,\cdots,K;b=1,\cdots,P),
$$
$$
\lambda_{lb} \sim f(\lambda_{lb} ),
\tau_{l} \sim g(\tau_{l}),
 \omega_{l}^{2}  \sim h( \omega_{l}^{2} ), \quad (l=1,\cdots,K;b=1,\cdots,P),
$$
where $f$, $g$, and $h$ are priors for $\lambda_{lb}$, $\tau_{l}$, and $ \omega_{l}^{2}$, respectively, supported on $(0,\infty)$. Here, $\lambda_{lb}$ and $\tau_{l}$ are referred to as local-scale and global-scale parameters, respectively. The choices of $f$ and $g$ play a key role in controlling the effective sparsity and concentration of the prior and posterior distributions \citep{polson2010shrink, pati2014posterior,song2017nearly,martin2017empirical,bai2018high, zhang2019ultra}. 
\end{itemize}
Roughly speaking, the roles of the $\tau_{l}$ in both prior frameworks are similar in the sense that they control the degree of the sparsity \citep{polson2010shrink}. Slab density and local-scale prior density are expected to put a sufficient mass on the tail regions of the densities to detect signal coefficients and produce a robust Bayes estimator for $\bm{\beta}_{l}$ \citep{lee2020tail,carvalho2010horseshoe}. For that reason, heavy-tailed densities (for e.g., double generalized Pareto distribution, Cauchy distributions \citep{lee2019exponentiated,armagan2013generalized}) are preferably used. Refer to \citep{o2009review,bhadra2019lasso} for comprehensive surveys on Bayesian variable selection. 
\subsection{Priors for covariance matrix}\label{subsec:Priors for covariance}
Consider a nonlinear function $f(t ; \bm{\theta})$ indexed by a $K$-dimensional model parameter $\bm{\theta} = (\theta_{1},\cdots,$ $\theta_{l},\cdots,\theta_{K})^{\top}$ to describe an individual trajectory. A basic assumption is that all the components $\theta_{l}$ are unrelated across $l$ ($l=1,\cdots,K$), which is referred to as \emph{uncorrelated design setting}. In many practical problems, this setting is reasonably accepted since one of the fundamental assumptions on $f$ is that each component $\theta_{l}$ has its own role in modifying a functional shape of $f$. A central goal of researcher in using nonlinear mixed models is to examine these roles mathematically, endowed with interpretations by domain experts in terms of physiology, epidemiology, or pharmacology, etc. The basic model (\ref{eq:Stage 1: Individual-Level Model}) -- (\ref{eq:Stage 3: Prior}) that we illustrated so far is designed under this assumption; recall that the covariance matrix $\bm{\Omega}\in \mathbb{R}^{K \times K}$ on Stage 2 was assumed to be diagonal, $\bm{\Omega} = \text{diag}(\omega_{1}^{2},\cdots,\omega_{l}^{2},\cdots, \omega_{K}^{2})$. Under this uncorrelated design setting, possible options for priors for the scale components $\omega_{l}^{2}$ (or $\omega_{l}$) are (i) Jeffreys prior $\pi(\omega_{l}^{2}) \propto 1/\omega_{l}^{2}$ \citep{jeffreys1946invariant}; (ii) inverse-gamma prior $\pi(\omega_{l}^{2}) = \mathcal{IG}(a_{\omega_{l}^{2}},b_{\omega_{l}^{2}})$ with shape $a_{\omega_{l}^{2}}>0$ and scale $b_{\omega_{l}^{2}}>0$; and (iii) half-Cauchy prior $\pi(\omega_{l})=\mathcal{C}^{+}(0,b_{\omega_{l}}) = \{2/(\pi b_{\omega_{l}})\} \cdot 1/\{ 1 + (\omega_{l}/b_{\omega_{l}})^{2}\}$ with scale $b_{\omega_{l}}>0$. See discussion by \citep{gelman2006prior} for the prior options implemented on \emph{8-schools example}.

We discuss Bayesian inference about a population covariance matrix $\bm{\Omega}$ under a \emph{correlated design setting}, where the off-diagonal entries of $\bm{\Omega}$ (\ref{eq:Omega_Cov}) are allowed to be nonzeros:
\begin{align}
\label{eq:Omega_Cov}
\bm{\Omega}
&=
\begin{bmatrix}
\omega_{11} & \cdots & \omega_{1l}  & \cdots & \omega_{1K} \\
\vdots && \vdots  & &\vdots \\
\omega_{l1} & \cdots & \omega_{ll}  & \cdots & \omega_{Kl} \\
\vdots && \vdots  & &\vdots \\
\omega_{K1} & \cdots & \omega_{Kl}  & \cdots & \omega_{KK} 
\end{bmatrix}\in \mathbb{R}^{K \times K}.
\end{align}
In (\ref{eq:Omega_Cov}), the $(l_{1},l_{2})$-th entry is denoted by
$\bm{\Omega}[l_{1}, l_{2}] = \omega_{l_{1} l_{2}} = \text{cov}(
\eta_{l_{1} i},\eta_{l_{2} i}
)$ ($l_{1},l_{2} = 1,\cdots,K; i = 1, \cdots,N$). With $l_{1} = l_{2}$, we have $\omega_{l_{1} l_{2}} = \omega_{l_{1}}^{2}$ ($l_{1}=1,\cdots,K$). 

Researchers often wish to work with the correlated design setting to examine whether any pair of model parameters, $\theta_{l_{1}}$ and $\theta_{l_{2}}$, are physiologically (or epidemiology, pharmacology, financially, etc) associated or not. Taking the term structure modeling discussed in Subsection \ref{subsec:Example 3: Yield curve modeling} as an example, it is a valid question whether the Nelson-Siegel parameters are correlated or not as they are all associated with the interest rate \citep{diebold2006forecasting}. Statistically, having a well-designed covariance structure can also improve the model fitting and produce reliable estimators for the model parameters compared to uncorrelated designs. To make a fully Bayesian inference about the $\bm{\Omega}$ (\ref{eq:Omega_Cov}), we need to specify an appropriate prior $\pi(\bm{\Omega})$ which we will discuss shortly. After that, we operate the Gibbs sampler in Subsection \ref{subsec:Gibbs sampling algorithm}, with some modifications, if necessary. For instance, to implement the parallel computation in \emph{\textbf{Step 1}}, we recommend to sample from the joint density $\pi(\bm{\theta}^{i}|-)$ across $i$, instead of sampling from the individual $\pi(\theta_{li}|-)$. Especially, among the five steps of the Gibbs sampler, implementation of the \emph{\textbf{Step 5}} needs special care in sampling from the full-conditional posterior density $\bm{\Omega}$. This step can be highly complicated depending on the chosen prior.

A challenge in choosing a workable prior $\pi(\bm{\Omega})$ is briefly mentioned. Researches regarding this subject are vast, growing, and deep. As similar to the obstacles encountered in classical covariance estimation \citep{fan2016overview,bickel2008covariance,lam2009sparsistency,el2010high,stein1975estimation}, there are three major aspects, among many others, in the consideration of a thoughtful prior $\pi(\bm{\Omega})$ to produce a reliable Bayes estimator of $\bm{\Omega}$: (1) sample size $N$; (2) the number of model parameters $K$; and (3) positive-definiteness of $\bm{\Omega}$ \citep{pourahmadi2013high}. The first two aspects are related to theoretical constraints. In general, it is  known that the estimation of the covariance $\bm{\Omega}$ can be distorted unless the ratio $K/N$ is sufficiently small (see, e.g.,  \citep{ledoit2004well,rajaratnam2008flexible,won2013condition}). The third one is germane to the modeling consideration and computation strategies to estimate $\bm{\Omega}$, typically resolved via principal component analysis, Cholesky decomposition, and Gaussian graphical models, etc \citep{liu1993bartlett,barnard2000modeling}.

For any prior $\pi(\bm{\Omega})$, the full conditional density of $\bm{\Omega}$ is analytically expressed as follows (see \citep{geisser1965bayesian} for a similar derivation): 
\begin{align}
\label{eq:full_condi_Omega_cov_design}
\pi(\bm{\Omega} &| \bm{\Theta},\sigma^2,  \bm{\alpha}, \bm{B},\textbf{y}_{1:N})\propto
\pi(\bm{\Theta} |\bm{\alpha}, \bm{B}, \bm{\Omega} ) \cdot 
\pi(\bm{\Omega})
\propto
\bigg\{
\prod_{i =1}^{N}
\mathcal{N}_{K} (\bm{\theta}^{i}|\bm{\alpha} + \bm{B} \textbf{x}_{i} , \bm{\Omega})
\bigg\}
 \cdot
\pi( \bm{\Omega})\\
\nonumber
&\propto
(\text{det}\bm{\Omega})^{-N/2}
\exp
\left(
-\frac{1}{2}
\text{tr}
\left[
\bm{\Omega}^{-1}
\left\{
(N-1)
\textbf{G}
+
N
(\bar{\bm{\phi}}
-\bm{\alpha}
)
(\bar{\bm{\phi}}
-\bm{\alpha} 
)^{\top}
\right\}
\right]
\right)
\cdot 
\pi(\bm{\Omega}),
\end{align}
where $\bm{\phi}^{i} = \bm{\theta}^{i}  - \bm{B}\textbf{x}_{i}\, (i=1,\cdots,N)\,,\bar{\bm{\phi}} \in \mathbb{R}^{K}$, and $\textbf{G} \in \mathbb{R}^{K \times K }$ are defined by
\begin{align*}
&\bar{\bm{\phi}} =
N^{-1} \sum_{i=1}^{N}  \bm{\phi}^{i}=
 N^{-1} \sum_{i=1}^{N} (\bm{\theta}^{i}  - \bm{B}\textbf{x}_{i} ), \quad
\textbf{G} = (N-1)^{-1} 
\sum_{i=1}^{N} (\bm{\phi}^{i} - \bar{\bm{\phi}})
(\bm{\phi}^{i} - \bar{\bm{\phi}})^{\top};\\
&\sum_{i=1}^{N} ( \bm{\phi}^{i}
-\bm{\alpha})
(\bm{\phi}^{i} -\bm{\alpha})^{\top} 
=
(N-1)
\textbf{G}
+
N
(\bar{\bm{\phi}}
-\bm{\alpha}
)
(\bar{\bm{\phi}}
-\bm{\alpha} 
)^{\top}.
\end{align*}
In (\ref{eq:full_condi_Omega_cov_design}), we used the vector-form (b) (i.e., $i$-indexing) to express the prior for $\bm{\Theta}$. Notations $\text{det}(\textbf{A})$ and $\text{tr}(\textbf{A})$ denote the determinant and trace of a square matrix $\textbf{A}$, respectively. Matrix $\textbf{G}$ is the `latent' covariance matrix based on the model matrix $\bm{\Theta} \in \mathbb{R}^{K \times K}$ (\ref{eq:model_matrix}) and coefficient matrix $\bm{G} \in \mathbb{R}^{K \times P}$, whose form resembles sample covariance matrix assuming $\bm{\phi}^{i}$ are observed \citep{won2013condition}.

Traditionally used priors for the covariance matrix $\bm{\Omega}$ (\ref{eq:Omega_Cov}) are the Jefferys prior and the conjugate inverse Wishart prior (see \citep{lin1985monte,brown1994inference} for the reviews of the earlier works):
\begin{itemize}
\baselineskip=15pt
\item[] $\bullet$ \textbf{Jeffreys prior}. The common non-informative prior has been the Jeffreys improper prior
\begin{align*}
\pi(\bm{\Omega}) \propto (\text{det}\bm{\Omega} )^{-(K+1)/2}.
\end{align*}
This prior was originally derived from an invariance argument by \citep{jeffreys1998theory} for the case $K = 1,2$; and it was considered for arbitrary $K$ by \citep{geisser1963posterior,geisser1965bayesian,villegas1969priori} to develop Bayesian multivariate theory. 
\item[] $\bullet$ \textbf{Inverse-Wishart prior}. The common informative prior is the inverse-Wishart prior \citep{schervish2012theory}
\begin{align*}
\pi(\bm{\Omega}) = \mathcal{IW}(\textbf{V},d)
=
\frac{ (\text{det}\textbf{V})^{d/2} }{2^{d K/2} \Gamma_{K}(d/2)}
(\text{det}\bm{\Omega})^{-(d + K +1)/2}
\exp
\left(
-\frac{1}{2}
\text{tr}
\left[
\bm{\Omega}^{-1}
\textbf{V}
\right]
\right),
\end{align*}
where $\bm{\Omega}$ and $\textbf{V}$ are $K$-by-$K$ positive definite matrices, and $\Gamma_{K}(\cdot)$ is the multivariate gamma function \citep{james1964distributions}. $\textbf{V}$ is a scale matrix, and $d (> K-1)$ is the number of degrees of freedom. Conventionally, $d$ is chosen to be as small as possible to reflect vague prior knowledge. A univariate specialization ($K=1$) is the inverse-gamma distribution. 
\end{itemize}

The success of Bayesian computation and MCMC in the late 1980s opened up the potential of using more flexible non-conjugate priors for covariance matrices \citep{yang1994estimation,daniels2001shrinkage,wong2003efficient}. Limitations of the traditional priors studied by many statisticians also motivated them to develop a new prior. For example, some of them argued that the Jeffreys prior may not be really non-informative, particularly in high dimensional setting \citep{sun2007objective,barnard2000modeling}, and inverse Wishart prior is very restrictive and lacks flexibility \citep{daniels2002bayesian}. Among many new priors developed for particular applications \citep{smith2002parsimonious}, a combination of \emph{separation strategy} developed by \citep{barnard2000modeling} and LKJ prior \citep{lewandowski2009generating} has been successful, heavily used in a variety of industrial problems, and relevant software has been developed, including \textsc{R} package  \textsc{stan} \citep{stan,carpenter2017stan}.

We illustrate a central idea of using the separation strategy \citep{barnard2000modeling} to estimate the population covariance matrix $\bm{\Omega} \in \mathbb{R}^{K \times K}$. First, we decompose the matrix $\bm{\Omega}$ (\ref{eq:Omega_Cov}) into two components, $K$ standard deviations $\omega_{l}  = \sqrt{\omega_{ll}}  = \sqrt{\bm{\Omega}[l,l] }$ $(l=1,\cdots,K)$ and correlation matrix $\textbf{R}\in \mathbb{R}^{K \times K}$:
\begin{align}
\label{eq:Omega_Cov_Separate}
\bm{\Omega}
&=
\text{diag}(\bm{\omega})
\,
\textbf{R}
\,
\text{diag}(\bm{\omega})
\in \mathbb{R}^{K \times K},
\end{align}
where $\bm{\omega} = (\omega_{1},\cdots,\omega_{K})^{\top}$ is the $K$-dimensional vector of standard deviations, $\text{diag}(\bm{\omega})$ is the diagonal matrix with diagonal elements $\bm{\omega}$, and $\textbf{R}$ is the $K$-by-$K$ correlation matrix. Second, we specify priors independently for $\bm{\omega}$ and $\textbf{R}$, denoted as $\pi(\bm{\omega})$ and $\pi(\textbf{R})$, respectively, so that we have the joint prior for $(\bm{\omega},\textbf{R})$, $\pi(\bm{\omega},\textbf{R}) = \pi(\bm{\omega}) \cdot \pi(\textbf{R})$. Following notation from \citep{barnard2000modeling}, let $\mathcal{R}^{K}$ denote the correlation matrix space. Then, the priors $\pi(\bm{\omega}) $ and $\pi(\textbf{R})$ are supported on $(0,\infty)^{K}$ and $\mathcal{R}^{K}$, respectively. Finally, draw sample from each of the full conditional posterior distributions $\pi(\bm{\omega}|\textbf{R},-)$ and $\pi(\textbf{R}|\bm{\omega},-)$ at time in \textbf{\emph{Step 5}} within the Gibbs sampler in Subsection \ref{subsec:Gibbs sampling algorithm}. (This means that, we are not directly sampling the covariance matrix $\bm{\Omega}$ from $\pi(\bm{\Omega}|-)$ (\ref{eq:full_condi_Omega_cov_design}) as we would do when Jeffreys or inverse-Wishart prior were used for the prior.)

Standardly used prior options for the scale vector $\bm{\omega} = (\omega_{1},\cdots,\omega_{l},\cdots,\omega_{K})^{\top}$ are (i) $\log (\bm{\omega}) \sim \mathcal{N}_{K}(\textbf{a}_{\omega},\textbf{B}_{\omega})$, where $\log (\bm{\omega})=(\log \omega_{1},\cdots,\log \omega_{l},\cdots, \log \omega_{K})^{\top}$, with hyperparameters, mean $\textbf{a}_{\omega}\in \mathbb{R}^{K}$ and covariance $\textbf{B}_{\omega}\in \mathbb{R}^{K \times K}$ which is often diagonal \citep{barnard2000modeling}; and (ii) $\omega_{l} \sim \mathcal{C}^{+}(0,b_{\omega_{l}})$ with the scale hyperparameter $b_{\omega_{l}}>0$ \citep{stan}. As for the prior distribution for the correlation $\textbf{R}\in \mathbb{R}^{K \times K}$, the LKJ prior proposed by \citep{lewandowski2009generating} is popularly used:
\begin{itemize}
\baselineskip=15pt
\item[] $\bullet$ \textbf{LKJ prior}. LKJ prior is supported over the correlation matrix space $\mathcal{R}^{K}$, or equivalently over the set of $K \times K$ Cholesky factors of real symmetric positive definite matrces
\begin{small}
\begin{align}
\label{eq:LKJ_prior}
\pi(\textbf{R})
=
\left[
2^{\sum_{q=1}^{Q-1}
(2\gamma - 2
+
Q-q)
(Q-q)
}
\prod_{q= 1}^{Q-1}
\mathcal{B}
\bigg(
\gamma + 
\frac{Q - q-1}{2}
,
\gamma + 
\frac{Q - q-1}{2}
\bigg)^{Q - q	}
\right]
(\text{det} \, \textbf{R})^{\gamma - 1},
\end{align}
\end{small}
with the shape parameter $\gamma>0$. The function $\mathcal{B}(\alpha,\beta)$ is the beta function. If $\gamma=1$, the density is uniform over the space $\mathcal{R}^{K}$; for $\gamma>1$, the density increasingly concentrates mass around the identity matrix $\textbf{I}\in \mathbb{R}^{K \times K}$ (i.e., favoring less correlation); for $\gamma<1$, the density increasingly concentrates mass in the other direction, and has a trough at the identity matrix (i.e., favoring more correlation). 
\end{itemize}
Note that the normalizing constant of the LKJ prior (\ref{eq:LKJ_prior}) is constant with respect to $\gamma$, therefore, we have $\pi(\textbf{R}) \propto (\text{det} \, \textbf{R})^{\gamma - 1}$, with the shape hyperparameter $\gamma>0$. The behavior of LKJ prior with $\gamma=1$ (i.e., $\pi(\textbf{R}) \propto 1$) was studied by \citep{barnard2000modeling}, where the author found that as $K$ increases the marginal correlations tend to concentrate around zeros (see Figure 1 from \citep{barnard2000modeling}), hence, model matrix $\bm{\Theta}$ (\ref{eq:model_matrix}) are more likely to be treated as in the uncorrelated design setting.

As for the hyperparameter specification for the LKJ prior, Stan Development Team \citep{stan} recommends to use $\gamma \geq 1$. This suggestion is also well-aligned with the original intention of using the separation strategy to make a variance-correlation structure by \citep{barnard2000modeling} in that: (1) the authors intend to choose a diffuse prior for $\pi(\textbf{R})$ to reflect weak knowledge about the correlation $\textbf{R}$, while (2) prior knowledge, possibly informative, shall be put on the scale parameters by specifying $\pi(\bm{\omega})$, as most statisticians are normally trained to do. Computational algorithm and theory concerning the LKJ prior can be found in \citep{lewandowski2009generating,ghosh2003behavior,joe2006generating}.


\section{Model selection}\label{sec:Model selection}
\subsection{Setting}\label{subsec:Setting}
The recent development of MCMC methods has made it possible to fit enormously large classes of models with the aim of exploring real world complexities of data \citep{gilks1995markov}. This ability naturally led us to wish to compare several candidate models that vary substantially in the model complexities and choose the best model out of them. For example, authors \citep{lee2021bayesian} tried to compare four different rate decline curves to fit the production data from the 360 wells in Figure \ref{fig:Shale_Oil_Production}. Indeed, upstream petroleum engineers endeavor to find a rate decline curve describing the production trajectories as accurately as possible so that EUR can be accurately estimated. Another application of model selection can be found in PK analysis. Taking the theophylline data in Figure \ref{fig:Theophylline_data} as an example, PK modelers may debate whether they need to use a two or three-compartment model with a nonlinear clearance to describe the PK exposure, or just one-compartment model with a linear clearance is sufficient.

In the current section, our primary focus is to illustrate a Bayesian approach to compare multiple Bayesian nonlinear mixed effects models explaining the data $\{(\textbf{y}_{i},\textbf{t}_{i},\textbf{x}_{i})\}_{i=1}^{N}$ introduced in Subsection \ref{subsec:Statistical problem}. To that end, we want to lay out the set-up that underlies our model selection procedure. Consider Stage 1 and 2 of the basic model (i.e., the hierarchy (\ref{eq:Stage 1: Individual-Level Model}) -- (\ref{eq:Stage 2: Population Model})), endowed with a joint prior $\pi(\sigma^2,  \bm{\alpha}, \bm{B}, \bm{\Omega})$. Very importantly, we do \emph{not} assume the independent prior assumption as we did in Stage 3 (\ref{eq:Stage 3: Prior}): any prior assumption on the parameters $(\sigma^2,  \bm{\alpha}, \bm{B}, \bm{\Omega})$ works fine in our framework for the model comparison. Therefore, the basic model (\ref{eq:Stage 1: Individual-Level Model}) -- (\ref{eq:Stage 3: Prior}) is considered as a subclass of candidate models that we want to compare. In our framework, we can also consider the correlated design setting discussed in Subsection \ref{subsec:Priors for covariance}, where the covariance matrix $\bm{\Omega}$ (\ref{eq:Omega_Cov}) is allowed to be any positive-definite matrix (i.e., does not need to be a diagonal matrix), as one of candidates. Eventually, in our framework, modelers have freedom to choose (i) the nonlinear function $f$ to describe the temporal profile $\textbf{y}_{i}$ and (ii)  prior distribution $\pi(\sigma^2,  \bm{\alpha}, \bm{B}, \bm{\Omega})$. 

Assume that a researcher wants to consider $H$ functions, denoted as $f(t; \bm{\theta}_{[\mathcal{M}_{1}]})$, $f(t; \bm{\theta}_{[\mathcal{M}_{2}]})$, $\cdots$, $f(t; \bm{\theta}_{[\mathcal{M}_{H}]})$, as a possible option for the use of $f$ in Stage 1. Here, the subscript `$[\mathcal{M}_{h}]$' on the model parameter $\bm{\theta}$ is noted to indicate for the model index ($h=1,\cdots,H$). Obviously, the candidate functions $f(t; \bm{\theta}_{[\mathcal{M}_{h}]})$ ($h=1,\cdots,H$) can have different functional forms dictated by different dimensions for the model parameters $\bm{\theta}_{[\mathcal{M}_{h}]} \in \mathbb{R}^{K_{[\mathcal{M}_{h}]}}$. This will consecutively change the dimensions of the parameter blocks, $\bm{\alpha}_{[\mathcal{M}_{h}]} \in \mathbb{R}^{K_{[\mathcal{M}_{h}]}}, \bm{B}_{[\mathcal{M}_{h}]} \in \mathbb{R}^{K_{[\mathcal{M}_{h}]} \times P}$, and $\bm{\Omega}_{[\mathcal{M}_{h}]} \in \mathbb{R}^{K_{[\mathcal{M}_{h}]} \times K_{[\mathcal{M}_{h}]}} $, accordingly, yet the support of the $\sigma^{2}$ remains the same with $(0,\infty)$ because we still consider the additive error model. After that, she now has the freedom to choose a prior $\pi(\sigma^2,  \bm{\alpha}_{[\mathcal{M}_{h}]}, \bm{B}_{[\mathcal{M}_{h}]}, \bm{\Omega}_{[\mathcal{M}_{h}]})$ ($h=1,\cdots,H$). There are infinitely many choices for the prior, and one can use priors discussed in Section \ref{sec:Prior options}.

Now, the fundamental question naturally arising at this point is "what is the best model among the $H$ candidate models?" To illustrate, we write the situation more technically. With above specifications, Stage 1 of each of the $H$ candidate models is given as
\begin{align*}
\mathcal{M}_{1}: \quad 
y_{ij} &=
f(t_{ij} ; (\bm{\theta}_{[\mathcal{M}_{1}]})^{i} ) + \epsilon_{ij}, \quad \epsilon_{ij}\sim \mathcal{N}(0,\sigma^{2}), \quad (i=1,\cdots,N; j  = 1,\cdots,M_{i});\\
&\vdots \\
\mathcal{M}_{h}: \quad 
y_{ij} &=
f(t_{ij} ; (\bm{\theta}_{[\mathcal{M}_{h}]})^{i}) + \epsilon_{ij}, \quad \epsilon_{ij}\sim \mathcal{N}(0,\sigma^{2}), \quad (i=1,\cdots,N; j  = 1,\cdots,M_{i});\\
&\vdots \\
\mathcal{M}_{H}: \quad 
y_{ij} &=
f(t_{ij} ; (\bm{\theta}_{[\mathcal{M}_{H}]})^{i}) + \epsilon_{ij}, \quad \epsilon_{ij}\sim \mathcal{N}(0,\sigma^{2}), \quad (i=1,\cdots,N; j  = 1,\cdots,M_{i}),
\end{align*}
where each of the models has the corresponding  model matrix
\begin{align*}
\mathcal{M}_{1}: \quad 
\bm{\Theta}_{[\mathcal{M}_{1}]} 
&=
\begin{bmatrix}
(\bm{\theta}_{[\mathcal{M}_{1}]})^{1} 
&\cdots &
(\bm{\theta}_{[\mathcal{M}_{1}]})^{i} 
&
\cdots 
&
(\bm{\theta}_{[\mathcal{M}_{h}]})^{N} 
\end{bmatrix} \in \mathbb{R}^{K_{[\mathcal{M}_{1}]}   \times N};\\
&\vdots\\
\mathcal{M}_{h}: \quad 
\bm{\Theta}_{[\mathcal{M}_{h}]} 
&=
\begin{bmatrix}
(\bm{\theta}_{[\mathcal{M}_{h}]})^{1} 
&\cdots &
(\bm{\theta}_{[\mathcal{M}_{h}]})^{i} 
&
\cdots 
&
(\bm{\theta}_{[\mathcal{M}_{h}]})^{N} 
\end{bmatrix} \in \mathbb{R}^{K_{[\mathcal{M}_{h}]}   \times N};\\
&\vdots\\
\mathcal{M}_{H}: \quad 
\bm{\Theta}_{[\mathcal{M}_{H}]} 
&=
\begin{bmatrix}
(\bm{\theta}_{[\mathcal{M}_{H}]})^{1} 
&\cdots &
(\bm{\theta}_{[\mathcal{M}_{H}]})^{i} 
&
\cdots 
&
(\bm{\theta}_{[\mathcal{M}_{h}]})^{N} 
\end{bmatrix} \in \mathbb{R}^{K_{[\mathcal{M}_{H}]}   \times N},
\end{align*}
obtained by stacking individual-based vector horizontally as we did to obtain $\bm{\Theta}$ (\ref{eq:model_matrix}). Again, the number of rows of the matrix $\bm{\Theta}_{[\mathcal{M}_{h}]}$, that is, $K_{[\mathcal{M}_{h}]}$, depends on the choice of the function $f$. Stage 2 of each of the $H$ models will then be
\begin{align*}
\mathcal{M}_{1}: \quad 
(\bm{\theta}_{[\mathcal{M}_{1}]})^{i} & = \bm{\alpha}_{[\mathcal{M}_{1}]} + \bm{B}_{[\mathcal{M}_{1}]} \textbf{x}_{i} +  (\bm{\eta}_{[\mathcal{M}_{1}]})^{i}, \quad (\bm{\eta}_{[\mathcal{M}_{1}]})^{i} \sim \mathcal{N}_{K}(\textbf{0}, \bm{\Omega}_{[\mathcal{M}_{1}]}),\quad (i=1,\cdots,N);
\\
&\vdots
\\
\mathcal{M}_{h}: \quad 
(\bm{\theta}_{[\mathcal{M}_{h}]})^{i} & = \bm{\alpha}_{[\mathcal{M}_{h}]} + \bm{B}_{[\mathcal{M}_{h}]} \textbf{x}_{i} +  (\bm{\eta}_{[\mathcal{M}_{h}]})^{i}, \quad (\bm{\eta}_{[\mathcal{M}_{h}]})^{i} \sim \mathcal{N}_{K}(\textbf{0}, \bm{\Omega}_{[\mathcal{M}_{h}]}),\quad (i=1,\cdots,N);
\\
&\vdots
\\
\mathcal{M}_{H}: \quad 
(\bm{\theta}_{[\mathcal{M}_{H}]})^{i} & = \bm{\alpha}_{[\mathcal{M}_{H}]} + \bm{B}_{[\mathcal{M}_{H}]} \textbf{x}_{i} +  (\bm{\eta}_{[\mathcal{M}_{H}]})^{i}, \quad (\bm{\eta}_{[\mathcal{M}_{H}]})^{i} \sim \mathcal{N}_{K}(\textbf{0}, \bm{\Omega}_{[\mathcal{M}_{H}]}),\quad (i=1,\cdots,N).
\end{align*}
Finally, Stage 3 of each of the $H$ models is comprised of the prior:
\begin{align*}
\mathcal{M}_{1}: \quad 
(\sigma^2,  \bm{\alpha}_{[\mathcal{M}_{1}]}, \bm{B}_{[\mathcal{M}_{1}]}, \bm{\Omega}_{[\mathcal{M}_{1}]}) &\sim
\pi(\sigma^2,  \bm{\alpha}_{[\mathcal{M}_{1}]}, \bm{B}_{[\mathcal{M}_{1}]}, \bm{\Omega}_{[\mathcal{M}_{1}]})
;
\\
&\vdots
\\
\mathcal{M}_{h}: \quad 
(\sigma^2,  \bm{\alpha}_{[\mathcal{M}_{h}]}, \bm{B}_{[\mathcal{M}_{h}]}, \bm{\Omega}_{[\mathcal{M}_{h}]}) &\sim
\pi(\sigma^2,  \bm{\alpha}_{[\mathcal{M}_{h}]}, \bm{B}_{[\mathcal{M}_{h}]}, \bm{\Omega}_{[\mathcal{M}_{h}]})
;
\\
&\vdots
\\
\mathcal{M}_{H}: \quad
(\sigma^2,  \bm{\alpha}_{[\mathcal{M}_{H}]}, \bm{B}_{[\mathcal{M}_{H}]}, \bm{\Omega}_{[\mathcal{M}_{H}]}) &\sim
\pi(\sigma^2,  \bm{\alpha}_{[\mathcal{M}_{H}]}, \bm{B}_{[\mathcal{M}_{H}]}, \bm{\Omega}_{[\mathcal{M}_{H}]})
.
\end{align*}

We describe three model comparison criteria that are popularly used in the literature: deviance information criterion (DIC) \cite{spiegelhalter2002bayesian,gelman2004bayesian}, widely applicable information criterion (WAIC) \cite{watanabe2010asymptotic}, and posterior predictive loss criterion (PPLC) \cite{gelfand1998model}. As in frequentist information criteria \citep{akaike1998information,efron1986biased,burnham1998practical}, formulation of the DIC, WAIC, and PPLC also takes the two terms into a consideration: \emph{goodness-of-fit} and \emph{penalty for model complexity}. Because increasing (or decreasing) model complexity is accompanied by the risk of over-fitting (or under-fitting), models should be compared by trading-off these two terms. Particularly, as we are currently discussing about a Bayesian hierarchical model, these criteria are obviously depending on what part of the model specification is considered to be part of the likelihood, and what is not. \citep{spiegelhalter2002bayesian} refer to this as the \emph{focus issue}. For example, in the general form of a Bayesian hierarchical model consisting of a top-level likelihood $p(\textbf{y}|\bm{\Psi})$ for data $\textbf{y}$, a prior model $\pi(\bm{\Psi}|\eta)$, and a hyperprior $\pi(\eta)$, one might choose as the likelihood either the conditional density $p(\textbf{y}|\bm{\Psi})$, or the marginal density $p(\textbf{y}|\eta) = \int p(\textbf{y}|\bm{\Psi}) \pi(\bm{\Psi}|\eta) d\bm{\Psi}$. Based on \citep{spiegelhalter2002bayesian}, the former situation is referred to as ``focus on $\bm{\Psi}$", while the latter situation is referred to as ``focus on $\eta$", respectively. 

In our case, we shall ``focus on parameters $(\bm{\Theta}_{[\mathcal{M}_{h}]}, \sigma^{2})$" $(h=1,\cdots,H)$ used in the conditional density in Stage 1. For notational simplicity, we denote $\bm{\Psi}_{[\mathcal{M}_{h}]} = (\bm{\Theta}_{[\mathcal{M}_{h}]}, \sigma^{2})$. Then, likelihood of each of the $H$ models based on $N$ observations $\{(\textbf{y}_{i},\textbf{t}_{i},\textbf{x}_{i})\}_{i=1}^{N}$ is
\begin{align*}
\nonumber
\mathcal{M}_{1}: \quad
\mathcal{L}(\bm{\Psi}_{[\mathcal{M}_{1}]}|\textbf{y}_{1:N})
& =
\prod_{i=1}^{N}
\mathcal{L}((\bm{\Psi}_{[\mathcal{M}_{1}]})^{i} |\textbf{y}_{i})
=
\prod_{i=1}^{N}
\mathcal{N}_{M_{i}}(\textbf{y}_{i}|
\bm{f}_{i}(\textbf{t}_{i}, (\bm{\theta}_{[\mathcal{M}_{1}]})^{i})
,
\sigma^{2}  \textbf{I} 
);\\
\nonumber
&\vdots
\\
\nonumber
\mathcal{M}_{h}: \quad
\mathcal{L}(\bm{\Psi}_{[\mathcal{M}_{h}]}|\textbf{y}_{1:N})
& =
\prod_{i=1}^{N}
\mathcal{L}((\bm{\Psi}_{[\mathcal{M}_{h}]})^{i} |\textbf{y}_{i})
=
\prod_{i=1}^{N}
\mathcal{N}_{M_{i}}(\textbf{y}_{i}|
\bm{f}_{i}(\textbf{t}_{i}, (\bm{\theta}_{[\mathcal{M}_{h}]})^{i})
,
\sigma^{2}  \textbf{I} 
);\\
\nonumber
&\vdots
\\
\nonumber
\mathcal{M}_{H}: \quad
\mathcal{L}(\bm{\Psi}_{[\mathcal{M}_{H}]}|\textbf{y}_{1:N})
& =
\prod_{i=1}^{N}
\mathcal{L}((\bm{\Psi}_{[\mathcal{M}_{H}]})^{i} |\textbf{y}_{i})
=
\prod_{i=1}^{N}
\mathcal{N}_{M_{i}}(\textbf{y}_{i}|
\bm{f}_{i}(\textbf{t}_{i}, (\bm{\theta}_{[\mathcal{M}_{H}]})^{i})
,
\sigma^{2}  \textbf{I} 
),
\end{align*}
where $\mathcal{L}((\bm{\Psi}_{[\mathcal{M}_{h}]})^{i} |\textbf{y}_{i})$ is the likelihood (i.e., data distribution) based on an individual data with parameter $(\bm{\Psi}_{[\mathcal{M}_{h}]})^{i} = ( (\bm{\theta}_{[\mathcal{M}_{h}]})^{i}, \sigma^{2})$ ($i=1,\cdots,N$). One caveat of the practical calculation to obtain the three criteria is that, we mainly need the posterior samples of $\bm{\Psi}_{[\mathcal{M}_{h}]}$ drawn from the joint posterior density $\pi(\bm{\Psi}_{[\mathcal{M}_{h}]}, 
\bm{\alpha}_{[\mathcal{M}_{h}]}, \bm{B}_{[\mathcal{M}_{h}]}, \bm{\Omega}_{[\mathcal{M}_{h}]}
|\textbf{y}_{1:N}) $ and the explicit form the likelihood function $\mathcal{L}(\bm{\Psi}_{[\mathcal{M}_{h}]}|\textbf{y}_{1:N})$ from each of the models $\mathcal{M}_{h}$ ($h=1,\cdots,H$) due to our assumption of the focus.

In the next subsections, we provide some brief summaries of the criteria and then adapt them to our context. In what follows, to simplify the notation, we suppress the arguments $[\mathcal{M}_{h}]$ in the parameters. For a detailed explanation of the criteria, refer to \cite{banerjee2014hierarchical,gelman2014understanding}. 
\subsection{Deviance information criterion}\label{subsec:Deviance information criterion}
\cite{gelman2004bayesian} suggested $\text{DIC}$, a generalized version of  Akaike information criterion \citep{akaike1998information} for a Bayesian hierarchical model, given by
\begin{align}
\label{eq: DIC_Gelman}
\text{DIC}
& = 
D(\overline{\bm{\Psi}})
+ 2\cdot p_{\textrm{D}}.
\end{align}
In (\ref{eq: DIC_Gelman}), the function $D(\bm{\Psi}) = -2 \log\ \mathcal{L}(\bm{\Psi}|\textbf{y}_{1:N})$ is referred to as deviance. Deviance is a goodness-of-fit statistics whose lower value indicates a better fitting \citep{celeux2006deviance}. Goodness-of-fit term of DIC (i.e., $D(\overline{\bm{\Psi}})$) is the value of deviance evaluated at the posterior mean of $\bm{\Psi}$, denoted by $\overline{\bm{\Psi}} =\mathbb{E}[\bm{\Psi}|\textbf{y}_{1:N}]
=
\int \bm{\Psi} \pi(\bm{\Psi}|\textbf{y}_{1:N}) \bm{\Psi}$, where $\pi(\bm{\Psi}|\textbf{y}_{1:N})$ represents the posterior distribution of $\bm{\Psi}=(\bm{\Theta}, \sigma^{2})$. Effective number of parameters (i.e, penalty term for model complexity) of DIC in (\ref{eq: DIC_Gelman}) is obtained by $p_{\textrm{D}} = \text{Var}[D(\bm{\Psi})| \textbf{y}_{1:N}]/2 = 2 \cdot \text{Var}[\log\ \mathcal{L}( \bm{\Psi}|\textbf{y}_{1:N})| \textbf{y}_{1:N}]$. A model with a smaller value for $\text{DIC}$ indicates a better predictive performance among considered models.

Some intuition behind having two competing additive terms in (\ref{eq: DIC_Gelman}) is as follows. Typically, complex models get rewards in terms of the deviance than simple models: therefore, over-fitted models normally have a preference over under-fitted models when only the deviance is considered in model comparison, which is undesirable. By adding a penalty term for the model complexity to the deviance term, we hope that the resulting criterion produces a reasonable value based on fair comparison regardless of model complexity. Roughly speaking, this principle (i.e., a trade-off between the goodness-of-fit and penalty terms) is commonly manifested in the three criteria, DIC, WAIC, and PPLC.

Going back to our examples, we can obtain the $\text{DIC}$ corresponding to each of the $H$ candidate models $\mathcal{M}_h$ ($h=1,\cdots,H$):
\begin{align*}
D(\bm{\Psi})  = 
-2 \sum_{i=1}^{N} 
\log\ \mathcal{N}_{M_{i}}
(
\textbf{y}_{i}
|
\bm{f}_{i}(\textbf{t}_{i}, \bm{\theta}^{i})
, \sigma^{2} \textbf{I}),
\end{align*}
where only mean function $\bm{f}_{i}(\textbf{t}_{i}, \bm{\theta}^{i})$ differs across $H$ models $\mathcal{M}_{h}$, $h=1,\cdots,H$. In practice, posterior mean $\overline{\bm{\Psi}}$ and effective number of parameters $p_{\textrm{D}}$ are not expressed in closed-forms, hence, the $\text{DIC}$ (\ref{eq: DIC_Gelman}) is stochastically approximated though MCMC techniques \citep{robert2013monte}.
\subsection{Widely applicable information criterion}\label{subsec: Widely applicable information criterion}
\cite{watanabe2010asymptotic} introduced WAIC which is regarded as a fully Bayesian version of the DIC (\ref{eq: DIC_Gelman}) in the sense that a goodness-of-fit term exploits the entire posterior distribution. Note that the goodness-of-fit term of the DIC (\ref{eq: DIC_Gelman}) is obtained by plugging the posterior mean $\overline{\bm{\Psi}}$ into the deviance $D(\bm{\Psi})$, which lacks a fully Bayesian sense. It is known that WAIC is asymptotically equivalent to Bayesian cross-validation \citep{vehtari2014waic}, and also applicable to singular models. 

WAIC is defined by
\begin{align}
\label{eq:WAIC}
\text{WAIC}  
& =
-2 \cdot \text{LPPD} + 2 \cdot p_{\textrm{W}},
\end{align}
where the goodness-of-fit term is called the log posterior predictive density (LPPD), which is defined as
$
\text{LPPD} = 
\sum_{i=1}^{N}
\log\ \mathbb{E}  [\mathcal{L}(\bm{\Psi}^{i}|\textbf{y}_{i})|\textbf{y}_{1:N}]
$, and the effective number of parameter in the penalty term is defined by $p_{\textrm{W}} = \sum_{i=1}^{N}
\text{Var}[\log\ \mathcal{L}(\bm{\Psi}^{i}|\textbf{y}_{i})|\textbf{y}_{1:N}]$.  

In practice, as similar to DIC (\ref{eq: DIC_Gelman}), WAIC (\ref{eq:WAIC}) is obtained by stochastic approximations. Given posterior samples $\{(\bm{\Psi})^{(s)} \}_{s=1}^{S} \sim \pi(\bm{\Psi} | \textbf{y}_{1:N})$, the $\text{LPPD}$ and $p_{W}$ terms may be approximated by
\begin{align}
\label{eq: LPPD}
\widehat{\text{LPPD}}
 & =
 \sum_{i=1}^{N}
 \log\ \bigg(
 \frac{1}{S}
  \sum_{s=1}^{S}
  \mathcal{L}((\bm{\Psi}^{i})^{(s)}|\textbf{y}_{i}  )
  \bigg),\\
  \label{eq: p_{W}}
 \widehat{p_{\textrm{W}}}&=
  \sum_{i=1}^{N}
    \bigg\{
   \frac{1}{S-1}
  \sum_{s=1}^{S}
  \bigg(
  \log\ \mathcal{L}((\bm{\Psi}^{i})^{(s)} |\textbf{y}_{i}   )
  -
   \frac{1}{S}\sum_{s=1}^{S}
     \log\ \mathcal{L}((\bm{\Psi}^{i})^{(s)}|\textbf{y}_{i}   )
  \bigg)^{2}
  \bigg\}.
 \end{align}
 
Returning to our examples, we can approximate the value of $\text{WAIC}$ corresponding to each of the $H$ models as follows. First, replace $\mathcal{L}(\bm{\Psi}^{i}|\textbf{y}_{i}  )$ in (\ref{eq: LPPD}) and (\ref{eq: p_{W}}) with the individual-based data distribution $p(\textbf{y}_{i}
|\bm{\Psi}^{i}) = \mathcal{N}_{M_{i}}(\textbf{y}_{i} | \bm{f}_{i}(\textbf{t}_{i}, \bm{\theta}^{i}), \sigma^{2} \textbf{I})$ $(i=1, \cdots, N)$, where only the mean function $\bm{f}_{i}(\textbf{t}_{i}, \bm{\theta}^{i})$ differs across the $H$ candidate models, and second, approximate $\widehat{\text{LPPD}}$ and $\widehat{p_{\textrm{W}}}$ by using a MCMC method, and finally, obtain an approximation of $\text{WAIC}$ (\ref{eq:WAIC}) corresponding to each model.

\subsection{Posterior predictive loss criterion}\label{subsec:Posterior predictive loss criterion }
\cite{gelfand1998model} introduced PPLC as an alternative to DIC (\ref{eq: DIC_Gelman}) or WAIC (\ref{eq:WAIC}). A notable feature of PPLC different from DIC and WAIC is its use of replicated observations, denoted by $\textbf{y}_{i}^{rep}
=
(y_{i1}^{rep},y_{i2}^{rep},\cdots,$ $y_{iM_{i}}^{rep})^{\top}\in \mathbb{R}^{M_{i}}
$, corresponding to the actual observations $\textbf{y}_{i} = (y_{i1},$ $y_{i2},$ $\cdots,$ $y_{iM_{i}})^{\top}$ $\in \mathbb{R}^{M_{i}}$, for each $i = 1,\cdots,N$. Here, the replicate $\textbf{y}_{i}^{rep}$ for the subject $i$ is drawn from its posterior predictive density
 \begin{align}
\label{eq: predictive distribution of replicate}
f(
\textbf{y}_{i}^{rep}
|
\textbf{y}_{1:N}
)
&
=
\int
p(
\textbf{y}_{i}^{rep}
|
\bm{\Psi}^{i}
)
\cdot
\pi
(\bm{\Psi}
|
\textbf{y}_{1:N}
)d \bm{\Psi}
, \quad (i = 1, \cdots, N),
\end{align}
where $p(
\textbf{y}_{i}^{rep}
|
\bm{\Psi}^{i}
)$ is the data density for the $i$-th subject and $\pi(\bm{\Psi}|\textbf{y}_{1:N})$ is posterior distribution. The idea of using replicates $\{\textbf{y}_{i}^{rep}\}_{i=1}^{N}$ for a criticism of the model in light of the observed data $\{\textbf{y}_{i}\}_{i=1}^{N}$ is also purported by \cite{box1980sampling}.

A general rule of the PPLC is principled on a balanced loss function \citep{zellner1994bayesian}. Given any loss function $l(\cdot)$ and a positive real number $k$, a balanced loss function is defined by
\begin{align}
\label{eq: balanced loss function}
l(
\textbf{y}_{i}^{rep}
,
\textbf{a}_{i}
;
\textbf{y}_{1:N}
)
& =
l(\textbf{y}_{i}^{rep},\textbf{a}_{i})
+
k
\cdot
l(\textbf{y}_{i},\textbf{a}_{i}), \quad k >0,\, i = 1, \cdots, N,
\end{align}
where $\textbf{a}_{i}$ is a non-stochastic action vector, $k$ is a weight, and $\textbf{y}_{i}^{rep}$ is a replicate for its observed counterpart $\textbf{y}_{i}$. Conceptually, the role of action vector $\textbf{a}_{i}$ is to accommodate both $\textbf{y}_{i}$, and what we predict for $\textbf{y}_{i}^{rep}$. Note that the loss function on the left-hand side of (\ref{eq: balanced loss function}) penalizes actions $\textbf{a}_{i}$ both for departure from the corresponding observed value (fit) as well as for departure from what we expect the replicate to be (smoothness) \citep{banerjee2014hierarchical}. A generic version of PPLC is defined by
$D_{k}  = 
\sum_{i=1}^{N}
\min_{\textbf{a}_{i}}
\mathbb{E}
[
l(
\textbf{y}_{i}^{rep}
,
\textbf{a}_{i}
;
\textbf{y}_{1:N}
)
|
\textbf{y}_{1:N}
]
$,
where the expectation $\mathbb{E}[\cdot|\textbf{y}_{1:N}]$ is taken with respect to the predictive density $f(
\textbf{y}_{i}^{rep}
|
\textbf{y}_{1:N}
)$ (\ref{eq: predictive distribution of replicate}) for some specified $k\geq 0$. Note that the resulting value $D_{k}$ does not depend on the action vector $\textbf{a}_{i}$ and replicates $\{\textbf{y}_{i}^{rep}\}_{i=1}^{N}$ as they are marginalized out by the minimization and expectation, respectively, but is dependent on the constant $k>0$.

By choosing the quadratic loss $l(\textbf{y},\textbf{a}) = \|\textbf{y} - \textbf{a} \|_2^{2}$ in (\ref{eq: balanced loss function}), the generic PPLC $D_{k}$ may be simplified as
\begin{align}
\label{eq: PPLC equation}
D_{k} & = \frac{k}{k+1} G + P, \quad k \geq 0,
\end{align}
where $G
= \sum_{i=1}^{N}  \| \bm{\nu}_{i} - \textbf{y}_{i} \|_2^{2}$ and $P
= 
\sum_{i=1}^{N}
\bm{\varsigma}_{i}^{2}$ represent the goodness-of-fit and penalty terms, respectively, with $\bm{\nu}_{i} = \mathbb{E} [\textbf{y}_{i}^{rep} | \textbf{y}_{1:N} ]$ and $\bm{\varsigma}_{i}^{2}=\mathbb{E} [ \| \textbf{y}_{i}^{rep}-\bm{\nu}_{i} \|_2^{2}| \textbf{y}_{1:N}]$, $i=1, \cdots, N$. Eventually, a model with a smaller value for the $D_{k}$ (\ref{eq: PPLC equation}) is preferable. It is known that ordering of models is insensitive to the particular choice of $k$ \citep{banerjee2014hierarchical}.

Finally, we adapt the PPLC (\ref{eq: PPLC equation})  to our examples. Due to the definition of notation 
$\bm{\Psi}^{i} = (\bm{\theta}^{i}, \sigma^{2})$ ($i=1,\cdots,N$), the posterior predictive distribution of $\textbf{y}_{i}^{rep}$ (\ref{eq: predictive distribution of replicate}) can be detailed as follows
\begin{align*}
f(
\textbf{y}_{i}^{rep}
|
\textbf{y}_{1:N}
)
&
=
\int
\mathcal{N}_{M_{i}}
(
\textbf{y}_{i}^{rep}
|
\bm{f}_{i}(\textbf{t}_{i}, \bm{\theta}^{i})
, \sigma^{2} \textbf{I})
\cdot
\pi
(
\bm{\Psi}^{i}
|
\textbf{y}_{1:N}
)d \bm{\Psi}^{i}
, \quad (i = 1, \cdots, N).
\end{align*}
To approximate $D_k$ (\ref{eq: PPLC equation}) for each model, first, choose a number $k$, saying $k=1$, and second, approximate $\bm{\nu}_{i}$ and $\bm{\varsigma}_{i}^{2}$ through replicates $\textbf{y}_{i}^{rep}$ drawn from the predictive density $f(
\textbf{y}_{i}^{rep}
|
\textbf{y}_{1:N}
)$ (\ref{eq: predictive distribution of replicate}) for each $i=1,\cdots,N$, and finally, complete the $G$ and $P$ to get an approximation to the $D_k$ (\ref{eq: PPLC equation}).


\section{Extensions and recent developments}\label{sec:Extensions and recent developments}
\subsection{Residual error models}\label{subsec:Residual error models}
In the basic version of the Bayesian nonlinear mixed effects model (\ref{eq:Stage 1: Individual-Level Model}) -- (\ref{eq:Stage 3: Prior}), we assume that residual errors in the individual-level model are additive to the mean function $f$ across all subjects and times. Under this assumption, the (conditional) variance of the $i$-th subject's trajectory $\mathbb{V}[y_{ij}| \bm{\theta}^{i}]$ is constant with $\sigma^{2}$ over time $t_{ij}$ ($j=1,\cdots,M_{i}$). The additive error model is the most standard assumption used in a variety of problems arising from many industrial and academic researches \citep{vonesh1992non,muller1997bayesian,lee2021bayesian,lee2020estimation,wakefield1996bayesian}. However, when there exists some systematic temporal trend in the volatility of individual trajectories, for instance, the variance $\mathbb{V}[y_{ij}| \bm{\theta}^{i}]$ seems to decrease over times $t_{ij}$ as shown in Figure \ref{fig:Theophylline_data} and \ref{fig:Shale_Oil_Production}, the additive residual assumption may not be adequate to fully account for the reality of the data.

List in Table \ref{table:Error model} are popularly used residual error models that can be used in Stage 1 (\ref{eq:Stage 1: Individual-Level Model}). Some of them are deployed as options for user to choose in industrial software such as \textsc{Monolix} \citep{lavielle2005monolix} and \textsc{NONMEM} \citep{beal1992nonmem,bauer2019nonmem}, and open source \textsc{R} package such as \textsc{nlmixr} \citep{fidler2019nonlinear}. If we assume $\epsilon_{ij}=\varepsilon_{ij}=0$, then all the error models leads to the same deterministic equation $y_{ij}=f(t_{ij} ; \bm{\theta}^{i})$. That being said, if the variances of the residuals $\epsilon_{ij}\sim \mathcal{N}(0,\sigma^{2}) $ and $\varepsilon_{ij} \sim \mathcal{N}(0,\varsigma^{2})$, that is, $\sigma^{2}$ and $\varsigma^{2}$, are quite small, then the inference outcome based on each of the error models will be similar each other.

\begin{table}[H] 
\begin{footnotesize}
\caption{List of residual error models that can be used in Stage 1 (\ref{eq:Stage 1: Individual-Level Model}).\label{table:Error model}}
\begin{tabular}{llll}
\toprule
\textbf{Type}	& \textbf{Individual-level model}& \textbf{Mean} $\mathbb{E}[y_{ij} | \bm{\theta}^{i}]$ & \textbf{Variance} $\mathbb{V}[y_{ij}| \bm{\theta}^{i}]$\\
\midrule
Additive		&  
$y_{ij} =
f(t_{ij} ; \bm{\theta}^{i}) + \epsilon_{ij}$
			&  $f(t_{ij} ; \bm{\theta}^{i})$ & $\sigma^{2}$\\
Proportional		& 
$y_{ij} =
f(t_{ij} ; \bm{\theta}^{i})  \cdot (1 + \epsilon_{ij})$
			& 
$f(t_{ij} ; \bm{\theta}^{i})$			
			 & $\{ f(t_{ij} ; \bm{\theta}^{i}) \}^{2} \cdot \sigma^{2} $			 
			 \\
Exponential		& 
$y_{ij} =
f(t_{ij} ; \bm{\theta}^{i}) \cdot \exp(\epsilon_{ij}) $
			& 
$f(t_{ij} ; \bm{\theta}^{i})  \cdot \exp(\sigma^{2}/2)$						
			 & $\{ f(t_{ij} ; \bm{\theta}^{i}) \}^{2} \cdot (\exp(\sigma^{2})-1) \cdot \exp(\sigma^{2})$			 
			 \\
Add. and prop.
		& $y_{ij} =
f(t_{ij} ; \bm{\theta}^{i})  \cdot (1 + \epsilon_{ij}) + \varepsilon_{ij}$		& 
$f(t_{ij} ; \bm{\theta}^{i})$			
			 & $\{ f(t_{ij} ; \bm{\theta}^{i}) \}^{2}\cdot \sigma^{2}  + \varsigma^{2}$			 \\
		Add. and exp.
		& $y_{ij} =
f(t_{ij} ; \bm{\theta}^{i}) \cdot \exp(\epsilon_{ij}) + \varepsilon_{ij} $			& 
$f(t_{ij} ; \bm{\theta}^{i})  \cdot \exp(\sigma^{2}/2)$						
&
$\{ f(t_{ij} ; \bm{\theta}^{i}) \}^{2} \cdot ( \exp(\sigma^{2}) -1) \cdot \exp(\sigma^{2})
+\varsigma^{2}
$		 
\\
\bottomrule
\end{tabular}
\noindent{\textsuperscript{1} Random errors are assumed to be distributed according to
	$\epsilon_{ij}\sim \mathcal{N}(0,\sigma^{2})$ and $\varepsilon_{ij} \sim \mathcal{N}(0,\varsigma^{2})$ ($i=1,\cdots,N ; j = 1,\cdots, M_{i}$), with independence between $\epsilon_{ij}$ and $\varepsilon_{ij}$.}
	\end{footnotesize}
\end{table}

The exponential error model (i.e., $y_{ij} =
f(t_{ij} ; \bm{\theta}^{i}) \cdot \exp(\epsilon_{ij})$) is a routine option, which is used when the ranges of the response $y_{ij}$ and mean function $f$ are positive real numbers, while we want to systematically describe the temporal volatility. In the practical implementation, we take the natural logarithm on the both sides of equation of the error model so that the model is converted to an additive error model in log-scale (i.e., $\log y_{ij} = \log (f(t_{ij} ; \bm{\theta}^{i}))  + \epsilon_{ij}$). That way, relevant Bayesian computation becomes much straightforward. \citep{lee2021bayesian} analyzed the shale oil data shown in Figure \ref{fig:Shale_Oil_Production} in this formulation. 

\subsection{Bayesian nonparametric methods}\label{subsec:Bayesian Nonparametric methods}
Recently, the use of the Bayesian nonparametric (BNP) statistical models has received increasing attention in the statistical literature because they allow modelers to gain model flexibility and robustness compared to its parametric counterpart \citep{muller2004nonparametric,hjort2010bayesian}. BNP methods can be applied to the formulation of the basic model (\ref{eq:Stage 1: Individual-Level Model}) --  (\ref{eq:Stage 3: Prior}), when the parametric specification for the error distributions is too restrictive to achieve certain purpose of the analysis, or inference results lead to poor performance due to the inappropriate parametric form. Typically, BNP methods are applied to the population-level model, by extending or relaxing the parametric assumption on the random errors $\eta_{li}$, while retaining the individual-level model as fully parametric \citep{walker1998population}. A Gaussian process prior \citep{mackay1998introduction,rasmussen2003gaussian} or a Dirichlet process prior \citep{ferguson1974prior,escobar1994estimating,escobar1995bayesian} is popularly used for such extension and relaxation. Mathematical concepts of the processes are explained in \citep{rasmussen2003gaussian,ferguson1974prior}.

To illustrate some motivation behind the application of BNP methods, we take the shale oil production data in Figure \ref{fig:Shale_Oil_Production} researched by \citep{lee2021bayesian} as an example. Their goal was to predict EUR at a new location before the actual drilling takes place. Because the geological location is not stochastically incorporated into the basic model (\ref{eq:Stage 1: Individual-Level Model}) -- (\ref{eq:Stage 3: Prior}), authors extended the linear regression in Stage 2 into a spatial linear regression as follows
\begin{align*}
\theta_{li} &= \theta_{l}(\textbf{s}_{i})= \alpha_{l} + \textbf{x}_{i}^{\top} \bm{\beta}_{l} +
\upsilon_{l}(\textbf{s}_{i}) + \eta_{l}(\textbf{s}_{i}),  \quad (i=1,\cdots,N; l=1,\cdots,K),
\end{align*}
where $\eta_{l}(\cdot)\sim \mathcal{GP}(0,\omega_{l}^{2} I(\cdot,\cdot))$ represents a Gaussian white noise process with the indicator function $I(\cdot,\cdot)$ with variance $\omega_{l}^{2}$. The stochastic process $\upsilon_{l}(\cdot)\sim \mathcal{GP}(0,\mathcal{K}(\cdot,\cdot))$ is the newly introduced Gaussian process with a radial basis function kernel $\mathcal{K}(\textbf{s}_{i_{1}},\textbf{s}_{i_{2}})) = \gamma_{l}^{2} \exp [ - \|\textbf{s}_{i_{1}} - \textbf{s}_{i_{2}} \|_{2}^{2}/\{2 \rho_{l}^{2}\}]$ with variance $\gamma_{l}^{2}$ and range parameter $\rho_{l}^{2}$, and $\textbf{s}_{i}$ represents the (longitude, latitude) of the $i$-th shale oil well. The existence of $\upsilon_{l}(\cdot)$ enables spatial prediction of EUR at a new location, taking an advantage of the geological proximity information, which is called the latent kriging technique. 

An another motivation on the use BNP methods is the situation when there exists multimodality in the distribution $\mathcal{P}$ of model parameter vector $\{\bm{\theta}^{i}\}_{i=1}^{N} \sim \mathcal{P}$. Note that, in the basic model (\ref{eq:Stage 1: Individual-Level Model}) -- (\ref{eq:Stage 3: Prior}), the distribution $\mathcal{P}$ is assumed to be a single multivariate normal distribution $\mathcal{N}_{K}(\bm{\alpha} + \bm{B} \textbf{x}_{i} , \bm{\Omega})$. In the multimodality case, the population may consist of disparate subpopulations, and the single multivariate normal distribution of the basic model can produce a poor model performance due to the lack of flexibility. A natural generalization to accommodate such multimodality is an extension to a finite mixture of multivariate normal distributions \citep{mclachlan2019finite}, or furthermore, to a countably infinite number of mixtures of multivariate normal distributions \citep{rasmussen1999infinite}. Particularly, in the latter case, if a Dirichlet process prior \citep{ferguson1974prior} is placed on the mixture components, then the resulting infinite mixture models are generally called Dirichlet process mixture (DPM) model \citep{antoniak1974mixtures,teh2006hierarchical}. DPM model is one of the most studied topics in BNP methods in recent years \citep{hjort2010bayesian}. See \citep{jara2017theory} for a survey of the posterior computations of using DPM models. 

A number of authors have studied DPM models under the basic model (\ref{eq:Stage 1: Individual-Level Model}) -- (\ref{eq:Stage 3: Prior}) or similar forms with their own specifications \citep{muller1997bayesian,walker1998population,rosner1997bayesian,bush1996semiparametric}. For example, \citep{walker1998population} placed a DPM model only for the model parameter vector $\bm{\theta}^{i}$ ($i=1,\cdots,N$), while the covariates $\textbf{x}_{i}$ are incorporated into the base measure of Dirichlet process. In contrast, \citep{muller1997bayesian,rosner1997bayesian} used a DPM model jointly for the model parameter vector and covariates, $(\bm{\theta}^{i},\textbf{x}_{i})$ ($i=1,\cdots,N$), to induce a nonparametric regression function $\mathbb{E}[\bm{\theta}^{i}|\textbf{x}_{i}]$. \citep{bush1996semiparametric} used a Dirichlet process prior only for a certain component $\theta_{li}$ ($i=1,\cdots,N$) corresponding to a block indicator in an analysis-of-variance setup. Refer to the \citep{muller2004method} for a review and references therein for more specifications. 





\subsection{Software development}\label{subsec:Software development}
Recent years have seen the great success of Bayesian nonlinear mixed effects models, or more generally, Bayesian hierarchical models (BHM), in a variety of disciplines such as biology, medical research, physics, social, and educational sciences \citep{brown2015applied,lee2021optimization,lee2021bayesian,lee2020estimation,congdon2010applied}. This was partly due to the widespread introduction of non-commercial software packages that enabled applied researchers to answer substantive research questions through applications of BHM \citep{plummer2003jags,lunn2009bugs,margossiantorsten,stan,burkner2017brms,beal1992nonmem}. Most of the Bayesian software such as \textsc{JAGS} \citep{plummer2003jags}, \textsc{BUGS} \citep{lunn2009bugs}, and \textsc{Stan} \citep{stan} are designed to require a reasonable understanding of the MCMC sampling scheme. From the perspective of implementation, spirits of most Bayesian software are similar in that, researchers only need to designate a DAG structure \citep{jordan2004graphical,lauritzen1990independence} of a BHM. Such a DAG structure can be abstractly represented as the collection $\{$ data $y$ , likelihood $p(y|\theta)$, prior $\pi(\theta|\eta)$, hyperprior $\pi(\eta)$ $\}$ that should be programmed by textually or graphically, after which Bayesian software prints out simulated Markov chains from the posterior distribution $\pi(\theta,\eta|y)$. See \citep{lunn2000winbugs,lunn2009bugs} for an overall idea about how Bayesian software operates.

From the algorithmic perspective, the performance of Bayesian software may highly depend on two aspects: (i) whether the program has been designed to exploit a conditional independence structure arising from the hierarchy; and (ii) what sampling algorithms have been deployed to simulate Markov chains from a non-closed form distribution, possibly of high-dimensional. As discussed in Subsection \ref{subsec:Parallel computation for model matrix}, conditional independence is inherent in the formulation of BHM, of which proper exploitation can greatly improve the computational efficiency \citep{geman1984stochastic,lee2021gibbs}. This can be mostly done by constructing a Gibbs sampling algorithm with a blocking strategy into the consideration \citep{liu1994collapsed}. A general rule is that the convergence of the Gibbs sampler can be improved by grouping correlated latent variables as a single parameter block to sample from as a whole \citep{park2021improving}. On the other hand, in the task of sampling from a non-closed form distribution, we know that a naive MH algorithm \citep{metropolis1953equation} requires the specification of proposal density, which can be problematic in developing software. Therefore, fully automated sampling algorithms such as ESS (Algorithm \ref{alg:ESS}) \citep{murray2010elliptical}, NUTS \citep{hoffman2014no}, and slice sampler \citep{neal2003slice} are appreciated as general-purpose inference engines in the development of Bayesian software when conjugate-update is infeasible.

Most of the Bayesian software packages, for instance, \textsc{WinBUGS} \citep{lunn2000winbugs,spiegelhalter2003winbugs}, \textsc{OpenBUGS} \citep{spiegelhalter2007openbugs}, and \textsc{JAGS} \citep{plummer2003jags}, use three family of MCMC algorithms: Gibbs \citep{geman1984stochastic}, MH \citep{metropolis1953equation}, and slice sampling \citep{neal2003slice}. In contrast, \textsc{Stan} \citep{stan} implements HMC \citep{duane1987hybrid,neal2011mcmc} and its extension, NUTS \citep{hoffman2014no}. Perhaps, \textsc{Stan} is one of the most extensively used Bayesian software packages in recent years due to the fast converge of the inference engines regardless of whether the priors are conjugate or not. By that reason, and its great modeling flexibility, \textsc{Stan} has been used as a basic platform for other high-level packages like \textsc{brms} \citep{burkner2017brms} and \textsc{Torsten} \citep{margossiantorsten}.

\subsection{Future research topics}\label{subsec:Future research}
We briefly mention two topics that have generated great recent interest in the Bayesian statistical community. The first topic we want to bring out is the use of Bayesian optimization techniques, namely variational inference \citep{zhang2018advances,blei2017variational} and expectation propagation \citep{barthelme2014expectation,minka2013expectation}, for the Bayesian nonlinear mixed effects models. These methods received significant attention in the recent past because of their scalability to large-scale problems enabling `Big Bayesian Learning' \citep{zhu2017big,jordan2011message}. Essentially, the main goal of these methods is to approximate the joint posterior density (\ref{eq:joint_posterior}) via optimization, rather than via sampling such as MCMC
sampling which may cause scalability problems. The basic idea behind them is to first posit a family of densities and then to find a member of that family which is close to the target density, where the closeness is often measured by Kullback–Leibler divergence \citep{johnson2001symmetrizing}. See Chapter 10 in \citep{bishop2006pattern} for a general idea for the methods. Although there were published research works for a new algorithmic development of the methods for the application to (generalized) linear mixed effect models \citep{tan2013variational,ormerod2012gaussian,tan2014stochastic}, to our knowledge, there is no relevant published research for the application to nonlinear linear mixed effect models. 

Another topic untapped in the literature is the development of the Bayesian version of mixed effects machine learning models \citep{ngufor2019mixed,capitaine2021random,mandel2021neural,fu2015unbiased}. This is a relatively new branch in statistics and machine learning community, where most research works were published in the past five years. The central idea of the models is to estimate the nonlinear function $f$ in the individual-level model (\ref{eq:Stage 1: Individual-Level Model}) nonparametrically by using a verity of machine learning methods, rather than specifying a parametric function, while maintaining the mixed effects modeling framework. Therefore, the modeling framework will be similar to a nonparametric regression of which the primary purpose is to estimate the unknown function $f$ \citep{tsybakov2004introduction,rasmussen2003gaussian,schulz2018tutorial}. However, the main difference is that, in mixed effects machine learning models, (1) there exists some random variables to describe inter-subject variability, and (2) curve fitting mechanism (i.e., estimation of $f$) is mostly done by machine learning models, including deep learning \citep{goodfellow2016deep}, random forest \citep{breiman2001random}, gradient boosted machine \citep{friedman2001greedy}, etc. 
\section{Discussion}\label{sec:Discussion}
This review of Bayesian nonlinear mixed effects models is of necessity incomplete, as the literature is too vast to attempt even a moderate review. We have chosen to focus much of our attention on providing some of the most recent literature on the Bayesian analysis of the underlying basic model, with an emphasis on implementation of the model and introduction of recently developed prior distributions. We hope that this review can be read as a guideline to develop computational algorithms for complex and realistic Bayesian hierarchical nonlinear models. We wish that this review will offer readers familiar with frequentist analysis some pedagogical insight into the Bayesian approach, and provide those new to nonlinear mixed effects modeling a foundation of the implementation of Bayesian and frequentist computations for appreciating its idea and utility. We look forward to continuing methodological developments, software developments, and new applications of this rich class of models in industrial and academic research.

\bibliographystyle{chicago}
\baselineskip=10pt
\bibliography{BNONMEM_ref}
\end{document}